\pdfoutput=1
\documentclass[12pt]{article}
\usepackage{amsmath, amsthm, amssymb, amsfonts,mathrsfs}
\allowdisplaybreaks
\usepackage{graphicx}
\usepackage{enumerate}
\usepackage{natbib}
\usepackage{url} %
\usepackage{cleveref, bm, commath}
\usepackage[utf8]{inputenc}
\usepackage{bbm}
\usepackage{etoolbox}
\usepackage{natbib}
\usepackage{graphicx}
\usepackage{xcolor}
\usepackage{float}
\usepackage{authblk}
\usepackage{multirow}
\usepackage{caption}
\usepackage{mwe}
\usepackage{enumitem}
\usepackage{subfig}
\usepackage{transparent}
\usepackage{booktabs,siunitx}
\usepackage{xr}
\usepackage{threeparttable}
\usepackage[symbol]{footmisc}
\allowdisplaybreaks

\newcommand{\blind}{1}

\newcommand{\var}{\mathrm{Var}}
\newcommand{\cov}{\mathrm{Cov}}
\newcommand{\bmX}{\bm X}
\newcommand{\bmW}{\bm W}
\newcommand{\bmZ}{{Z}}
\def\bS{\bm \Sigma}

\def\ipw{\hat{\theta}_{jk}^{\rm \, ipw}}

\def\aipw{\hat{\theta}_{jk}^{\rm \, aipw}}
\def\sipw{\hat{\theta}_{jk}^{\rm \, sipw}}
\def\saipw{\hat{\theta}_{jk}^{\rm \, saipw}}

\def\var{{\rm Var}}
\def\cov{{\rm Cov}}

\newtheorem{theorem}{Theorem}

\newtheorem*{definition*}{Definition}
\newtheorem{assumption}{Assumption}
\newtheorem{corollary}{Corollary}

\begin{document}

\def\spacingset#1{\renewcommand{\baselinestretch}%
{#1}\small\normalsize} \spacingset{1}

\if1\blind
{
	 \begin{center} 
	\spacingset{1.7} 	{\LARGE\bf  
 From Estimands to Robust Inference of Treatment Effects in Master Protocol Trials
 } \\ \bigskip %
		\spacingset{1.3} 
		{\large Yuhan Qian$^1$, Yifan Yi$^{2}$, Jun Shao$^3$, Yanyao Yi$^4$, Gregory Levin$^5$\footnote{\scriptsize This article reflects the views of the author and should not be construed to represent FDA's views or policies}, \\ Nicole Mayer-Hamblett$^{6,7,1}$,  Patrick J. Heagerty$^1$, Ting Ye$^1$\footnote[2]{\scriptsize  Correspond to tingye1@uw.edu.}}  \\ \medskip
	 {\spacingset{1.1} $^1$Department of Biostatistics, University of Washington\\
$^2$Department of Biostatistics, The University of Texas MD Anderson Cancer Center\\
$^3$Department of Statistics, University of Wisconsin-Madison\\
$^4$Global Statistical Sciences, Eli Lilly and Company \\
$^5$Food and Drug Administration \\
$^6$Seattle Children’s Research
Institute \\
$^7$Department of Pediatrics, University of Washington
    }
	\end{center}
} \fi

\if0\blind
{
  \bigskip
  \bigskip
  \begin{center}
    \spacingset{1.7}
    {\Large\bf   From Estimands to Robust Inference of Treatment Effects in Master Protocol Trials}
\end{center}
  \medskip
} \fi

\begin{abstract}
    \spacingset{1}
    Master protocol trials use a single overarching protocol to evaluate multiple interventions, diseases, or disease subtypes, where individuals are often randomized to different subsets of intervention arms based on individual characteristics, enrollment timing, and intervention availability. While offering increased flexibility, this \emph{constrained} and \emph{non-uniform} intervention assignment poses two fundamental inferential challenges:  the precise definition of treatment effects and robust, efficient inference on these effects. {These challenges arise primarily because some commonly used analysis approaches may target estimands defined on populations that inadvertently depend on the intervention allocation ratio, making them impossible to fully pre-specify,  thereby undermining interpretability and opening the door to ambiguity, post-hoc decisions, and potential bias.} This article, for the first time, presents a formal estimand framework for master protocol trials with precise specification of the population. The proposed entire concurrently eligible (ECE) trial population not only preserves the integrity of randomized comparisons but also remains invariant to the randomization ratio.
    Then, we develop weighting and post-stratification methods to estimate treatment effects under the same minimal assumptions used in traditional randomized trials. We also consider model-assisted covariate adjustment to fully unlock the efficiency potential of master protocol trials while maintaining robustness against model misspecification.
    The SIMPLIFY trial, a master protocol assessing continuation versus discontinuation of two common therapies in cystic fibrosis, is utilized to highlight the practical significance of this research. All analyses are conducted using the \textsf{R} package \textsf{RobinCID}. 
\end{abstract}

\spacingset{1} 
\noindent%
\vfill

\newpage
\spacingset{1.7} %

\section{Introduction}
\label{sec: intro}
\vspace{-6mm}

\subsection{Background} \vspace{-3mm}
The traditional paradigm of conducting a separate clinical trial for every investigational intervention has become ever more expensive and challenging \citep{woodcock2017master}. As a result, the landscape of clinical trials has been evolving, 
to make clinical trials faster and more efficient while continuing to provide reliable information on safety and effectiveness  \citep{ostp}. 

Master protocol trials have emerged as innovative designs to accelerate the evaluation of potentially effective interventions \citep{woodcock2017master}. These trials evaluate one or more interventions across one or more diseases within an overarching study structure. Three major types of master protocol designs are umbrella trials, basket trials, and platform trials. Umbrella trials evaluate multiple therapies for a single disease \citep{rashdan2016going, renfro2017statistical}, basket trials evaluate a medical product for multiple diseases or disease subtypes \citep{redig2015basket, hobbs2022basket},  
while platform trials offer the added flexibility of adding or removing intervention arms over time \citep{berry2015platform, saville2016efficiencies, park2020overview, gold2022platform, burki2023platform, pitre2023methodology}. These three types are not mutually exclusive, and a master protocol trial can incorporate features of more than one type.

Master protocols originated in oncology and remain widely used in this field, where many new biomarker-driven therapies require efficient evaluation \citep{park2019systematic,park2020overview,ouma2022design}. By utilizing a common screening platform to efficiently screen for biomarkers and identify all interventions for which a patient is eligible, these trial designs significantly improve operational efficiency, accelerate patient accrual, and enable the study of rare tumor subtypes within one trial \citep{woodcock2017master}. In addition to streamlining operations, these designs also have the potential to improve statistical efficiency by facilitating data sharing across
different intervention evaluations. Notable examples include the BATTLE umbrella trials \citep{kim2011battle, papadimitrakopoulou2016battle}, the 
Lung-MAP platform trial \citep{herbst2015lung}, and the I-SPY 2 platform trial \citep{barker2009spy}. The utility of master protocol trials is much broader. They are gaining prominence in other diseases areas, including infectious diseases \citep{normand2021recovery}, neurodegenerative diseases such as amyotrophic lateral sclerosis 
\citep{quintana2023design} and Alzheimer’s disease \citep{bateman2017dian}, as well as inflammatory bowel disease \citep{honap2024basket}. Section 2 presents two trials to provide concrete illustrations. \vspace{-8mm}

\subsection{Two fundamental gaps} \vspace{-3mm}
In master protocol trials, interventions often have different eligibility criteria and can be added or removed over time, so the set of interventions available to any given individual and the corresponding assignment probabilities may vary across enrollment time and individual characteristics. This gives rise to \emph{constrained} and \emph{non-uniform} intervention assignments, which marks a key distinction from traditional parallel-group trials and poses significant challenges for both the interpretability of and statistical inference for the treatment effect, some of which are highlighted in a recent FDA draft guidance \citep{fda:2023platform}. Although master protocols are increasingly conducted, two fundamental questions remain unresolved.

\vspace{1mm}

\noindent \emph{1. How to define a clinically meaningful estimand?} \\
An estimand is ``a precise description of the treatment effect reflecting the clinical question posed by a given clinical trials objective'' and is of ultimate importance \citep{ICHE9-R1}. Defining an estimand in master protocol trials is more challenging than in traditional trials, as the treatment effect must be interpreted in the context of restricted eligibility, variable assignment probabilities, and evolving trial structure over time. In particular, among the five estimand attributes (\cite{ICHE9-R1}), the population requires special attention. Many researchers \citep{lee2020including, dodd2021platform} and \cite{fda:2023platform} have recommended using a population of individuals who are concurrently randomized and meet the eligibility criteria, as this preserves the integrity of randomized comparisons and avoids systematic differences between intervention arms due to measured and unmeasured factors.
However, this general recommendation does not uniquely define the population. For instance, Section 3.2 shows that a seemingly reasonable population (and thus the estimand) may depend arbitrarily on randomization ratios and can be problematic, even though it technically follows the general recommendation in \cite{fda:2023platform}. %

To address this question, Section 3.2  introduces a population for comparing two interventions (typically with one as a control) that not only follows the general recommendations in \cite{fda:2023platform} but also respects the eligibility criteria while avoiding arbitrary dependencies on randomization ratios. We refer to this as the \emph{entire concurrently eligible (ECE) trial population}. %
The corresponding estimand is defined in Section 3.3. Our proposal is the first in the literature to offer a systematic framework for defining clinically meaningful estimands with clear specification of the population of interest. 

\vspace{1mm}

\noindent
  \emph{2. How to provide robust estimation and inference regarding the defined treatment effect?} \\
  Although randomization is preserved, the standard estimator used in traditional trials—comparing outcome means between intervention arms—can produce biased estimates in master protocol trials, because intervention assignment is often non-uniform, creating confounding that can lead to issues such as Simpson's paradox \citep{collignon2020current, dodd2021platform, lee2021statistical, meyer2021systematic}. Therefore, master protocol trials require specialized statistical approaches for unbiased treatment effect estimation.

Two main approaches can address this bias and provide robust treatment effect estimation: inverse probability weighting and post-stratification.
Crucially, because trials use known randomization probabilities, these methods achieve full robustness without relying on modeling assumptions, a key advantage over observational studies. Model-assisted covariate adjustment provides an additional strategy to unlock the efficiency potential of master protocol trials, while maintaining robustness against model misspecification. To guide practitioners in selecting among these methods, we derive the asymptotic variance and relative efficiency of these estimation methods.

\vspace{-8mm}
\subsection{Prior work}
\vspace{-4mm}

The increasing popularity of platform trials has led to a surge of research addressing their
statistical challenges in both design and analysis. Many studies have focused on adaptive designs, which aim to efficiently screen candidate interventions and adaptively randomize individuals to the most promising interventions \citep{yuan2016midas,kaizer2018multi, ventz2018adding}. Additionally, \cite{bofill2024optimal} examined the optimal allocation ratios in these trials. A key concern in the analysis of platform trials is the  potential presence of time trends, which can introduce bias when using nonconcurrent controls \citep{bofill2023use}. To address this, various strategies have been proposed to account for time trends in the analysis \citep{elm_flexible_2012, saville_bayesian_2022,roig2022model, jiang2023nonconcurrent, wang_bayesian_2023, guo2024treatment}. However, all of these statistical methods have predominantly relied on correct modeling and are sensitive to model misspecification.

The platform trial literature on non-model-based approaches for concurrently eligible individuals remains limited.
\cite{marschner2022analysis} proposed stratification by enrollment windows. \cite{huang2023improved} developed inference methods for time-to-event outcomes under a strong constancy assumption that, conditional on a discrete baseline covariate, the efficacy of interventions is invariant across enrollment windows. 
More recently, \cite{santacatterina2025identification} defined an estimand based on the concurrent period and established identification and estimation results using only concurrent controls. They also proposed strategies to incorporate nonconcurrent controls under a strong conditional exchangeability assumption between concurrent and nonconcurrent controls, given observed covariates.

Related lines of work have emerged in basket and umbrella trials. In basket trials, many studies have emphasized adaptive designs \citep{cunanan2017efficient} and Bayesian approaches to borrowing information across molecularly or clinically defined subgroups \citep{chu2018bayesian, jin2020bayesian}. In contrast, the statistical literature on umbrella trials remains sparse \citep{ouma2022design}.

Despite these advances, existing studies do not consider the broader master protocol trial setting where interventions may have distinct eligibility criteria and availabilities. Nor do they provide a rigorous framework for defining clinically meaningful estimands within a general setting that is independent of specific study designs or modeling assumptions.

\vspace{-8mm}
\subsection{Outline of the article}
\vspace{-3mm}

 Section 2 presents the SIMPLIFY trial and a
 stylized trial used to emphasize key issues. Section 3 details the ECE trial population and corresponding estimand. In Section 4 and \ref{sec: theory}, we develop robust estimation methods without or with covariate adjustment, and establish the asymptotic distributions and relative efficiency of the proposed  estimators, respectively. We highlight the practical significance through the SIMPLIFY trial in Section \ref{sec: real data} and conclude with a summary and recommendations in Section \ref{sec: discussion}.

\vspace{-10mm}
\section{Real and Stylized Trials}
\vspace{-6mm}

\subsection{The SIMPLIFY trial}\label{subsec: simplify}
\vspace{-3mm}

 We begin by presenting the real master protocol trial that motivates this work: the SIMPLIFY trial \citep{mayer2023discontinuation}. Despite its seemingly simple design, the trial involves both constrained and non-uniform intervention assignments, making it ideal for highlighting the two fundamental gaps and the underlying challenges.

Cystic fibrosis (CF) is caused by mutations in the CF transmembrane conductance regulator (CFTR) gene. CFTR modulator therapies are a class of small-molecule drugs designed to address the underlying cause of CF,  leading to transformative improvements in patient outcomes. A landmark example is the triple-combination therapy elexacaftor/tezacaftor/ivacaftor (ETI),  approved for individuals with at least one F508del allele, a mutation present in over 85\% of people with CF. Given the potential overlap in therapeutic effects between ETI and pre-existing treatments such as hypertonic saline (HS) and dornase alfa (DA) and the need to reduce treatment burden, it is important to evaluate the impact of discontinuing these pre-existing therapies in individuals taking ETI. The SIMPLIFY trial was conducted within this context. It  includes two concurrently run multi-center randomized controlled studies at 80 participating clinics across the United States in the CF Therapeutics Development Network. The primary objective for each study is to determine whether discontinuing either HS or DA, respectively, is non-inferior to continuing therapy in individuals using the
highly eﬀective modulator ETI. The outcome is the 6-week mean absolute change in percent-predicted forced expiratory volume in one second (ppFEV$_1$).

\begin{figure}[!h]
\spacingset{1.2} 
    \centering
    \includegraphics[scale=.65]{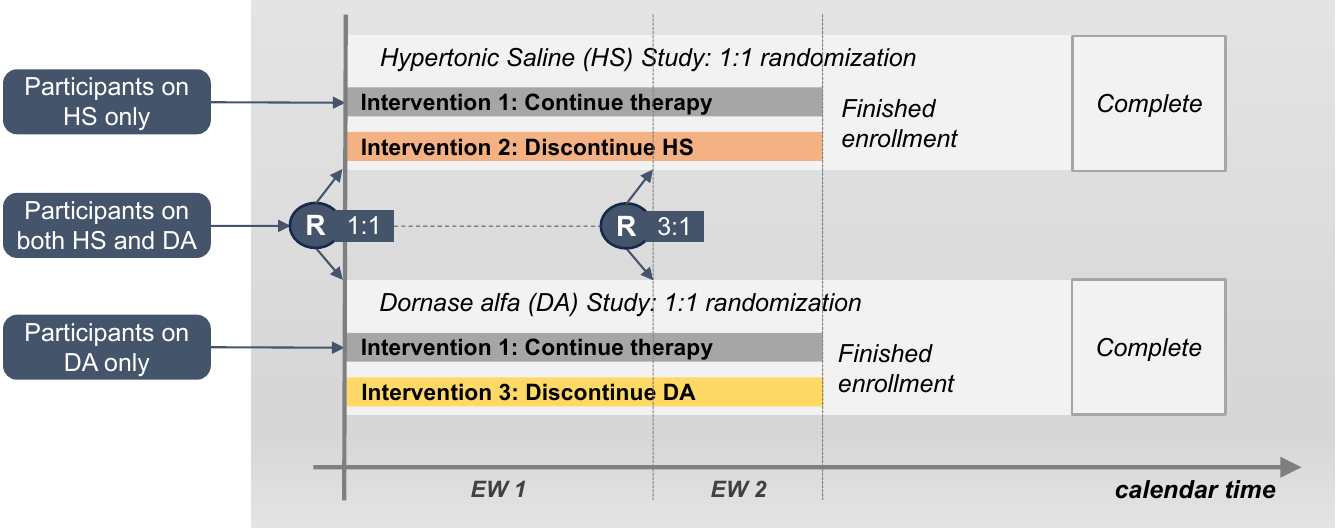}
     \caption{\label{fig: simplify} SIMPLIFY design schema. Individuals on a single therapy enter into the corresponding study with probability 1, while those on both are randomized to the HS or DA study, with ratio of 1:1 during enrollment window (EW) 1, then 3:1 during EW 2. Within HS (or DA) study, individuals are randomized 1:1 to continue or discontinue HS (or DA).}
\end{figure}

 Figure \ref{fig: simplify} depicts the SIMPLIFY study design. All enrolled individuals have initiated ETI for at least 90 days and belong to one of three groups based on medication use when entering the trial: those on HS only, DA only, or both therapies. Individuals on a single therapy (HS only or DA only) are assigned to the corresponding study with probability 1. Those on both therapies are randomized to the HS or DA study, with 1:1 ratio during initial enrollment window 1 (before January 14, 2022), and 3:1 ratio during enrollment window 2 (after January 14, 2022) to assign more individuals to the HS study, as the DA study was enrolling more quickly. This swift shift in the randomization ratio exemplifies how the flexibility of master protocol trials can accelerate clinical trials. Within each study (HS or DA), individuals were then randomized 1:1 to either continue or discontinue their respective therapy. Table \ref{table: prob simplify} summarizes assignment probabilities of the SIMPLIFY trial. It illustrates both constrained and non-uniform intervention assignment: individuals receiving only HS or DA have zero probability of being assigned to certain intervention arms, and the non-zero probabilities of receiving any given pair of interventions may vary based on baseline characteristics and enrollment windows.

\begin{table}[h]
\spacingset{1.2} 
\centering
\caption{Assignment probabilities in the SIMPLIFY trial. Intervention 1: continue therapy; Intervention 2: discontinue HS; Intervention 3: discontinue DA. 
\label{table: prob simplify}}
{\small  \begin{tabular}{cccccccccc}
\hline
\multicolumn{2}{c}{Baseline variables} & EW && \multicolumn{2}{c}{Studies} &  & \multicolumn{3}{c}{Interventions}                 \\ \cline{1-3} \cline{5-6} \cline{8-10} 
HS     & DA     & EW    &  & HS study         & DA   study       &  & 1  & 2  &3\\ \cline{1-3} \cline{5-10} 
1      & 1      & EW1        &  & 0.50        & 0.50        &  & 0.50          & 0.25          & 0.25          \\
1      & 1      & EW2        &  & 0.75        & 0.25        &  & 0.50          & 0.375         & 0.125         \\
1      & 0      & any      &  & 1           & 0           &  & 0.50          & 0.50          & 0             \\
0      & 1      & any      &  & 0           & 1           &  & 0.50          & 0             & 0.50          \\ \hline
\end{tabular}}
\end{table}

\vspace{-8mm}
\subsection{A stylized trial}
\label{subsec: example}
\vspace{-4mm}

\begin{figure}[!h]
\spacingset{1.2} 
    \centering
    \includegraphics[width=\textwidth]{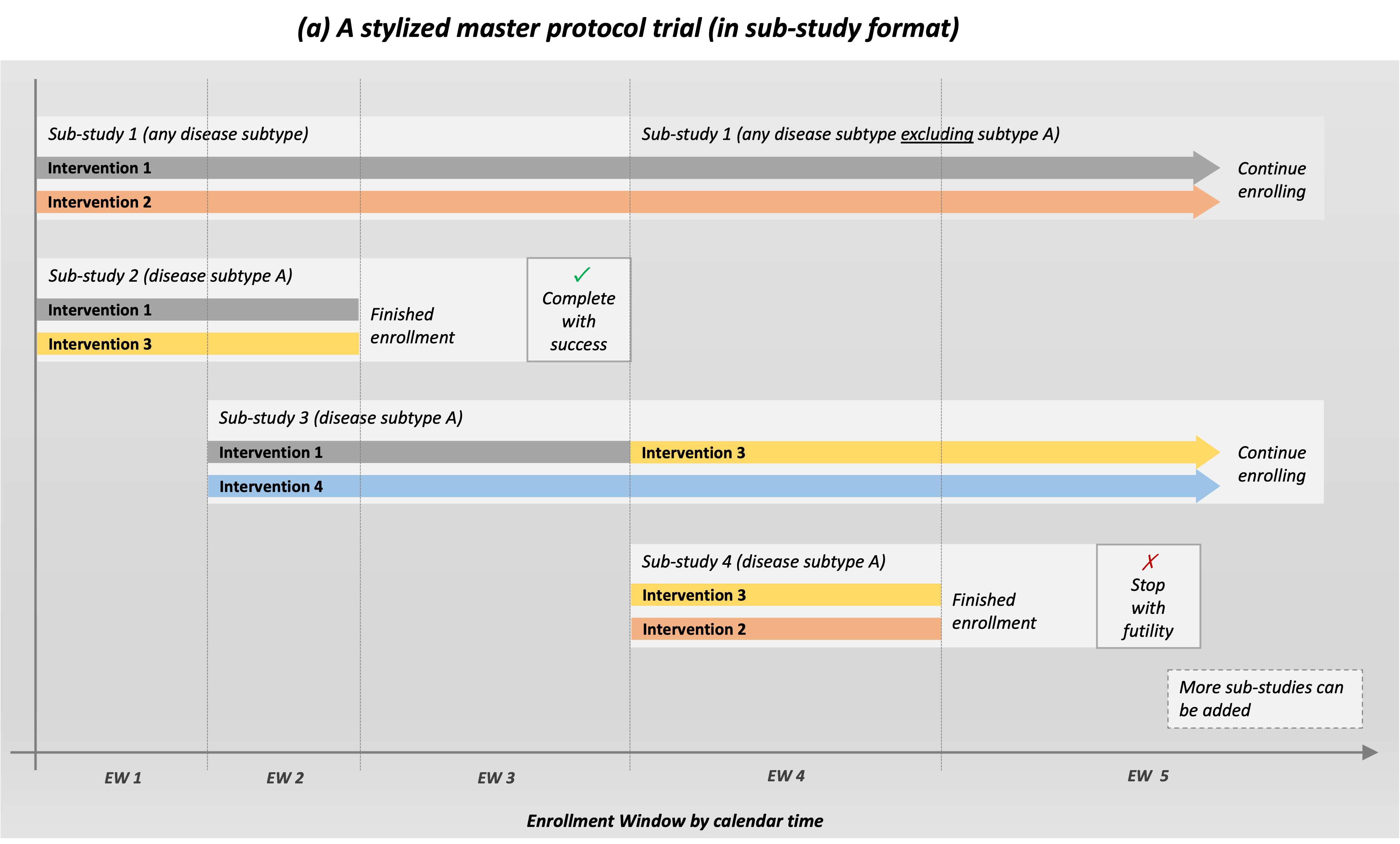}
    \includegraphics[width=\textwidth]{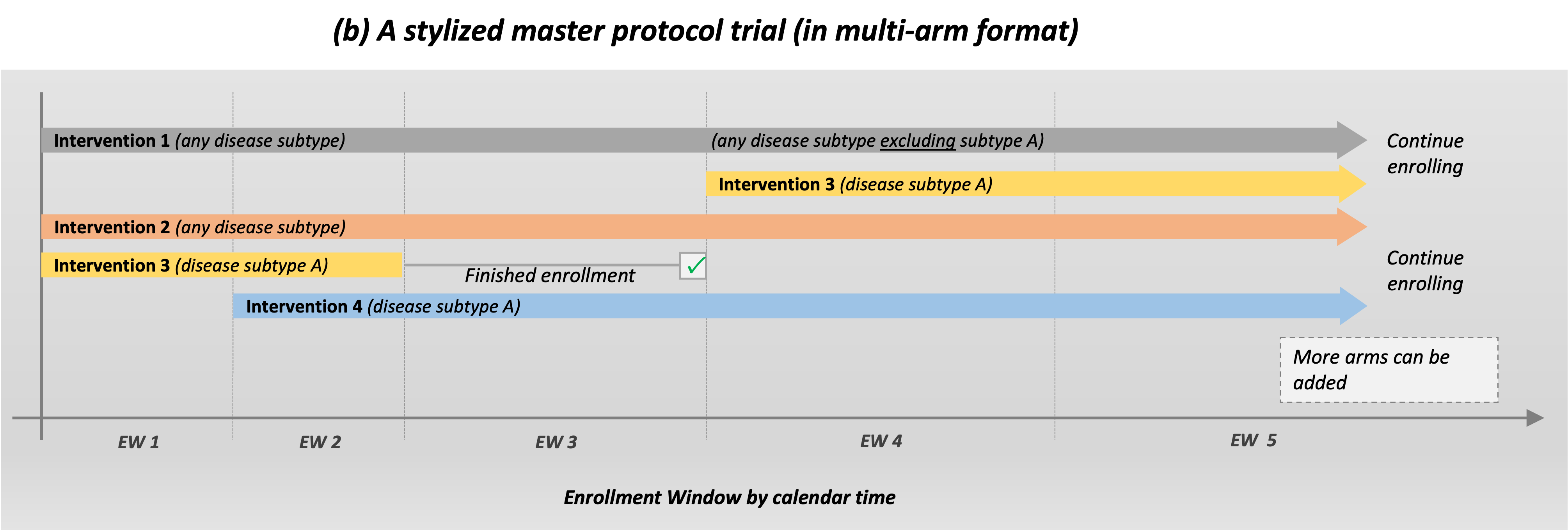}
    \caption{A stylized master protocol trial in two operating formats:
    (a) sub-study format, (b) multi-arm format. Each box in (a) represents a sub-study, along with its eligibility criteria and status.
    Each striped block within a box in (a) or standalone in (b) represents the enrollment period for an intervention evaluated.
    The horizontal axis shows calendar time, divided into discrete windows based on changes in available interventions for enrollment.} 
    \label{fig: schema}
\end{figure}

To motivate a general statistical framework for master protocol trials, we introduce a stylized trial in two operational formats, both can be unified under our proposed framework. %
Figure \ref{fig: schema}(a) presents the stylized trial in a sub-study format, where interventions are evaluated in separate sub-studies. Figure \ref{fig: schema}(b) shows a multi-arm format, where each intervention is studied in a separate arm without dividing the trial into sub-studies. 

The master protocol trial (in sub-study format) in Figure \ref{fig: schema}(a) comprises four
sub-studies,  each containing a randomized controlled comparison of two interventions. Initiating or stopping a sub-study is governed by a master protocol. Additionally, each sub-study has an intervention-specific appendix (ISA) outlining protocol elements specific to that sub-study. 
 Individuals are randomized first to eligible sub-studies and subsequently to intervention arms within a sub-study, which gives explicit intervention allocation probabilities for every subject,  sufficient for the identification of treatment effects and statistical analysis. 
Real trials resembling this structure include Lung-MAP for lung cancer \citep{herbst2015lung,redman2020biomarker} and an ongoing trial for chronic pain (ClinicalTrials.gov, NCT05986292). A narrative of 
Figure \ref{fig: schema}(a) is included in Section S1 of the Supplement.

In Figure \ref{fig: schema}(b), we reframe the same master protocol trial into a multi-arm format. In enrollment window 1, there are three initial intervention arms: investigational interventions 2 and 3 and intervention 1 serving as the common control. In this window, individuals with subtype A can be randomized to any of the three arms, while those with subtypes other than A can only be randomized to interventions 1 or 2. Descriptions for subsequent enrollment windows are omitted due to their similarity with Figure \ref{fig: schema}(a). Examples of real trials that follow this structure include the STAMPEDE trial for prostate cancer \citep{sydes2012flexible} and the RECOVERY trial for COVID-19 therapies \citep{normand2021recovery}.

\vspace{-8mm}
\section{Estimand}
\label{sec: setup}

\vspace{-6mm}
\subsection{Randomization procedure}
\label{subsec: general setup}

\vspace{-3mm}

Consider a master protocol trial with $J$ interventions. For each individual, let $\bmW$ be a vector of all observed baseline variables (e.g., baseline covariates and enrollment window by calendar time) not affected by the intervention,  $A$ be the intervention arm assignment that equals $j$ if the individual is assigned to arm $j$, 
 $Y^{(j)}$ be the potential outcome 
under intervention $j $, $j=1,...,J$ \citep{Neyman:1923a,Rubin:1974},  and $Y$ be the observed outcome. Throughout, $Y$ denotes a continuous or discrete outcome measured at a fixed follow-up time; time-to-event outcomes are outside the current scope and will be considered in future work. We make the consistency assumption, that is, $Y=Y^{(A)}$. We assume throughout that  $(\bmW_i, Y_i^{(1)}, \dots, Y_i^{(J)},  A_i)$, $i=1,\dots, n$, is an  independent and identically distributed sample of size $n$ from $(\bmW, Y^{(1)}, \dots, Y^{(J)}, A)$ with finite second-order moments. The following assumption describes the randomization procedure in master protocol trials. \vspace{-2mm}

\begin{assumption}[Constrained and non-uniform randomization] \label{assump: randomization}
   There exists an observed baseline variable $\bmZ $  such  that $    A\perp (\bmW, Y^{(j)}) \mid \bmZ $ for all $j$, and $ P(A = j\mid \bmZ )  = \pi_{j}(\bmZ), $
   where $\perp$ stands for independence, $\mid$ stands for conditioning,   $0\leq \pi_{j} (\bmZ) \leq 1 $ with known $\pi_j(\cdot)$ for $j=1,...,J$, and $\sum_{j=1}^J \pi_{j}(\bmZ)=1$.
\end{assumption}  \vspace{-2mm}

 Assumption \ref{assump: randomization}  captures two key features of randomization in master protocol trials and holds by design. First, conditional on $\bmZ$, intervention assignment is independent of all baseline variables and potential outcomes, which is the standard condition that justifies causal inference in randomized trials. Second, the intervention assignment probabilities $\pi_j(\bmZ)$ depend on $\bmZ$ and may differ across individuals. This contrasts with traditional trials with simple randomization, where randomization probabilities are fixed and equal across individuals. In master protocol trials, $\bmZ$ is usually discrete and may include enrollment window (defined by calendar time), site, and individual characteristics. $\pi_j(\bmZ)$ may even be zero for some interventions. For example, if intervention $j$ is not yet available, has closed enrollment, is not offered at a site, or the individual is ineligible. 

 Figure \ref{fig: simplify}
 and Table \ref{table: prob simplify} illustrate these dynamics in the SIMPLIFY trial, where $Z = (Z_{\rm EW}, Z_{\rm HS}, Z_{\rm DA})$ includes the enrollment window ($Z_{\rm EW}$: EW1 or EW2) and baseline use of HS and DA ($Z_{\rm HS}, Z_{\rm DA}$: 0 = no, 1 = yes). For example, an individual $i$ using only HS is always assigned to the HS study regardless of enrollment window and then randomized equally to  1 or 2, yielding $\pi_{1}(\bmZ_i)= 0.5$, $\pi_{2}(\bmZ_i) = 0.5$, $\pi_{3}(\bmZ_i) = 0$. In contrast, an individual $i'$ using both HS and DA who enters during EW1 ($\bmZ_{i'} = (\text{EW1}, 1, 1)$) is first randomized to the HS or DA study with equal probability. If assigned to the HS study, the individual is further randomized to Intervention 1 or 2; if assigned to the DA study, to Intervention 1 or 3. Therefore, $\pi_{1}(\bmZ_{i'})= 0.5\times 0.5+0.5\times 0.5= 0.5$, $\pi_{2}(\bmZ_{i'}) = 0.5\times 0.5=0.25$,  $\pi_{3}(\bmZ_{i'}) = 0.5\times 0.5=0.25$.

\vspace{-5mm}
\subsection{Population}
\label{subsec: population}
\vspace{-3mm}

Building on the discussion in Section 1.2, we now formally define the population for comparing two interventions  $j$ and $k$  (one of which is typically a control). In this paper, population specifically refers to the population of individuals used to define the estimand.

\vspace{-2mm}
\begin{definition*}[ECE trial population]
 For the given interventions $j$ and $k$, 
the entire concurrently eligible (ECE) trial population is a population of all individuals who meet the eligibility criteria for both interventions and could potentially be enrolled during a time period when both interventions are available, and therefore could have been randomized to either intervention. Formally, the ECE trial population is a population of all individuals with $\pi_j(\bmZ) >0 $ and $ \pi_k(\bmZ)>0 $. 
\end{definition*} 
\vspace{-2mm}

For each pair of interventions $(j,k)$ under consideration, we assume $
P\{\pi_j(\bmZ)>0,\ \pi_k(\bmZ)>0\}>0,$
which guarantees that the ECE trial population for that pair has positive probability and that the corresponding estimand is well defined. Throughout, we
consider a finite set of categorical intervention arms; extending the framework to
continuous or dose-valued interventions is beyond the scope of this paper.

{The ECE trial population extends the standard entire trial population concept to the master protocol setting. It is the entire trial population with additional restrictions on concurrency and eligibility to preserve the integrity of randomized comparison, and can therefore be understood as the subpopulation of the entire trial population for whom both arms under comparison are accessible. Crucially, it represents the population of individuals who could \emph{potentially} be assigned to interventions $j$ and $k$, rather than those \emph{actually} assigned to these interventions (i.e., the population represented by $A \in \{j, k\}$). In traditional trials, this distinction is often overlooked because the potential and actual populations coincide. Master protocol trials, however, bring this issue to the forefront.

Despite its importance, many master protocol trial analyses fail to consider the population holistically.  Instead, they simply consider a population represented by individuals actually assigned to the interventions of interest (in the multi-arm format) or to the specific sub-study containing those interventions (in the sub-study format). While commonly implemented in practice, this approach ties the population to the realized intervention assignments rather than to pre-randomization characteristics, thus yields estimands that depend arbitrarily on randomization ratios. Moreover, since randomization ratios may change during the trial, for example when a new therapy is added to the platform, such a population cannot be fully pre-specified, leaving the estimand vulnerable to change after the trial has begun and opening the door to post-hoc manipulation and potential bias.}

We illustrate through two basic cases: Case I examines how time-varying randomization ratios complicate population definition when all individuals are eligible for all interventions, while Case II examines how varying eligibility criteria raise similar issues when interventions enroll concurrently. We then revisit the SIMPLIFY trial.

\begin{figure}[h]
\spacingset{1.2} 
    \centering
        \includegraphics[width=\textwidth]{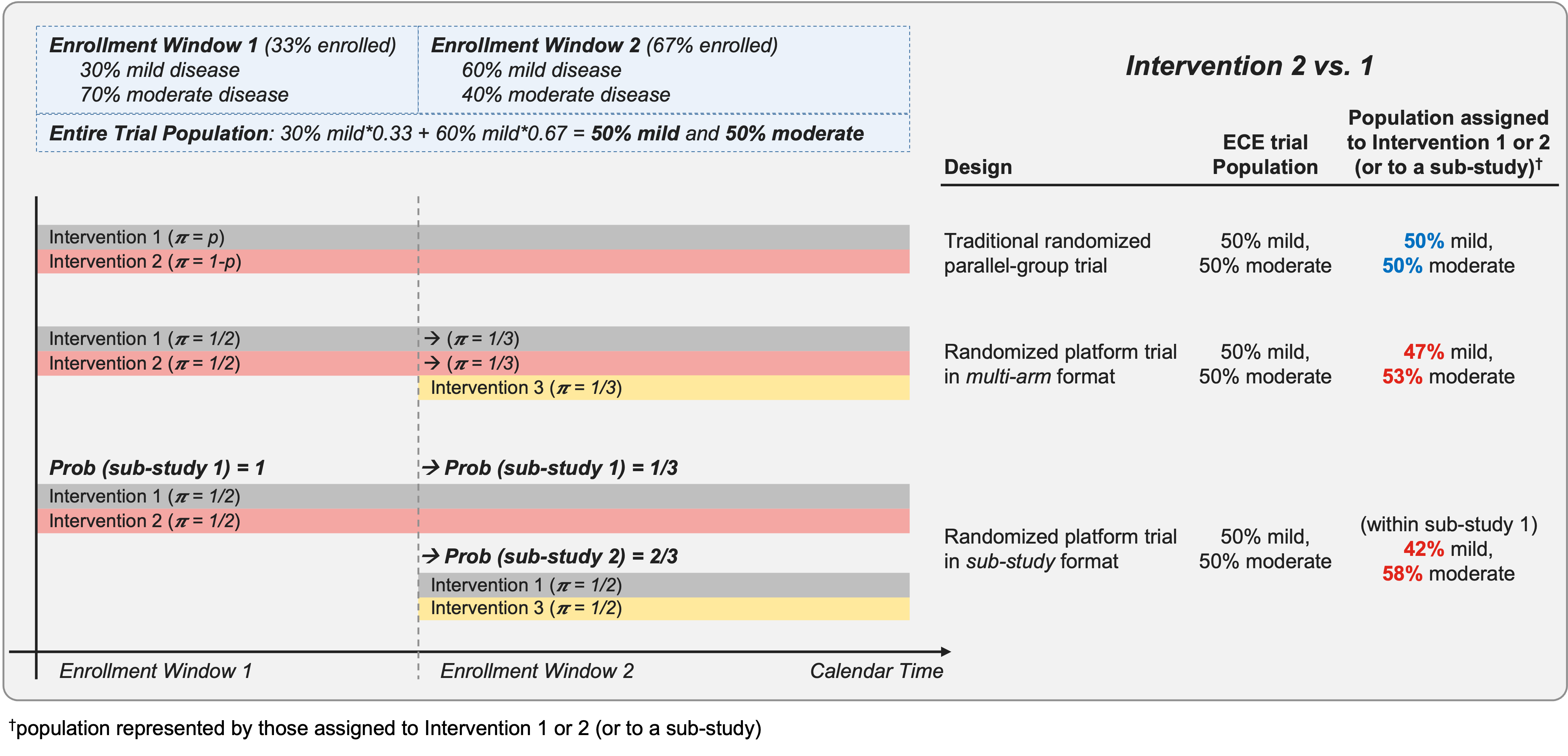}
    \caption{Comparison of traditional and platform trial designs in a hypothetical mild to moderate disease population (Case I). {Since Interventions 1 and 2 enroll individuals throughout the entire enrollment period and all individuals are eligible for both interventions, the ECE trial population for comparing these two interventions is the same as the entire trial population.} In contrast, the population represented by those assigned to specific interventions or sub-studies can vary arbitrarily depending on randomization ratios.}
    \label{fig: example table}
\end{figure}

{{\bf Case I} in Figure \ref{fig: example table} shows various trial designs evaluating Interventions 1 and 2, with all individuals eligible for all interventions. The entire trial population is 50\% mild and 50\% moderate, arising naturally from the enrollment composition across the two time windows: enrollment window 1 represents one third of the population with  30\% mild and 70\% moderate, while enrollment window 2 comprises the remaining two thirds with 60\% mild and 40\% moderate. Since Interventions 1 and 2 are available throughout the entire enrollment period and all individuals are eligible for both, the ECE trial population is the same as the entire trial population. 

In contrast, the population actually assigned to Intervention 1 or 2 coincides with the ECE trial population only in the traditional parallel-group trial. In platform trials, the assigned population depends on randomization ratios and shifts arbitrarily for reasons unrelated to the scientific question. In the multi-arm format, the addition of Intervention 3 with $\pi = 1/3$ in enrollment window 2 means that only $2/3$ of individuals in that window are assigned to Intervention 1 or 2, shifting the assigned population to 47\% mild and 53\% moderate. In the sub-study format, a proportion of individuals in enrollment window 2 is assigned to sub-study 2, distorting the composition of sub-study 1 to 42\% mild and 58\% moderate. These shifts arise purely from change in randomization ratios. Since it is generally unknown at the outset when new interventions will be added and how they will affect randomization ratios, the composition of the assigned population cannot be fully pre-specified, leaving the population and estimand vulnerable to change after the trial has begun and opening the door to post-hoc manipulation and potential bias.}

\begin{figure}[ht]
\spacingset{1.2} 
    \centering        \includegraphics[width=\textwidth]{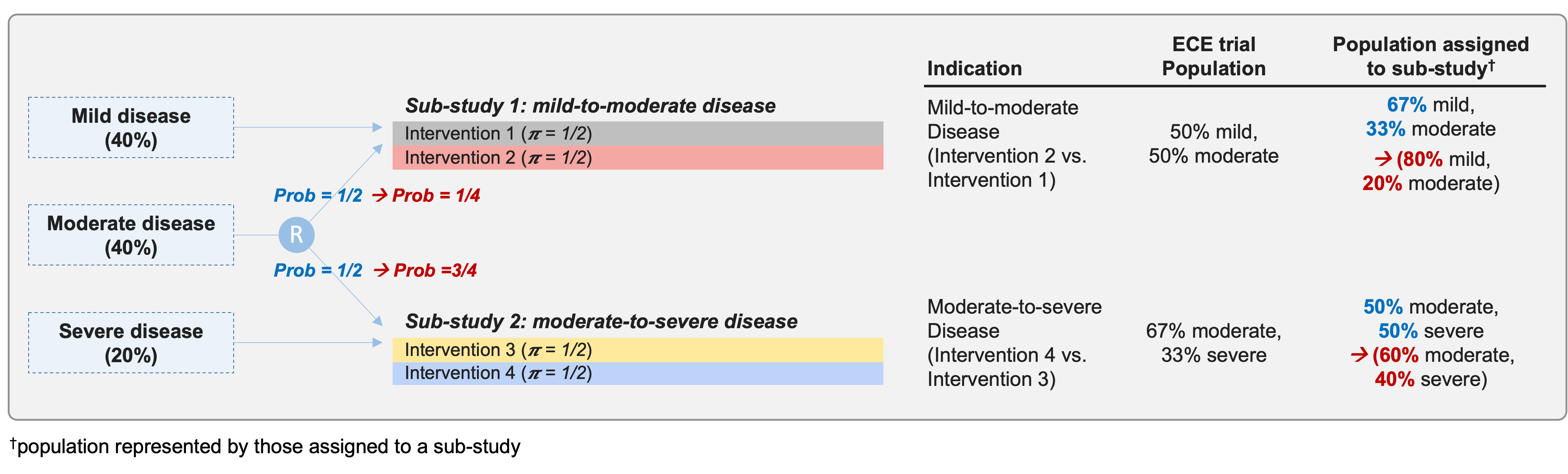}
    \caption{
    A hypothetical umbrella trial (Case II) enrolling ulcerative colitis patients of any disease severity, naturally comprising 40\% mild, 40\% moderate, and 20\% severe disease. The ECE trial population for comparing Interventions 2 and 1 includes all mild-to-moderate disease patients (50\% mild and 50\% moderate), reflecting the natural patient mix within this disease group and representing a clinically meaningful population for this comparison. In contrast, the population represented by those actually assigned to each sub-study depends arbitrarily on how moderate disease patients are allocated between sub-studies.} \label{fig: case 2}
\end{figure}

\textbf{Case II} in Figure \ref{fig: case 2} considers an umbrella trial for ulcerative colitis in which all interventions enroll concurrently. As it is standard in ulcerative colitis to develop interventions separately for mild-to-moderate and moderate-to-severe disease \citep{us2022ulcerative}, the trial includes two sub-studies: Sub-study 1 evaluates Intervention 2 against Intervention 1 in mild-to-moderate disease, and Sub-study 2 evaluates Intervention 4 against Intervention 3 in moderate-to-severe disease. Patients with moderate disease, who are eligible for both sub-studies, are randomized between them with a 1:1 allocation ratio.

Suppose the umbrella trial enrolls ulcerative colitis patients of any disease severity, naturally yielding 40\% mild, 40\% moderate, and 20\% severe disease, reflecting the underlying distribution among patients presenting to the trial sites. The ECE trial population for comparing Interventions 1 and 2 includes all individuals with mild or moderate disease, since only they have positive probability of being assigned to either intervention, yielding a composition of 50\% mild and 50\% moderate that exactly reflects the natural patient mix within this disease group.
In contrast, the population represented by those actually assigned to Sub-study 1 depends arbitrarily on the allocation ratio between sub-studies. For example, if the allocation probability to Sub-study 1 for moderate disease patients decreases from 1/2 to 1/4, the composition of Sub-study 1 shifts from 67\% mild and 33\% moderate to 80\% mild and 20\% moderate. This shift may be driven purely by logistical or statistical efficiency considerations, and has no bearing on the scientific question of interest, yet it materially changes the population being studied.

As introduced in Section 2.1, the SIMPLIFY trial has two sub-studies: one compares Interventions 1 and 2 in patients on HS only or on both HS and DA; the other compares Interventions 1 and 3 in patients on DA only or on both HS and DA. Patients on both HS and DA are first randomized between the two sub-studies using a 1:1 ratio, which was later adjusted to 3:1. The ECE trial population for comparing Interventions 1 and 2 includes all individuals on HS only or on both HS and DA, since only they have positive probability of being assigned to either intervention. This composition is determined entirely by the eligibility criteria and remains invariant to the sub-study allocation ratio. In contrast, the population actually assigned to the HS sub-study weights the HS-only and HS+DA groups in proportions determined by the sub-study randomization ratio and the timing of any changes to that ratio.

\vspace{-7mm}
\subsection{Treatment effect for the ECE trial population}

\vspace{-3mm}

With the population now clearly defined, we propose the following estimand to define the treatment effect for two given interventions $j$ and $k$ within the ECE trial population: 
\begin{equation}
{\bm \vartheta}_{jk} = \left( \begin{array}{c}
  \theta_{jk}   \\
    \theta_{kj}   
\end{array}\right)  
   =   \left( \begin{array}{c}
 E ( Y^{(j)} \mid \pi_{j}(\bmZ) >0, \pi_{k}(\bmZ)>0 ) \\
  E ( Y^{(k)} \mid \pi_{j}(\bmZ)>0, \pi_{k}(\bmZ)>0 )
\end{array}\right) ,  
    \label{effect1}
\end{equation}
where $\theta_{jk}$ and $\theta_{kj}$ depend on $j$ and $k$ because the ECE trial population depends on them. Note that the estimand is defined over the ECE trial population regardless of actual intervention assignment, and therefore includes individuals assigned to arms other than $j$ or $k$. A contrast of \eqref{effect1} is an unconditional treatment effect for the ECE trial population, defined as an expectation over all eligible individuals and over time during which the enrollment window is open. For example, the treatment effect of $j$ versus $k$ can be defined as the linear contrast $\theta_{jk} - \theta_{kj}$, the ratio $\theta_{jk}/\theta_{kj}$, or some other function of ${\bm \vartheta}_{jk}$. In the context of the SIMPLIFY trial,  $\theta_{21}-\theta_{12}$ is the effect of discontinuing HS versus continuing in a population of all individuals taking only HS or both HS and DA. A similar interpretation applies to $\theta_{31}-\theta_{13}$.

{Since the estimand is the expected outcome integrated over a population, $\theta_{jk}$ and $\theta_{kj}$ in \eqref{effect1} differ from $E(Y^{(j)})$ and $E(Y^{(k)})$: the former integrates over the ECE trial population while the latter integrates over the entire master protocol trial population and may include individuals who are non-concurrent or ineligible for interventions $j$ and $k$. This distinction is not important in traditional trials since the ECE and entire population match, so the condition $\pi_j(\bmZ)>0, \pi_k(\bmZ)>0$ can be omitted; however, this becomes critical for master protocol trials. Additionally, the estimand in \eqref{effect1} differs from the treatment effect for those assigned to specific arms (e.g., $E\{Y^{(j)}-Y^{(k)} \mid A \in\{j, k\}\}$), which integrates over the assigned population and can depend on the distribution of $A$. The estimand in \eqref{effect1} similarly differs from treatment effects for those assigned to specific sub-studies.
}

\vspace{-10mm}
\section{Estimation Methods} 
\label{sec: estimation}
\vspace{-6mm}

\subsection{Analysis set} \vspace{-3mm}

The \emph{analysis set}, distinct from the population used to define the estimand, refers to the sample used in statistical analysis. While the ECE trial population is uniquely specified, the analysis set is not unique and can include multiple options. To maximize efficiency, we consider the analysis set that includes all concurrently eligible individuals in the index set ${{{\cal I}}}_{jk} = \{ i : \pi_j(\bmZ_i) > 0, \pi_k(\bmZ_i) > 0 \}$. However, practical considerations may lead to using a subset of ${{{\cal I}}}_{jk}$. For example, when applying a partial blinding strategy \citep{fda:2023platform} or if other sub-studies remain blinded at the time of analysis, it may be preferable to restrict the analysis set to individuals within a specific sub-study. In these cases, the analysis set is a biased sample of the ECE trial population and requires appropriate corrections. A general discussion of analysis sets and the conditions they must satisfy is provided in Section S1 of the Supplement.

After defining the estimand, the next question is how to robustly estimate each component of ${\bm \vartheta}_{jk}$ in (\ref{effect1}), because even among the individuals within ${{\cal I}}_{jk}$, the non-zero probabilities $\pi_{j}(\bmZ)$ and $ \pi_{k}(\bmZ)$ may still vary over enrollment time and individual characteristics, leading to confounding.

\vspace{-6mm}
\subsection{Naive method}
\label{subsec: naive method}
\vspace{-3mm}

A naive method is to estimate the component $\theta_{jk} $ by the sample mean $\bar{Y}_{jk}$ of $Y_i$'s from individuals with $\pi_j(\bmZ_i)>0$, $\pi_k(\bmZ_i)>0$, and $A_i=j$. However, 
$\bar{Y}_{jk}$ is typically biased for $\theta_{jk}$ under master protocol trials as it overlooks confounding by $\bmZ$, i.e., the fact that $Z$ affects both intervention and outcome. {Specifically, 
as $n \to \infty$, 
\(
 \bar{Y}_{jk} \to \frac{E\{ I( \pi_j(\bmZ)>0, \pi_k(\bmZ)>0, A = j) Y \}}{ E\{ I( \pi_j(\bmZ)>0, \pi_k(\bmZ)>0, A = j)\}}\ \mbox{in probability}.
\)
Under the consistency assumption and Assumption~\ref{assump: randomization}, the limit above equals $  \frac{E\{  \pi_j(\bmZ) E( Y^{(j)} \mid \bmZ) \mid \pi_j(\bmZ)>0, \pi_k(\bmZ)>0 \}}{ E\{  \pi_j(\bmZ)\mid  \pi_j(\bmZ)>0, \pi_k(\bmZ)>0\}} $,  as shown in the Supplement.} This limit does not equal $\theta_{jk}$
in (\ref{effect1}) unless $\pi_j(\bmZ)$ does not vary with $\bmZ$ or $Y^{(j)} \perp \bmZ$. 
 Take the trial described in Figure \ref{fig: schema} as an example, where $\pi_j(\bmZ)$ varies with $\bmZ$,  the condition $Y^{(j)} \perp \bmZ$ unrealistically rules out the influence of disease subtype on the outcome, as well as potential temporal effects, such as later recruited individuals being healthier than those recruited earlier. %

\vspace{-6mm}
\subsection{Inverse probability weighting}
\vspace{-3mm}

 Under Assumption \ref{assump: randomization}, since the true probability of receiving intervention, $\pi_j(\bmZ)$, is known by design, this confounding can be readily addressed through inverse probability weighting (IPW). 
Notably, weighting by $\pi_j(\bmZ)$ also helps correct for sample bias, as IPW uses only those assigned to intervention $j$, which might otherwise be a biased sample from the ECE trial population in master protocol trials (see the discussion at the end of Section 3.2).

The IPW identification formula is (the proof is  in the Supplement): %
\begin{align} 
    \theta_{jk} = E\bigg\{  \frac{I(A= j) Y }{\pi_{j}(\bmZ) }  \ \Big| \ \pi_{j}(\bmZ)>0, \pi_{k}(\bmZ)>0\bigg\} , \label{eq: IPW identification} %
\end{align}
where $I( \cdot )$ denotes the indicator function.
 Formula (\ref{eq: IPW identification}) naturally suggests the following IPW estimator of $\theta_{jk}$, which is widely used in survey sampling \citep{horvitz1952generalization} and causal inference \citep{rosenbaum1983central}, %
\begin{align*}
    \ipw = \frac{1}{n_{jk}}  \sum_{i \in {{{\cal I}}}_{jk}}  \frac{ I(A_i=j)Y_i }{\pi_{j}(\bmZ_i)}, %
\end{align*}
where ${{{\cal I}}}_{jk} = \{ i: \pi_j(\bmZ_i) >0 , \pi_k(\bmZ_i)>0 \} $ 
and
$n_{jk}$ is the number of elements in ${{{\cal I}}}_{jk}$.

This IPW estimator may be improved by the following stabilized IPW (SIPW) 
estimator \citep{hajek1971comment}, which normalizes the weights $I(A_i=j) / \pi_j(\bmZ_i)$ for $ i \in {{{\cal I}}}_{jk}$ to sum to 1:\vspace{-3mm}
\begin{align*}
        \sipw  =  \bigg\{ \sum_{i \in {{{\cal I}}}_{jk}} \frac{ I(A_i=j) }{\pi_j(\bmZ_i)}\bigg\}^{-1} \sum_{i \in {{{\cal I}}}_{jk}} \frac{ I(A_i=j) Y_i  }{\pi_j(\bmZ_i)} .
\end{align*} 
Both the IPW and SIPW estimators are asymptotically normal (Theorem 1 of Section 5).%

The IPW and SIPW methods utilize some covariate information through weighting with $\pi_j(\bmZ)$. 
To improve efficiency, we can adjust for $\bmZ$ and other baseline covariates using a model-assisted approach \citep{Tsiatis:2008aa, ye2021better, bannick2023general}, which incorporates these covariates through a \emph{working model} between the outcome and covariates and produces an asymptotically unbiased estimator even if the working model is incorrect. This robustness makes model-assisted approaches increasingly popular and recommended by the regulatory agencies \citep{fda:2019aa}.

Specifically, let $\bmX \subset \bmW$ denote the observed vector of baseline covariates for adjustment, which may  overlap with $\bmZ$. {We estimate 
 $ E(Y \mid \bm X,  \pi_j(\bmZ)>0, \pi_k(\bmZ)>0, A= j) $} by $\hat{\mu}_{jk} (\bmX)$, 
using a working model that may be misspecified. For continuous outcomes,  $ \hat \mu_{jk}(\bmX) $ is usually from least squares regression of {$Y$} on $\bmX$ using data with $\pi_j(\bmZ)>0, \pi_k(\bmZ)>0$ and $ A = j$. For binary outcomes,
 $\hat \mu_{jk}(\bmX) $ is often obtained from a logistic working model. 
After $\hat{\mu}_{jk} (\cdot) $ is obtained, we can adjust for $\bm X$ using the augmented inverse probability weighting (AIPW) \citep{robins1994estimation}, resulting in the following AIPW estimator of $\theta_{jk}$, 
\begin{align*}
    \aipw = & \ \frac{1}{n_{jk}}  \sum_{i \in {{{\cal I}}}_{jk}}  \frac{ I(A_i=j) \{Y_i - \hat \mu_{jk}(\bmX_i) \} }{\pi_j(\bmZ_i)}   + \frac{1}{n_{jk}}  \sum_{i \in {{{\cal I}}}_{jk}}   \hat \mu_{jk}(\bmX_i) . \vspace{-3mm}
\end{align*}
Interestingly, in its last term, the AIPW estimator leverages covariate information from all individuals within ${{{\cal I}}}_{jk}$,  including those not receiving interventions $j$ or $k$.  For example, in Figure \ref{fig: schema}(a), when estimating the effect of intervention 3 versus intervention 1 at the end of enrollment window 2, the AIPW also incorporates covariate data from subtype A individuals assigned to intervention 2 (in sub-study 1) and intervention 4 (in sub-study 3) during enrollment windows 1 and 2.
These covariate data are not used in the IPW or SIPW estimators.  Thus, AIPW offers a robust method to leverage data from all individuals within ${{{\cal I}}}_{jk}$, including those not assigned to intervention $j$, thereby further extracting the potential of
efficiency gain through master protocol trials. 

Similarly, one can  use the stabilized AIPW (SAIPW) estimator by normalizing the weights $I(A_i=j) / \pi_j(\bmZ_i)$ for $ i \in {{{\cal I}}}_{jk}$, resulting in the following SAIPW estimator of $\theta_{jk}$, 
$$
    \saipw  =   \bigg\{ \sum_{i \in {{{\cal I}}}_{jk}} \frac{ I(A_i=j) }{\pi_j(\bmZ_i)}\bigg\}^{-1} \sum_{i \in {{{\cal I}}}_{jk}} \frac{ I(A_i=j) \{ Y_i -  \hat \mu_{jk} (\bmX_i)\}  }{\pi_j(\bmZ_i)} + \frac{1}{n_{jk}}  \sum_{i \in {{{\cal I}}}_{jk}} \hat \mu_{jk}(\bmX_i). 
$$
Both AIPW and SAIPW estimators are asymptotically normal (Theorem 1 in Section 5). 

\vspace{-6mm}
\subsection{Post-stratification}
\vspace{-3mm}
When $\bmZ$ takes discrete values, an alternative method to account for the varying $\pi_j(\bmZ)$ is to stratify individuals  based on the values that $\bmZ$ can take. This approach is known as post-stratification (PS) in sample surveys \citep{fuller2009}. Specifically, 
for a given pair $j$ and $k$, 
we  divide all concurrently eligible individuals in ${{{\cal I}}}_{jk} $ into a finite number of strata ${{\cal I}}_{jk}^{(1)},...,{{\cal I}}_{jk}^{(H_{jk})}$,  where $H_{jk}$ is the number of strata, such that within each stratum ${{\cal I}}_{jk}^{(h)}$, the values of $\pi_j(\bmZ_i)$ and $\pi_k(\bmZ_i)$ are constant, denoted as $\pi_j({\cal I}_{jk}^{(h)})$ and $\pi_k({\cal I}_{jk}^{(h)})$, respectively, for all $i$ in ${{\cal I}}_{jk}^{(h)}$.
The resulting PS estimator of $\theta_{jk}$ is
$$
    \hat\theta_{jk}^{\rm \, ps} = \frac{1}{n_{jk}} 
    \sum_{h=1}^{H_{jk}} \frac{ n({{\cal I}}_{jk}^{(h)})}{n_j({{\cal I}}_{jk}^{(h)})} \sum_{i \in {{{\cal I}}}_{jk}^{(h)}}I(A_i=j)Y_i = 
    \frac{1}{n_{jk}} 
    \sum_{h=1}^{H_{jk}}  n({{\cal I}}_{jk}^{(h)})\bar{Y}_j({\cal I}_{jk}^{(h)})
    ,  
$$
where  $n({{\cal I}}_{jk}^{(h)})$ is the number of individuals in ${{\cal I}}_{jk}^{(h)}$, $n_{j} ({{\cal I}}_{jk}^{(h)})$ is the number of individuals in ${{\cal I}}_{jk}^{(h)}$  with $A_i= j$, and 
$\bar{Y}_j({\cal I}_{jk}^{(h)})$ is the sample mean of $Y_i$'s  for units in ${\cal I}_{jk}^{(h)}$ with $A_i = j$. This estimator does not need stabilizing, as the sum of weights in $\hat\theta_{jk}^{\rm \, ps}$ is equal to 1. 

 The idea of PS also appeared in 
\cite{marschner2022analysis}, but their approach of stratifying by enrollment window is inadequate unless $\bmZ$ consists solely of enrollment time.
 On the other hand, 
stratifying according to all joint levels of enrollment window and disease subtype is unnecessary. Our proposed method of stratification ensures that 
 $\pi_j(\bmZ)$  and $ \pi_k(\bmZ)$ 
remain constant within each post-stratum, which is exactly what is required to handle the issue of varying $\pi_j(\bmZ)$ and $\pi_k(\bmZ)$. It is important to note that PS with redundant strata may lead to excessively small strata, adversely affecting the finite sample performance of  PS estimator or its variance estimator; this is further explored through simulation results and discussed in Section S2 of the Supplement.

The PS approach utilizes covariate information by stratification. To adjust for additional covariates, we apply AIPW separately within each stratum and combine the results using weighted averages. This gives the  adjusted post-stratification (APS) estimator of $\theta_{jk}$,
\begin{align*}
        \hat\theta_{jk}^{\rm \, aps} =  \frac{1}{n_{jk}} 
    \sum_{h=1}^{H_{jk}} \frac{n({{\cal I}}_{jk}^{(h)})}{n_{j} ({{\cal I}}_{jk}^{(h)}) }  \sum_{i\in {{\cal I}}_{jk}^{(h)} } I(A_i =j ) \{ Y_i - \hat\mu_{jk}(\bmX_i)\}  +  \frac{1}{n_{jk}} \sum_{i\in {{\cal I}}_{jk} } \hat\mu_{jk}(\bmX_i) . 
\end{align*}

\vspace{-8mm}
\section{Asymptotic Theory}
\label{sec: theory}
\vspace{-6mm}

\subsection{Asymptotic normality} 
\vspace{-3mm}

  To develop the asymptotic theory for estimators of $\bm\vartheta_{jk}$ involving covariate adjustment, impose the following standard stability condition on the working model $\hat\mu_{jk}$. 

  \vspace{-2mm}

 \begin{assumption}[Stability] 
 For any given $j$ and $k$, there exists a function $\mu_{jk}(\cdot)$ with finite $E_X \{ \mu_{jk}^2(\bm X)\} $  such that, as $n\to\infty$, $E_X \{ \hat\mu_{jk}(\bm X) - \mu_{jk}(\bm X)\}^2 \to 0$ in probability with respect to the randomness of $\hat\mu_{jk}(\cdot)$ as a function of data, where $E_X$ is the expectation with respect to $\bm X$. 
If $ \hat\mu_{jk}(\cdot)$ is not from a finite-dimensional parametric model, then $\mu_{jk}  (\cdot) $ and $\hat\mu_{jk}(\cdot)$ also need to satisfy the Donsker condition stated in Section S1 of the Supplement.     \label{assump: stability} 
 \end{assumption}

  \vspace{-2mm}

This assumption requires that the fitted working model $\hat\mu_{jk}$ converges in mean square to a fixed limiting function $\mu_{jk}$ as the sample size grows, but places no restriction on whether $\mu_{jk}$ correctly specifies the true outcome model. The additional Donsker condition ensures that the function class is not too complex and guarantees the validity of asymptotic approximation \citep{vandervaartWeakConvergenceEmpirical1996}.

Theorem \ref{theo:1} shows that all six estimators described in Section \ref{sec: estimation} are fully robust in the sense that each estimator consistently estimates ${\bm \vartheta}_{jk}$ and is asymptotically normal under the stated assumptions,
without requiring correct specification of any working model. In what follows, $E_{jk}=E_{kj}$,  $\var_{jk}=\var_{kj}$, and $\cov_{jk}=\cov_{kj}$ denote respectively the expectation, variance, and covariance conditioned on $\pi_j(\bmZ)>0$ and $\pi_k(\bmZ)>0$. 
Similarly, $E_{jk}(\cdot \mid S) =E_{kj} (\cdot \mid S)$, $\var_{jk}(\cdot \mid S)=\var_{kj}(\cdot \mid S)$, and $\cov_{jk}(\cdot \mid S)=\cov_{kj}(\cdot \mid S)$ denote these quantities further conditioned on $S$, where $S$ is the covariate for post-stratification, i.e., for individual $i$, $S_i = {\cal I}_{jk}^{(h)}$ if and only if $i \in {\cal I}_{jk}^{(h)}$.

\vspace{-2mm}
\begin{theorem} \label{theo:1} 
 Under the consistency assumption, Assumption \ref{assump: randomization}, and Assumption 2 if  covariates are adjusted, for fixed $j$ and $k$, and $\star\in \{{\rm ipw, sipw, aipw, saipw, ps, aps}\}$,   $ n_{jk}^{1/2}({\hat{ \bm \vartheta}^\star_{jk}} - {\bm \vartheta}_{jk} )$ converges in distribution as $n \to \infty$ to the bivariate normal with mean vector 0 and covariance matrix $\bS^{\star}_{jk}$, where 
 $\hat{\bm{\vartheta}}_{jk}^{\star} = (\hat{\theta}_{jk}^{\star}, \hat{\theta}_{kj}^{\star})^T$,  $\bm{a}^T$ is the vector transpose of $\bm{a}$, and 
the explicit expression of $\bS^{\star}_{jk}$
is given in each of the following specific case. \\
(a) For the {\rm IPW} estimator  $\hat {\bm \vartheta}_{jk}^{\rm \, ipw} $,  \vspace{-2mm}
$$\bS_{jk}^{\rm \, ipw} =  E_{jk}  \Big[{{\rm diag} \left\{ {\textstyle  \frac{I(A=j)Y^2 }{\pi_j^2(\bmZ)}  , \,  \frac{I(A=k)Y^2 }{\pi_k^2(\bmZ)} }\right\} } \Big] - \bm{\vartheta}_{jk}\bm{\vartheta}_{jk}^T,  \vspace{-2mm}$$
where ${\rm diag}\{ a_1, a_2\}$ denotes the diagonal matrix ${a_1 \ \, 0 \ \choose \ 0 \ \, a_2 \, } $ throughout. \\
(b) For the {\rm SIPW} estimator  $\hat {\bm \vartheta}_{jk}^{\rm \, sipw} $,  \vspace{-2mm}
$$\bS_{jk}^{\rm sipw} = E_{jk} \Big[ {{\rm diag} \left\{ {\textstyle
\frac{I(A=j)(Y-\theta_{jk})^2}{\pi_j^2(\bmZ)} , \, \frac{I(A=k)(Y-\theta_{kj})^2}{\pi_k^2(\bmZ)} }\right\} }\Big] . \vspace{-2mm}
$$
(c) For the {\rm AIPW} estimator  $\hat {\bm \vartheta}_{jk}^{\rm \, aipw} $,  \vspace{-2mm}
$$\bS_{jk}^{\rm aipw} =  E_{jk} \Big[ {{\rm diag} \left\{ {\textstyle 
 \frac{I(A=j)\{ Y-\mu_{jk}(\bmX)\}^2}{\pi_j^2(\bmZ)}, \, 
  \frac{I(A=k)\{ Y-\mu_{kj}(\bmX)\}^2}{\pi_k^2(\bmZ)}}  \right\} }\Big]  +  
\bm{\Lambda}_{jk} - \bm{\delta}_{jk} \bm{\delta}_{jk}^T ,  \vspace{-2mm}
$$
where $\bm{\delta}_{jk} = (\delta_{jk}, \delta_{kj})^T$, 
${\delta_{jk} = E_{jk} \left[ I(A=j)\{ Y\! - \mu_{jk} (\bmX)\}/\pi_j(\bmZ)\right]}$,  
$\mu_{jk}$ is the limit of $\hat\mu_{jk} $ given in Assumption \ref{assump: stability}, and $\bm\Lambda_{jk}$ is a symmetric matrix of order 2 with two diagonal elements
${\lambda_{jk}  = 2 \,
\cov_{jk} \{ I(A=j)Y/\pi_j(\bmZ)  ,  \mu_{jk} (\bmX)  \} - \var_{jk} \{   \mu_{jk} (\bmX) \} }$ and $\lambda_{kj}$, and the off-diagonal element $c_{jk} = c_{kj} = $ ${\cov_{jk}\{I(A=j)Y/\pi_j(\bmZ) , \mu_{kj} ( \bmX) \} + \cov_{jk}\{I(A=k)Y/\pi_k(\bmZ) , \mu_{jk} ( \bmX)\} } - \cov_{jk}\{ \mu_{jk}(\bmX) , \mu_{kj}(\bmX)\} $. \\
 (d) For the {\rm SAIPW} estimator  $\hat {\bm \vartheta}_{jk}^{\rm \, saipw} $, %
 $$\bS_{jk}^{\rm saipw} =  E_{jk}  \Big[ {{\rm diag} \left\{ {\textstyle 
 \frac{I(A=j)\{ Y-\mu_{jk}(\bmX)-\delta_{jk} \}^2}{\pi_j^2(\bmZ)}  , \, 
 \frac{I(A=k)\{ Y-\mu_{kj}(\bmX) - \delta_{kj} \}^2}{\pi_k^2(\bmZ)} } \right\} }\Big]  + \bm\Lambda_{jk}  .$$
(e) For the {\rm PS} estimator  $\hat {\bm \vartheta}_{jk}^{\rm \, ps} $, \vspace{-2mm}
$$ \bS^{\rm ps}_{jk} =  E_{jk}  \Big[ {\rm diag} \left\{{{\textstyle  
  \frac{\var_{jk} ( Y \mid A=j,\, S )}{\pi_j(S)}  , \, 
\frac{\var_{jk} ( Y \mid A=k,\, S)}{\pi_k(S)} }}
\right\} \Big]  +   
\var_{jk} \{ {
E_{jk} (Y\mid A=j,\, S) , \, 
E_{jk} (Y\mid A=k,\, S)} \}, \vspace{-2mm}
$$
where $S$ is the covariate for post-stratification, i.e., for individual $i$, $S_i = {\cal I}_{jk}^{(h)}$ if and only if $i \in {\cal I}_{jk}^{(h)}$,  and $\pi_j(S)= P(A=j \mid S) = \pi_j(\bmZ)$ by Lemma S1 in the Supplement.  \\
(f) For the {\rm APS} estimator  $\hat {\bm \vartheta}_{jk}^{\rm \, aps} $,\vspace{-2mm}
\begin{align*}
    \bS^{\rm aps}_{jk} & =   E_{jk} \Big[ {\rm diag} \left\{ {{\textstyle  
  \frac{\var_{jk} \{Y - \mu_{jk}(\bmX ) \mid A=j,\, S \}}{\pi_j(S)} , \, 
 \frac{\var_{jk} \{Y - \mu_{kj}(\bmX ) \mid A=k,\, S\}}{\pi_k(S)}} }
\right\}\Big]  \\
& \quad +  E_{jk} \{ \bm\Lambda_{jk}(S)\} + \var_{jk} \{ {E_{jk} (Y\mid A=j,\, S) , \, 
E_{jk} (Y\mid A=k,\, S)} \} , \vspace{-4mm}
\end{align*} 
where  $\bm\Lambda_{jk}(S)$ is $\bm\Lambda_{jk}$ with $\lambda_{jk}$ and $c_{jk}$ replaced by $\lambda_{jk}(S) = {2\cov_{jk}\{ Y, \mu_{jk} (\bmX) \mid A=j,\, S\}} - \var_{jk} \{\mu_{jk}(\bmX)\mid S  \} $ and $c_{jk} (S)  = {\cov_{jk}\{Y,\ \mu_{kj} ( \bmX) \mid A=j,\, S \}} + \, {\cov_{jk}\{Y ,\mu_{jk} (\bmX) \mid A=k,\, S \}}  - \cov_{jk}\{ \mu_{jk}(\bmX) , \mu_{kj}(\bmX) \mid S\} $, respectively. 
\end{theorem} %

All technical proofs are given in the Supplement.
{Importantly, Theorem~1 holds regardless of whether the working outcome model is
correctly specified. Although the robustness of AIPW to misspecification of the working model when the randomization probabilities are known is well established \citep{robins1994estimation}, to our knowledge, the asymptotic characterization of the six estimators 
in master protocol trials and the accompanying efficiency comparisons in
Section~5.2 are new to the literature.} We also show that the AIPW and SAIPW estimators with
\( \mu_{jk} = E_{jk}(Y^{(j)} \mid \bm{X}) \) and
\( \mu_{kj} = E_{jk}(Y^{(k)} \mid \bm{X}) \) are asymptotically equivalent and both
attain semiparametric efficiency (proof in Section~S3 of the Supplement).
The asymptotic covariance matrices motivate the robust variance estimators presented
in Section~S1 of the Supplement.

Theorem~1 is stated for randomized master protocol trials, where the
$\pi_j(\bmZ)$ are prespecified and known. Extension to observational studies, where
these probabilities must generally be estimated and may be misspecified, is beyond
the scope of this paper. One special case is covered: when $\bmZ$ is discrete, the
PS/APS estimators are algebraically equivalent to the corresponding IPW/SIPW and
AIPW/SAIPW estimators using sample treatment proportions within post-strata
(Section~S2 of the Supplement). Thus, Theorem~1(e)--(f) also applies to these
estimators in observational studies under the required identification assumptions.

\vspace{-6mm}
\subsection{Efficiency comparison}
\vspace{-3mm}

\begin{figure}[htbp]
\spacingset{1.2} 
    \centering
    \includegraphics[width=.7\textwidth]{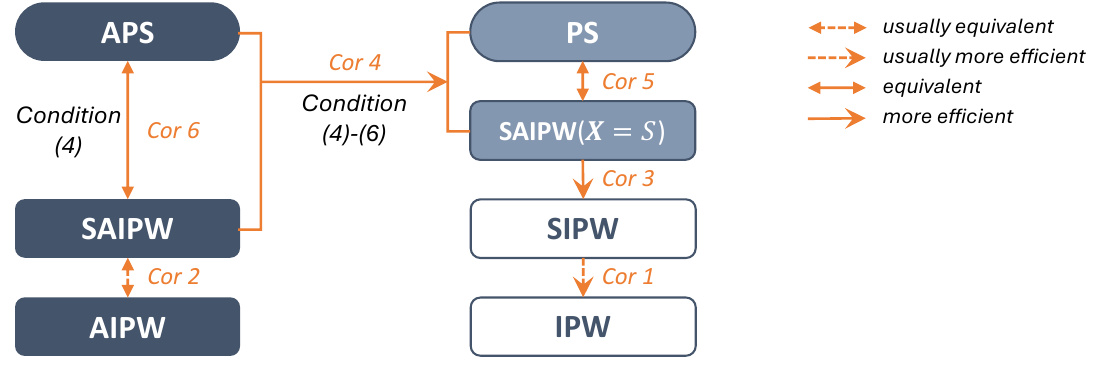}
    \caption{Efficiency comparisons among all robust estimators, with the corresponding corollary annotated for each comparison.
    }
    \label{fig:eff}
\end{figure}

{We compare the asymptotic relative efficiency between pairs of six estimators in the subsequent corollaries. Corollaries \ref{coro} and \ref{coro2} compare the efficiency of stabilized versus unstabilized estimators. Corollaries \ref{coro: SIPW and SAIPW} and \ref{coro: PS and APS} show when covariate adjustment improves efficiency. Corollaries \ref{coro5} and \ref{coro6} connect PS-type and IPW-type estimators by demonstrating asymptotic equivalence between PS and SAIPW with $\bmX=S$, and between APS and SAIPW with the same covariate adjustment. All the efficiency comparisons are summarized in Figure \ref{fig:eff}.}

\vspace{-2mm}
\begin{corollary}[Comparison of IPW and SIPW] \label{coro}
Under Assumption 1,  $\bS_{jk}^{\rm \, ipw} - \bS_{jk}^{\rm \, sipw} 
=  E_{jk} \Big[ {\rm diag} \Big\{ {\textstyle 
 \frac{\theta_{jk}^2  }{\pi_j(\bmZ)} , \,   \frac{\theta_{kj}^2 }{\pi_k(\bmZ)} } \Big\}  \Big] -\bm\vartheta_{jk} \bm\vartheta_{jk}^T 
  + 
 2 \, {\rm diag} \Big\{ {{\textstyle 
 \cov_{jk} \big\{ \frac{I(A=j)Y}{\pi_j(\bmZ)}, \frac{\theta_{jk}}{\pi_j(\bmZ)}\big\}, \, 
\cov_{jk} \big\{ \frac{I(A=k)Y}{\pi_k(\bmZ)}, \frac{\theta_{kj}}{\pi_k(\bmZ)}\big\} } } \Big\}$. 
\end{corollary}

Corollary \ref{coro} shows that neither IPW nor SIPW dominates the other in general: the asymptotic relative efficiency depends on the data-generating distribution and can favor either estimator (see Section S3.7 of the Supplement for further derivations). Nonetheless, SIPW is often more efficient in simulations (Section S2 of the Supplement). SIPW is also preferable due to  its invariance to outcome shifts for estimating $ \theta_{jk} - \theta_{kj}$, i.e.,  adding a constant to every outcome does not change  $\hat\theta_{jk}^{\rm sipw} - \hat\theta_{kj}^{\rm sipw} $ but does affect $\hat\theta_{jk}^{\rm ipw} - \hat\theta_{kj}^{\rm ipw} $ \citep{ding2023first}. For these reasons, we do not consider IPW in the remaining efficiency comparisons.

\vspace{-2mm}
 \begin{corollary}[Comparison of AIPW and SAIPW] \label{coro2}
  Under Assumptions 1-2, \linebreak $
    \bS_{jk}^{\rm \, aipw} - \bS_{jk}^{\rm \, saipw} 
  =  E_{jk} \Big[ {\rm diag} \Big\{ {\textstyle 
 \frac{\delta_{jk}^2  }{\pi_j(\bmZ)} , \,   \frac{\delta_{kj}^2 }{\pi_k(\bmZ)} } \Big\}  \Big] -\bm\delta_{jk} \bm\delta_{jk}^T 
 + 
 2 \, {\rm diag} \Big\{ {{\textstyle 
 \cov_{jk} \big\{ \frac{I(A=j)\{ Y- \mu_{jk}(\bmX)\}}{\pi_j(\bmZ)}, \frac{\delta_{jk}}{\pi_j(\bmZ)}\big\},} }$ \linebreak ${{\textstyle 
\cov_{jk} \big\{\frac{I(A=k)\{Y- \mu_{kj}(\bmX)\}}{\pi_k(\bmZ)}, \frac{\delta_{kj}}{\pi_k(\bmZ)}\big\} }  }\Big\} $, where $\delta_{jk}$ is defined in Theorem \ref{theo:1}.
\end{corollary}

Corollary 2 shows that the AIPW and SAIPW estimators are asymptotically equivalent under the condition $ \delta_{jk} = \delta_{kj} =0$, which holds for many widely-used regression models such as generalized linear models with canonical links, whose score equations set the mean of residuals to zero even under misspecification. Later discussion about Equation (\ref{adju}) shows that, even if the initial 
 $\hat\mu_{jk}(\bmX)$ does not meet this condition, it can be readily modified by (\ref{adju}) to satisfy this requirement. Therefore, AIPW and SAIPW are usually asymptotically equivalent and we do not include AIPW in the rest of the efficiency comparisons.

 \begin{corollary}[Comparison of SIPW and SAIPW] \label{coro: SIPW and SAIPW}
 Under Assumptions 1-2 and \vspace{-2mm} {\begin{align}
&  E_{jk}\{ Y - \mu_{jk}(\bmX) \mid A=j,\, S \} = \delta_{jk} , \ \    E_{jk}\{ Y - \mu_{kj}(\bmX) \mid A=k,\, S  \} = \delta_{kj}, \label{U} \\
& \cov_{jk} \{ Y - \mu_{jk}(\bmX) , \,
\mu_{jk}(\bmX) \mid A=j, \, S \} =
  \cov_{jk} \{ Y-\mu_{kj}(\bmX), \, \mu_{kj}(\bmX) \mid A=k,\, S \} = 0, \ \ \label{gura} \\
 & \cov_{jk} \{ 
\frac{I(A=j)\{Y - \mu_{jk}(\bmX)\}}{\pi_j(\bmZ)}, \, \mu_{kj}(\bmX)
\} =  \cov_{jk} \{ 
\frac{I(A=k)\{Y - \mu_{kj}(\bmX)\}}{\pi_k(\bmZ)}, \, \mu_{jk}(\bmX)
\} = 0 , \vspace{-2mm} \label{gura2}
\end{align}
}
where $S$ is the covariate for post-stratification as defined in Theorem 1(e), 
the {\rm SAIPW} estimator $\hat {\bm \vartheta}_{jk}^{\rm \, saipw} $ 
is asymptotically  more efficient than  the {\rm SIPW} estimator $\hat {\bm \vartheta}_{jk}^{\rm \, sipw} $, i.e., 
\begin{align*}
    \bS_{jk}^{\rm \, sipw} - \bS_{jk}^{\rm \, saipw}  = 
 E_{jk} \Big[ {\rm diag} \left\{ {\textstyle \frac{\{\mu_{jk}(\bmX) -\mu_{jk}\}^2}{\pi_j(S)} , \, 
\frac{\{\mu_{kj}(\bmX) -\mu_{kj}\}^2}{\pi_k(S)} } \right\} \Big] - \var_{jk} \{ \mu_{jk}(\bmX), \mu_{kj}(\bmX) \} 
\end{align*}
is positive definite,  where $\mu_{jk} = E_{jk}\{ \mu_{jk}(\bmX)\}$ and $\mu_{kj} = E_{jk}\{ \mu_{kj}(\bmX)\}$,  unless 
 either one of 
$\var_{jk}\{ \mu_{jk}(\bmX)\} $ and $\var_{jk}\{ \mu_{kj}(\bmX)\} $ is 0  or 
$   \pi_j(S)+\pi_k(S) =1 $ and the correlation between $\mu_{jk}(\bmX)$ and $\mu_{kj}(\bmX)$
is $\pm 1$, in which case  $  \bS_{jk}^{\rm \, sipw} - \bS_{jk}^{\rm \, saipw}$ is positive semidefinite. 
\end{corollary}\vspace{-2mm}

When covariate adjustment for $\bmX$ is applied,  SAIPW is not necessarily more efficient than SIPW if working models are incorrect.  Conditions (\ref{U})-(\ref{gura2}) in Corollary 3 provide sufficient conditions to guarantee efficiency gains from covariate adjustment when estimating any differentiable function of $\bm\vartheta$, except in trivial scenarios.
Condition (\ref{U})  holds without loss of generality since it can always be achieved by replacing $\hat\mu_{jk}(\bmX)$ with 
\begin{equation}
    \tilde \mu_{jk} (\bmX)  = \hat \mu_{jk}(\bmX)
+ \frac{1}{n_{jk}(S) } \sum_{i\in {\cal I}_{jk},  A_i=j, S_i = S} \{{Y_i} - \hat\mu_{jk}(\bmX_i) \} ,  \label{adju}
\end{equation} 
where $n_{jk}(S) $ is the number of elements 
in $\{ i \in {\cal I}_{jk}, A_i=j, S_i = S \}$. 
Conditions \eqref{gura}-\eqref{gura2} hold when working models are correct. Otherwise,  they require specific construction of 
$\hat{\mu}_{jk}(\bmX)$ 
and $\hat{\mu}_{kj}(\bmX)$ that embeds  \eqref{gura}-\eqref{gura2}. For example,  (\ref{U})-(\ref{gura2}) hold if applying a linear ANHECOVA working model that includes  $S$ and its interaction with $\bmX$ \citep{ye2021better}, or 
when applying joint calibration with non-linear working models \citep{bannick2023general}; see S3 in the Supplement for details.  

In fact,  \eqref{U}-\eqref{gura2} always hold for the SAIPW estimator with $\bmX = S$, where $\hat\mu_{jk}(\bmX) = \bar{Y}_{j}(S)$ is the sample mean of $Y_i$'s for $i \in {\cal I}_{jk}$, $A_i=j$, and $S_i = S $.  
This is  because, with  $S$ as the stratum indicator,
the limit of $\bar{Y}_{j}(S)$ is $ E_{jk}(Y^{(j)} \mid \bmX = S)$, which is always a correct working model, and consequently conditions \eqref{U}-\eqref{gura2}  hold.
Therefore, the SAIPW estimator with $\bmX = S$
is asymptotically more efficient than the SIPW estimator according to  Corollary 3, and we use it in later corollaries to bridge comparisons between IPW-type and PS-type estimators.

\vspace{-2mm}
 \begin{corollary}[Comparison of PS and APS]  \label{coro: PS and APS}
 Under Assumptions 1-2 and  \eqref{U}-\eqref{gura2}, the {\rm APS} estimator $\hat {\bm \vartheta}_{jk}^{\rm \, apw} $ 
is asymptotically  more efficient than  the {\rm PS} estimator $\hat {\bm \vartheta}_{jk}^{\rm \, ps} $, i.e., 
\begin{align*}
    \bS_{jk}^{\rm \, ps} - \bS_{jk}^{\rm \, aps} \! = \! 
     E_{jk} \Big[ {\rm diag} \left\{ {\textstyle \frac{\var_{jk}\{ \mu_{jk}(\bmX)\mid S\} }{\pi_j(S)} , \, 
\frac{\var_{jk}\{ \mu_{kj}(\bmX)\mid S\}}{\pi_k(S)} } \right\}  -  \var_{jk} \{ \mu_{jk}(\bmX), \mu_{kj}(\bmX) \mid S\} \Big]
\end{align*}
is positive definite, unless 
either one of $\var_{jk}\{ \mu_{kj}(\bmX)\mid S \}$ and $\var_{jk}\{ \mu_{kj}(\bmX)\mid S\}$ is 0  or  $\pi_j(S)+\pi_k(S)=1$ and  
the correlation between $\mu_{jk}(\bmX) $ and $\mu_{kj}(\bmX)$ conditioned on $S$ is $\pm 1$, in which case $ \bS_{jk}^{\rm \, ps} - \bS_{jk}^{\rm \, aps}$ is positive semidefinite. 
 \end{corollary}
\vspace{-2mm}

 Corollary \ref{coro: PS and APS} indicates a guaranteed efficiency gain after covariate adjustment on top of post-stratification,  except in some trivial scenarios.

\begin{corollary}[Comparison of PS and SAIPW with $\bmX =S$] \label{coro5}
 Under Assumption 1, $\hat{\bm\vartheta}_{jk}^{\rm \, ps} $ and $\hat{\bm\vartheta}_{jk}^{\rm \, saipw} $ with $\bmX=S$  have 
 the same asymptotic covariance matrix.
 \end{corollary} %
Corollary \ref{coro5} shows the asymptotic equivalence between the PS estimator and the SAIPW estimator with $\bmX = S$, two estimators that fully adjust for the discrete covariate $S$.

\begin{corollary}[Comparison of APS and SAIPW]\label{coro6}
 Under Assumptions 1-2 and condition (\ref{U}), 
 $\hat{\bm\vartheta}_{jk}^{\rm \, aps} $ and $\hat{\bm\vartheta}_{jk}^{\rm \, saipw} $  have 
 the same asymptotic covariance matrix.
 \end{corollary}
     Corollary \ref{coro6} compares APS and SAIPW with the same covariate adjustment.  It demonstrates that when both estimators adjust for $\bmX$ using a working model that satisfies condition \eqref{U}, the APS estimator is asymptotically equivalent to the SAIPW estimator.

It follows from Corollaries 4-6 that, 
  under (\ref{U})-(\ref{gura2}),  SAIPW with $\bmX \supset S$ is  more efficient than SAIPW with $\bmX = S$; see Section S3 in the Supplement for a proof.

Lastly, we discuss the data used by the estimation methods as an informal way to conceptualize the efficiency comparison. When comparing interventions $j$ and $k$, all methods can robustly utilize the outcome data from arms $j$ and $k$ within ${{{\cal I}}}_{jk}$, which may achieve substantial efficiency gains compared to traditional stand-alone trials. Furthermore, covariate adjustment methods (including PS, APS, AIPW, and SAIPW) can leverage additional strata and covariate information from all individuals within  ${{{\cal I}}}_{jk}$, even those assigned to arms other than $j$ and $k$. This can lead to further efficiency gains, especially in master protocol trials with many arms.

\vspace{-10mm}
\section{Results} 
\label{sec: real data}
\vspace{-4mm}
As reviewed in Section \ref{subsec: simplify} and shown in Figure \ref{fig: simplify}, the SIMPLIFY trial \citep{mayer2023discontinuation} consists of two sub-studies (the HS and DA studies) and a total of three interventions: continuing therapy (Intervention 1, used as the control) in both the HS and DA studies,  discontinuing HS (Intervention 2) in the HS study, and discontinuing DA (Intervention 3) in the DA study. After completing the first study, participants on both HS and DA therapies may re-enroll in the other study contingent on eligibility and willingness to consent. The outcome $Y$ is the 6-week mean absolute change in ppFEV$_1$ as described in Section 2.2.

\begin{figure}[!h]
\spacingset{1.2} 
    \centering
    \includegraphics[width=.9\textwidth]{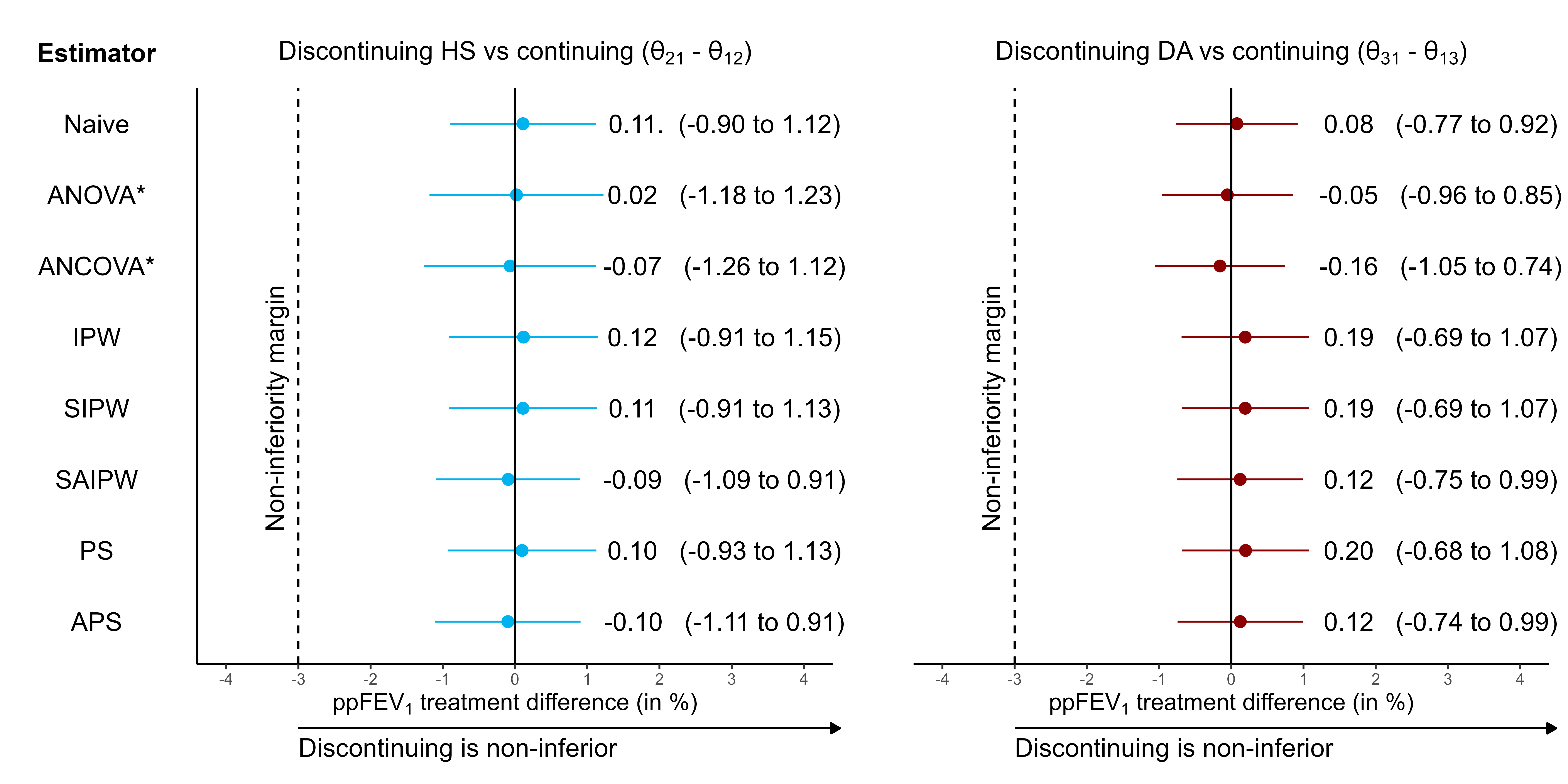}
    \caption{Estimates and  95\% Wald-type  confidence intervals (in \%) in the SIMPLIFY 
    trial. Naive refers to the naive sample mean difference described in Section 4.2. ANOVA* and ANCOVA* are estimates based on data from each individual sub-study. IPW, SIPW, SAIPW, PS, and APS are the correct and fully robust estimators for the estimand defined on the ECE trial population.}
    \label{fig: simplify result}
\end{figure}

In our analysis, the ECE trial population for evaluating the effect of discontinuing HS versus continuing is the population of all individuals taking only HS or both HS and DA. Accordingly, the treatment effect of interest, $\theta_{21} - \theta_{12}$, is defined on this ECE trial population. A similar interpretation applies to  $\theta_{31}-\theta_{13}$. Demographic and baseline characteristics for participants in the two ECE trial populations are presented in Tables S1-S2 of the Supplement. Note that our estimand differ slightly  from those in \cite{mayer2023discontinuation}. They analyzed the HS and DA studies separately according to their primary hypotheses of interest and included data collected after participants re-enrolled in an additional study. In contrast, we excluded data recorded after re-enrollment, as our methods currently do not account for repeated measures. This exclusion applies to all methods implemented in this section and explains why the confidence intervals from ANCOVA* in Figure \ref{fig: simplify result} are wider than those reported in \cite{mayer2023discontinuation}. We plan to address this limitation in future work. We also excluded 10 (1.7\%) participants due to missing outcomes, resulting in a sample size of $n=584$: 293 in Intervention 1, 119 in Intervention 2, and 172 in Intervention 3. We conducted a complete case analysis, which is valid under the assumption that outcomes are missing completely at random. The low missing rate (1.7\%) suggests that the impact of this assumption on the results is likely minimal.

 For the defined estimands,  we consider the IPW, SIPW, SAIPW,  PS, and APS estimators, as described in Section \ref{sec: estimation}, all of which can account for the different intervention assignment probabilities across $Z$. Point estimates and 95\% Wald-type confidence intervals for each estimator are evaluated in the simulation study in Section S2 of the Supplement at a similar sample size. Here, $Z$ includes baseline use of HS, DA, and enrollment windows.
For SAIPW and APS, $\hat\mu_{jk}$ is estimated by fitting a linear regression of $Y$ on covariates using data from arm $j$ in $\mathcal{I}_{jk}$. The covariates include $Z$ and several other baseline variables: two continuous covariates, age and baseline ppFEV$_1$; three baseline binary covariates, sex, 
race (white or non-white), and  
 ethnicity (Hispanic or Latino, or not Hispanic or Latino); and one three-category covariate:  genotype 
 (delta F508 homozygous, heterozygous, or other/unknown). 
 As shown in Corollary 2 and simulations, when adjusting for covariates using linear working models, the AIPW estimator is numerically nearly identical to the SAIPW
estimator and thus omitted.  We also include the naive ANOVA as described in Section \ref{subsec: naive method}, which is subject to confounding bias. Results are shown in Figure \ref{fig: simplify result}.

 For all five proposed estimators (IPW, SIPW, SAIPW, PS, and APS), all point estimates are similar, as they target the same estimand defined on the ECE trial population. Since all confidence intervals are on the right side of the pre-defined non-inferiority margin $-$3\% (the vertical dashed line in Figure \ref{fig: simplify result}), non-inferiority of discontinuing therapy can be clearly claimed for both HS and DA -- consistent with the findings reported in \cite{mayer2023discontinuation}. IPW and SIPW produce similar confidence intervals, likely because the true effect is near zero in this non-inferiority study. SIPW and SAIPW, as well as PS and APS, also yield similar confidence intervals, as most covariates are only weakly associated with the outcome (\( R^2 \) between 0.023 and 0.1 across three arms). This likely also explains why the naive estimator does not exhibit substantial bias in this analysis.

We also include ANOVA* (sample means) and ANCOVA* (using linear working models to adjust for the same covariates included in SAIPW
and APS separately for each arm) estimators for comparison, which are based solely on data from each individual sub-study.
As discussed in Section \ref{subsec: population}, these estimators target populations represented by those assigned to the HS or DA study.  As a result, the interpretations of these estimators differ from the five proposed estimators. Specifically, for the HS comparison, the HS study includes 18 HS-only and 222 HS+DA participants, while the sample from the ECE trial population includes 18 HS-only and 427 HS+DA participants. For the DA comparison, the DA study includes 139 DA-only and 205 HS+DA participants, whereas the sample from the ECE trial population includes 139 DA-only and 427 HS+DA participants. More detailed demographic and baseline characteristics for participants in the HS and DA studies are presented in Tables S3 and S4. The larger population difference in the DA comparison likely explains the more noticeable discrepancy in point estimates for that result (opposite sign relative to the five proposed estimators). 

Moreover, as expected, the ANOVA* and ANCOVA* estimators yield wider confidence intervals compared to the five proposed estimators. This is because, for the five proposed estimators, participants on both HS and DA assigned to Intervention 1 (\(n = 214\)) contribute to  both the Intervention 2 vs. 1 and Intervention 3 vs. 1 comparisons. The covariate-adjusted methods (SAIPW and APS) further leverage covariate information from all participants on both HS and DA (\(n = 427\)) in both comparisons.
In contrast, ANOVA* and ANCOVA* split these participants between the HS and DA sub-studies, reducing the sample size for each comparison. As a result, the five proposed estimators achieve approximately 5\%–27\% variance reduction relative to the sub-study-specific estimators.

Overall, the results from the SIMPLIFY trial highlight the critical importance of clearly defining the estimand in master protocol trials. Without this clarity, estimators may have interpretations that depend on the randomization ratio or trial design. An additional advantage of the five proposed estimators is the potential for efficiency gains through robust use of pooled data across sub-studies.

\vspace{-10mm}
\section{Summary and Recommendations}
\vspace{-5mm} \label{sec: discussion}
Master protocol trials have the potential to enhance statistical efficiency by facilitating the sharing of data across intervention evaluations. However, realizing this potential requires careful consideration of two fundamental challenges: defining meaningful estimands to reflect eligibility and concurrency, and developing robust estimation methods to accommodate the intervention assignment probabilities varying with enrollment window and individual characteristics.

In this article, we present a clear framework for constructing a clinically meaningful estimand with precise specification of the population. Our main discussion centers on comparing two interventions within the ECE trial population, as it preserves the integrity of randomization and avoids arbitrary dependence on randomization ratios. The framework is general in that it accommodates trials where interventions have different eligibility criteria and where new treatments may be added over time, and is adaptable to various contexts, such as comparisons across multiple interventions and evaluation of treatment effects within specific subgroups or a re-weighted population. Given the central importance of the estimand, our framework enhances the fundamental understanding of master protocol trials and lays the groundwork for future statistical research—both frequentist and Bayesian—on topics such as interim analysis and the use of nonconcurrent controls.

To estimate and infer this estimand with minimal statistical assumptions, we develop methods using weighting or post-stratification to account for varying intervention assignment probabilities. We also consider model-assisted approaches for covariate adjustment.
Theoretical and empirical results are given and a detailed relative efficiency comparison is in Figure \ref{fig:eff}.
We provide explicit formulas for robust variance estimation, which have been implemented in our R package, \textsf{RobinCID}.

\vspace{-3mm}
\spacingset{.7}
\bibliographystyle{apalike}
\bibliography{reference}

\end{document}


\def\spacingset#1{\renewcommand{\baselinestretch}%
{#1}\small\normalsize} \spacingset{1}

\if1\blind
{
	 \begin{center} 
	\spacingset{1.5} 	{\LARGE\bf Supplementary Material for \\ 
``From Estimands to Robust Inference of Treatment Effects in Master Protocol Trials''} \\ \bigskip \bigskip
		\spacingset{1} 
				{\large Yuhan Qian$^1$, Yifan Yi$^{2}$, Jun Shao$^3$, Yanyao Yi$^4$, Gregory Levin$^5$, \\ Nicole Mayer-Hamblett$^{6,7,1}$,  Patrick J. Heagerty$^1$, Ting Ye$^1$\footnote[2]{Correspond to tingye1@uw.edu.}} \\ \bigskip
	 {$^1$Department of Biostatistics, University of Washington\\
$^2$Department of Biostatistics, The University of Texas MD Anderson Cancer Center\\
$^3$Department of Statistics, University of Wisconsin-Madison\\
$^4$Global Statistical Sciences, Eli Lilly and Company \\
$^5$Food and Drug Administration \\
$^6$Seattle Children’s Research
Institute \\
$^7$Department of Pediatrics, University of Washington}
	\end{center}
} \fi

\if0\blind
{
  \bigskip
  \bigskip
  \bigskip
  \begin{center}
    \spacingset{1.8}
    {\LARGE\bf Supplementary Material for 
``From Estimands to Robust Inference of Treatment Effects in Master Protocol Trials''}
\end{center}
  \medskip
} \fi

\bigskip

\spacingset{1.9} %

\setcounter{equation}{0}
\setcounter{table}{0}
\setcounter{lemma}{0}
\setcounter{section}{0}
\renewcommand{\theequation}{S\arabic{equation}}
\renewcommand{\thetable}{S\arabic{table}}
\renewcommand{\thefigure}{S\arabic{figure}}
\renewcommand{\thesection}{S\arabic{section}}
\renewcommand{\thelemma}{S\arabic{lemma}}
\renewcommand{\thecondition}{S\arabic{condition}}

\section{Additional results}

\subsection{Additional data analysis results on the SIMPLIFY trial}
Table S1 presents the demographic and baseline characteristics of participants in the ECE population comparing TRT 1 (continued therapy) and TRT 2 (discontinued HS). The "Overall" column includes all individuals who were on either HS alone or both HS and DA. The two columns under "Before weighting" correspond to participants assigned to TRT 1 and TRT 2, respectively; these groups may not be comparable due to confounding (i.e., different intervention assignment probabilities across $Z$). The columns under "After weighting" show the characteristics after applying SIPW, which adjusts for confounding and ensures comparability between the treatment groups.  Table S2 presents the demographic and baseline characteristics of participants in the ECE population comparing TRT 1 (continued therapy) and TRT 3 (discontinued DA).

\begin{table}[htbp]
\spacingset{1.2} 
\centering
\caption{Demographic and baseline characteristics of participants in the ECE Population comparing Continuing therapy with Discontinuing HS}
\label{tab: demographics in hs in ECE}
\resizebox{\textwidth}{!}{%
\begin{tabular}{@{}ccccccc@{}}
\toprule
 & \multicolumn{6}{c}{ECE population} \\ \cmidrule(l){2-7} 
\multicolumn{1}{l}{} &  & \multicolumn{2}{c}{Before weighting} &  & \multicolumn{2}{c}{After weighting} \\ \midrule
 & \begin{tabular}[c]{@{}c@{}}Overall\\ (n=445)\end{tabular} & \begin{tabular}[c]{@{}c@{}}Continue\\ (n=223)\end{tabular} & \begin{tabular}[c]{@{}c@{}}Discontinue\\ (n=119)\end{tabular} &  & Continue & Discontinue \\ \midrule
\multicolumn{1}{l}{Baseline Age, yeas; mean (SD)} & 21.80 (10.24) & 22.80 (11.60) & 20.82 (9.35) &  & 22.80 (11.60) & 20.71 (8.96) \\
\multicolumn{1}{l}{Sex, \%} &  &  &  &  &  &  \\
Male & 52.6 & 50.7 & 54.6 &  & 50.7 & 55.7 \\
Female & 47.4 & 49.3 & 45.4 &  & 49.3 & 44.3 \\
\multicolumn{1}{l}{Race, \%} &  &  &  &  &  &  \\
White & 95.7 & 95.5 & 96.6 &  & 95.5 & 97.0 \\
Non-white & 4.3 & 4.5 & 3.4 &  & 4.5 & 3.0 \\
\multicolumn{1}{l}{Ethnicity, \%} &  &  &  &  &  &  \\
Hispanic or Latino & 6.1 & 4.5 & 7.6 &  & 4.5 & 7.6 \\
Not Hispanic or Latino & 93.9 & 95.5 & 92.4 &  & 95.5 & 92.4 \\
\multicolumn{1}{l}{Genotype category, \%} &  &  &  &  &  &  \\
Delta F508 Homozygous & 58.4 & 56.5 & 61.3 &  & 56.5 & 60.4 \\
Delta F508 Heterozygous & 40.0 & 41.3 & 37.8 &  & 41.3 & 38.7 \\
Other/Unkown & 1.6 & 2.2 & 0.8 &  & 2.2 & 0.9 \\
\multicolumn{1}{l}{Baseline HS use, \%} &  &  &  &  &  &  \\
Yes & 100.0 & 100.0 & 100.0 &  & 100.0 & 100.0 \\
No & 0.0 & 0.0 & 0.0 &  & 0.0 & 0.0 \\
\multicolumn{1}{l}{Baseline DA use, \%} &  &  &  &  &  &  \\
Yes & 96.0 & 96.0 & 92.4 &  & 96.0 & 95.9 \\
No & 4.0 & 4.0 & 7.6 &  & 4.0 & 4.1 \\
\multicolumn{1}{l}{Baseline ppFEV$_1$, \%; mean (SD)} & 96.96 (17.08) & 97.69 (16.79) & 96.78 (16.47) &  & 97.69 (16.79) & 96.66 (16.62) \\ \bottomrule
\end{tabular}%
}
\end{table}

\begin{table}[htbp]
\spacingset{1.2} 
\centering
\caption{Demographic and baseline characteristics of participants in the ECE Population comparing Continuing therapy with Discontinuing DA}
\label{tab: demographics in da in ECE}
\resizebox{\textwidth}{!}{%
\begin{tabular}{@{}ccccccc@{}}
\toprule
 & \multicolumn{6}{c}{ECE population} \\ \cmidrule(l){2-7} 
\multicolumn{1}{l}{} &  & \multicolumn{2}{c}{Before weighting} &  & \multicolumn{2}{c}{After weighting} \\ \midrule
 & \begin{tabular}[c]{@{}c@{}}Overall\\ (n=566)\end{tabular} & \begin{tabular}[c]{@{}c@{}}Continue\\ (n=284)\end{tabular} & \begin{tabular}[c]{@{}c@{}}Discontinue\\ (n=172)\end{tabular} &  & Continue & Discontinue \\ \midrule
\multicolumn{1}{l}{Baseline Age, yeas; mean (SD)} & 22.24 (10.22) & 23.11 (11.49) & 22.09 (8.82) &  & 23.11 (11.49) & 21.38 (8.37) \\
\multicolumn{1}{l}{Sex, \%} &  &  &  &  &  &  \\
Male & 52.3 & 50.4 & 52.9 &  & 50.4 & 54.0 \\
Female & 47.7 & 49.6 & 47.1 &  & 49.6 & 46.0 \\
\multicolumn{1}{l}{Race, \%} &  &  &  &  &  &  \\
White & 95.9 & 95.8 & 95.9 &  & 95.8 & 95.1 \\
Non-white & 4.1 & 4.2 & 4.1 &  & 4.2 & 4.9 \\
\multicolumn{1}{l}{Ethnicity, \%} &  &  &  &  &  &  \\
Hispanic or Latino & 6.7 & 6.0 & 7.0 &  & 6.0 & 7.0 \\
Not Hispanic or Latino & 93.3 & 94.0 & 93.0 &  & 94.0 & 93.0 \\
\multicolumn{1}{l}{Genotype category, \%} &  &  &  &  &  &  \\
Delta F508 Homozygous & 56.5 & 56.0 & 55.8 &  & 56.0 & 58.2 \\
Delta F508 Heterozygous & 41.2 & 41.2 & 41.9 &  & 41.2 & 40.1 \\
Other/Unkown & 2.3 & 2.8 & 2.3 &  & 2.8 & 1.7 \\
\multicolumn{1}{l}{Baseline HS use, \%} &  &  &  &  &  &  \\
Yes & 75.4 & 75.4 & 59.9 &  & 75.4 & 76.0 \\
No & 24.6 & 24.5 & 40.1 &  & 24.5 & 24.0 \\
\multicolumn{1}{l}{Baseline DA use, \%} &  &  &  &  &  &  \\
Yes & 100.0 & 100.0 & 100.0 &  & 100.0 & 100.0 \\
No & 0.0 & 0.0 & 0.0 &  & 0.0 & 0.0 \\
\multicolumn{1}{l}{Baseline ppFEV$_1$, \%; mean (SD)} & 96.99 (16.91) & 97.59 (16.51) & 96.14 (17.87) &  & 97.59 (16.51) & 96.40 (18.38) \\ \bottomrule
\end{tabular}%
}
\end{table}

Tables \ref{tab: demographics in hs}–\ref{tab: demographics in da} present the demographic and baseline characteristics of participants assigned to the two sub-studies, respectively. These tables describe the analysis populations corresponding to the ANOVA* and ANCOVA* estimators in Figure 6.

\begin{table}[htbp]
\spacingset{1.2} 
\centering
\caption{Demographic and baseline characteristics of participants assigned to the HS study\label{tab: demographics in hs}}
\resizebox{.8\textwidth}{!}{%
\begin{tabular}{@{}cccc@{}}
\toprule
 & \multicolumn{3}{c}{HS Study} \\ \midrule
 & \begin{tabular}[c]{@{}c@{}}Overall\\ (n=240)\end{tabular} & \begin{tabular}[c]{@{}c@{}}Continue\\ (n=121)\end{tabular} & \begin{tabular}[c]{@{}c@{}}Discontinue\\ (n=119)\end{tabular} \\ \midrule
\multicolumn{1}{l}{Baseline Age, yeas; mean (SD)} & 21.58 (10.00) & 22.32 (10.59) & 20.82 (9.35) \\
\multicolumn{1}{l}{Sex, \%} &  &  &  \\
Male & 53.3 & 52.1 & 54.6 \\
Female & 46.7 & 47.9 & 45.4 \\
\multicolumn{1}{l}{Race, \%} &  &  &  \\
White & 97.1 & 97.5 & 96.6 \\
Non-white & 2.9 & 2.5 & 3.4 \\
\multicolumn{1}{l}{Ethnicity, \%} &  &  &  \\
Hispanic or Latino & 6.2 & 5.0 & 7.6 \\
Not Hispanic or Latino & 93.8 & 95.0 & 92.4 \\
\multicolumn{1}{l}{Genotype category, \%} &  &  &  \\
Delta F508 Homozygous & 60.0 & 58.7 & 61.3 \\
Delta F508 Heterozygous & 38.3 & 38.8 & 37.8 \\
Other/Unkown & 1.7 & 2.5 & 0.8 \\
\multicolumn{1}{l}{Baseline HS use, \%} &  &  &  \\
Yes & 100.0 & 100.0 & 100.0 \\
No & 0.0 & 0.0 & 0.0 \\
\multicolumn{1}{l}{Baseline DA use, \%} &  &  &  \\
Yes & 92.5 & 92.6 & 92.4 \\
No & 7.5 & 7.4 & 7.6 \\
\multicolumn{1}{l}{Baseline ppFEV$_1$, \%; mean (SD)} & 97.35 (16.56) & 97.91 (16.70) & 96.78 (16.47) \\ \bottomrule
\end{tabular}%
}
\end{table}

\begin{table}[htbp]
\spacingset{1.2} 
\centering
\caption{Demographic and baseline characteristics of participants assigned to the DA study}
\label{tab: demographics in da}
\resizebox{.8\textwidth}{!}{%
\begin{tabular}{@{}cccc@{}}
\toprule
 & \multicolumn{3}{c}{DA study} \\ \midrule
 & \begin{tabular}[c]{@{}c@{}}Overall\\ (n=344)\end{tabular} & \begin{tabular}[c]{@{}c@{}}Continue\\ (n=172)\end{tabular} & \begin{tabular}[c]{@{}c@{}}Discontinue\\ (n=172)\end{tabular} \\ \midrule
\multicolumn{1}{l}{Baseline Age, yeas; mean (SD)} & 22.95 (10.50) & 23.82 (11.91) & 22.09 (8.82) \\
\multicolumn{1}{l}{Sex, \%} &  &  &  \\
Male & 50.9 & 48.8 & 52.9 \\
Female & 49.1 & 51.2 & 47.1 \\
\multicolumn{1}{l}{Race, \%} &  &  &  \\
White & 95.3 & 94.8 & 95.9 \\
Non-white & 4.7 & 5.2 & 4.1 \\
\multicolumn{1}{l}{Ethnicity, \%} &  &  &  \\
Hispanic or Latino & 6.7 & 6.4 & 7.0 \\
Not Hispanic or Latino & 93.3 & 93.6 & 93.0 \\
\multicolumn{1}{l}{Genotype category, \%} &  &  &  \\
Delta F508 Homozygous & 55.5 & 55.2 & 55.8 \\
Delta F508 Heterozygous & 41.9 & 41.9 & 41.9 \\
Other/Unkown & 2.6 & 2.9 & 2.3 \\
\multicolumn{1}{l}{Baseline HS use, \%} &  &  &  \\
Yes & 59.6 & 59.3 & 59.9 \\
No & 40.4 & 40.7 & 40.1 \\
\multicolumn{1}{l}{Baseline DA use, \%} &  &  &  \\
Yes & 100.0 & 100.0 & 100.0 \\
No & 0.0 & 0.0 & 0.0 \\
\multicolumn{1}{l}{Baseline ppFEV$_1$, \%; mean (SD)} & 96.75 (17.30) & 97.36 (16.74) & 96.14 (17.87) \\ \bottomrule
\end{tabular}%
}
\end{table}

\newpage

\subsection{A Narrative of Figure 2}

\noindent
{\em Enrollment Window} 1. 
There are two initial sub-studies. All individuals are eligible to sub-study 1, while sub-study 2 is restricted to individuals with disease subtype A. Individuals in sub-study 1 are randomized to  treatment 1 or  2. Individuals with subtype A are randomized to  sub-study 1 or  2, and then randomized to  treatment 1 or 3 if they are in sub-study 2.

\noindent
{\em Enrollment Window} 2. 
Sub-study 3 is added to the platform trial when a new intervention treatment 4 becomes available for disease subtype A. Following this addition, an individual with subtype A is randomly assigned to either sub-study 1, 2, or 3, and subsequently randomized to the treatments being assessed within the respective sub-study.

\noindent
{\em Enrollment Window} 3. 
After sub-study 2 reaches full enrollment, it closes its enrollment. Later during this period, treatment 3 meets the pre-specified success criteria.

\noindent
{\em Enrollment Window} 4. Since treatment 3 has been demonstrated to be superior to treatment 1 for subtype A, it replaces treatment 1 as the control in sub-study 3. Moreover, sub-study 1 excludes individuals with subtype A, and sub-study 4 is introduced to continue comparing treatment 2 with the new control treatment 3 for subtype A.

\noindent
{\em Enrollment Window} 5. Sub-study 4 has finished its planned enrollment and is no longer accepting new individuals. Later during this period, treatment 2 
is stopped with futility.

\subsection{More discussions of Figure \ref{fig: example table}}

We present two additional scenarios based on Figure \ref{fig: example table}:
\begin{itemize}
    \item In Figure 3, Design 2, if we compare TRT 1 vs. TRT 3, the ECE population corresponds to the population in Enrollment Window 2 (comprising 60\% mild and 40\% moderate disease), which is different from the overall trial population. This is expected, as TRT 3 is not available during Enrollment Window 1.
\item In Figure 3, Design 2, suppose TRT 4 is introduced in Enrollment Window 1 but discontinued in Enrollment Window 2 due to futility. In this case, the ECE population for comparing TRT 3 vs. TRT 4 would be empty, as no patients are concurrently eligible for both treatments within the same enrollment window. Thus, a direct randomized comparison is not possible. This is also supported by \cite{fda:2023platform}. Analogous to a traditional trial setting, one would conduct two separate studies for TRT 3 and TRT 4 at different times in distinct populations.  Any comparison between the two would then require indirect methods, which rely on strong assumptions. 

This scenario has implications for both the design and analysis of platform trials. At the design stage, if a particular comparison is of interest, the ECE framework can be used to ensure the existence of a non-empty ECE population, thereby enabling at least a subpopulation in which randomized comparison is feasible. At the analysis stage, one must proceed with extra caution and rely on additional assumptions — for example, by applying methods that incorporate non-concurrent controls or external controls \citep{pocock1976combination}. While this lies outside the scope of the current paper, which focuses on fully robust randomized comparisons, it represents an important area for future research.
\end{itemize}

\subsection{Robust Variance Estimation}
\label{sec: variance estimation}

To robustly assess variability and conduct large sample inference, it is essential to obtain consistent estimators of the asymptotic covariance matrices in Theorem \ref{theo:1} for various estimators of $\bm\vartheta_{jk}$, regardless of whether the working models $\hat \mu_{jk}$ and 
 $\hat \mu_{kj}$ are correctly specified. The consistency of the variance estimators is established in Section S3.

For IPW-type estimators, we propose 
the following estimators of covariance matrices: 
 \begin{align*}
   &   \hat \bS_{jk}^{\rm \, ipw}  \! = \! \frac{1}{n_{jk}} \sum_{i \in {\cal I}_{jk}} \! {\rm diag} \!
\left\{ {\textstyle 
\frac{I(A_i=j)Y_i^2}{\pi_j^2(\bmZ_i) }, \, 
\frac{I(A_i=k)Y_i^2}{\pi_k^2(\bmZ_i)}  } \right\} - 
\hat{\bm{\vartheta}}_{jk} \hat{\bm{\vartheta}}_{jk}^T ,  \\
 & \hat \bS_{jk}^{\rm \, sipw} \! = \!  
 \frac{1}{n_{jk}}\sum_{i \in {{{\cal I}}}_{jk}} \! {\rm diag} \! \left\{ {\textstyle 
\frac{I(A_i=j)(Y_i-\hat\theta_{jk})^2}{\pi_j^2(\bmZ_i)}, \, 
\frac{I(A_i=k)(Y_i-\hat\theta_{kj})^2}{\pi_k^2(\bmZ_i)}
 }    \right\}  ,\\
&  \hat \bS_{jk}^{\rm \, aipw}  \! =  \!
 \frac{1}{n_{jk}}\sum_{i \in {{{\cal I}}}_{jk}} 
 \! {\rm diag} \! \left\{ {\textstyle 
 \frac{I(A_i=j)\{ Y_i-\hat\mu_{jk}(\bmX_i)\}^2}{\pi_j^2(\bmZ_i)}  , \, 
 \frac{I(A_i=k)\{ Y_i-\hat\mu_{kj}(\bmX_i) \}^2 }{\pi_k^2(\bmZ_i)} }  \right\} + \hat{\bm\Lambda}_{jk}    - \hat{\bm\delta}_{jk} \hat{\bm\delta}_{jk}^T , 
\\
&    \hat \bS_{jk}^{\rm \, saipw} \! =  \!
 \frac{1}{n_{jk}}\sum_{i \in {{{\cal I}}}_{jk}} \! {\rm diag} \! \left\{ {\textstyle 
 \frac{I(A_i=j)\{ Y_i- \hat\mu_{jk}(\bmX_i)  - \hat\delta_{jk} \}^2}{\pi_j^2(\bmZ_i)}  , \, 
 \frac{I(A_i=k)\{ Y_i- \hat\mu_{kj}(\bmX_i) -\hat\delta_{kj} \}^2}{\pi_k^2(\bmZ_i)}   }  \right\}  + \hat{\bm\Lambda}_{jk} ,
 \end{align*}
 where  $\hat{\bm{\vartheta}}_{jk} = (\hat\theta_{jk}, \hat\theta_{kj})^T$,    $\hat\theta_{jk}$ is any estimator of $\theta_{jk}$,  
$ \hat{\bm\delta}_{jk} = (\hat\delta_{jk}, \hat\delta_{kj})^T$, $\hat\delta_{jk} =  
n_{jk}^{-1}\sum_{i \in {{{\cal I}}}_{jk}}\! I(A_i=j)\{Y_i - \hat\mu_{jk} (\bmX_i)\}/\pi_j(\bmZ_i) $,
  $\hat{\bm\Lambda}_{jk}$ is $\bm\Lambda_{jk}$ with $\lambda_{jk}$ and $c_{jk}$ respectively estimated by  $\hat\lambda_{jk} = 2 \hat q_{jk}^{(j)} - \hat{\sigma}^2_{jk}$ and $\hat{c}_{jk} = \hat q_{kj}^{(j)} + \hat q_{jk}^{(k)} - \hat{q}_{jk}$,   \begin{align*}
      \hat q_{jk}^{(j)} = \frac{1}{n_{jk}}\sum_{i\in {\cal I}_{jk}}\frac{I(A_i=j)Y_i\hat\mu_{jk}(\bmX_i)}{\pi_j(Z_i)}-\frac{1}{n_{jk}^2}\sum_{i\in{\cal I}_{jk}}\frac{I(A_i=j)Y_i}{\pi_j(Z_i)}\sum_{i\in{\cal I}_{jk}}\frac{I(A_i=j)\hat\mu_{jk}(\bmX_i)}{\pi_j(Z_i)},
  \end{align*}
  $\hat q_{kj}^{(j)}$ is $\hat q_{jk}^{(j)}$ with $\hat\mu_{jk}(\bmX_i)$ switched to $\hat\mu_{kj}(\bmX_i)$, $\hat q_{jk}^{(k)} $ is $\hat q_{jk}^{(j)}$ with 
$I(A_i = j) $ and $\pi_j(Z_i)$ replaced by $I(A_i = k) $ and $\pi_k(Z_i)$, respectively,
\begin{align*}
    \hat{\sigma}^2_{jk}=\frac{1}{n_{jk}}\sum_{i\in {\cal I}_{jk}}\frac{I(A_i=j)\hat\mu^2_{jk}(\bmX_i)}{\pi_j(Z_i)}-\bigg\{\frac{1}{n_{jk}}\sum_{i\in{\cal I}_{jk}}\frac{I(A_i=j)\hat\mu_{jk}(\bmX_i)}{\pi_j(Z_i)}\bigg\}^2,
\end{align*}
  $\hat{\sigma}^2_{kj}$ is $\hat{\sigma}^2_{jk}$ with $j$ and $k$ swapped, and
  \begin{align*}
      \hat q_{jk}&=\frac{1}{2}\bigg\{ \frac{1}{n_{jk}}\sum_{i\in {\cal I}_{jk}}\frac{I(A_i=j)\hat\mu_{jk}(\bmX_i)\hat\mu_{kj}(\bmX_i)}{\pi_j(Z_i)}-\frac{1}{n_{jk}^2}\sum_{i\in{\cal I}_{jk}}\frac{I(A_i=j)\hat\mu_{jk}(\bmX_i)}{\pi_j(Z_i)}\sum_{i\in{\cal I}_{jk}}\frac{I(A_i=j)\hat\mu_{kj}(\bmX_i)}{\pi_j(Z_i)}\bigg\}\\
      &+\frac{1}{2}\bigg\{ \frac{1}{n_{jk}}\sum_{i\in {\cal I}_{jk}}\frac{I(A_i=k)\hat\mu_{jk}(\bmX_i)\hat\mu_{kj}(\bmX_i)}{\pi_k(Z_i)}-\frac{1}{n_{jk}^2}\sum_{i\in{\cal I}_{jk}}\frac{I(A_i=k)\hat\mu_{jk}(\bmX_i)}{\pi_k(Z_i)}\sum_{i\in{\cal I}_{jk}}\frac{I(A_i=k)\hat\mu_{kj}(\bmX_i)}{\pi_k(Z_i)}\bigg\}.
  \end{align*}

For the PS and APS estimators,  we propose 
the following estimators of $\bS_{jk}^{\rm \, ps} $ and $\bS_{jk}^{\rm \, aps} $, 
\begin{align*}
    \hat \bS_{jk}^{\rm \, ps} & =  
 \sum_{h=1}^{H_{jk}}
  \frac{n({\cal I}_{jk}^{(h)})}{n_{jk}} \, {\rm diag} \! \left\{ {\textstyle 
\frac{\hat\sigma_{j}^2 ( {\cal I}_{jk}^{(h)})}{\hat\pi_j({\cal I}_{jk}^{(h)})} , \, 
\frac{\hat\sigma_{k}^2 ( {\cal I}_{jk}^{(h)})}{\hat\pi_k({\cal I}_{jk}^{(h)})} } \right\}  + \hat{\bm\Gamma}_{jk}  \\
 \hat \bS_{jk}^{\rm \, aps} & = \sum_{h=1}^{H_{jk}}
  \frac{n({\cal I}_{jk}^{(h)})}{n_{jk}} \, \Big[ {\rm diag} \! \left\{ {\textstyle 
\frac{\hat\tau_j^2 ( {\cal I}_{jk}^{(h)}) }{\hat\pi_j({\cal I}_{jk}^{(h)})}  , \, 
\frac{\hat\tau_k^2 ( {\cal I}_{jk}^{(h)}) }{\hat\pi_k({\cal I}_{jk}^{(h)})}  } \right\} + \hat{\bm\Lambda}_{jk}({\cal I}_{jk}^{(h)}) \Big] + \hat{\bm\Gamma}_{jk}   ,
\end{align*} 
where $\hat\pi_j({\cal I}_{jk}^{(h)}) = n_j({\cal I}_{jk}^{(h)})/ n({\cal I}_{jk}^{(h)})$, $\hat\sigma_{j}^2 ({\cal I}_{jk}^{(h)})$ is the sample variance of $Y_i$'s with $i\in {\cal I}_{jk}^{(h)}$ and $A_i=j$, $\hat\tau_j^2 ({\cal I}_{jk}^{(h)})$ is the sample variance of $\{ Y_i - \hat\mu_{jk}(\bmX_i)\}$'s with $i \in {\cal I}_{jk}^{(h)}$ and $ A_i =j$, $\hat{\bm\Lambda}_{jk}({\cal I}_{jk}^{(h)})$ is $\bm\Lambda_{jk}$ with $\lambda_{jk}$ and $c_{jk}$ estimated respectively by $\hat\lambda_{jk} ( {\cal I}_{jk}^{(h)}) = 2\hat q_{jk}^{(j)} ({\cal I}_{jk}^{(h)}) - \hat\sigma_{jk}^2({\cal I}_{jk}^{(h)})$ and $\hat{c}_{jk}({\cal I}_{jk}^{(h)})= \hat q_{kj}^{(j)} ({\cal I}_{jk}^{(h)}) +\hat q_{jk}^{(k)} ({\cal I}_{jk}^{(h)})  - \hat q_{jk}({\cal I}_{jk}^{(h)}) $, 
$\hat q_{jk}^{(j)}({\cal I}_{jk}^{(h)})$ is the sample covariance of $(Y_i, \hat\mu_{jk}(\bmX_i))$'s with $i \in {\cal I}_{jk}^{(h)}$ and $ A_i =j$,
$\hat q_{kj}^{(j)}({\cal I}_{jk}^{(h)}) $ is $\hat q_{jk}^{(j)}({\cal I}_{jk}^{(h)})$ with $\hat\mu_{jk}(\bmX_i)$ switched to $\hat\mu_{kj}(\bmX_i)$, $\hat q_{jk}^{(k)}({\cal I}_{jk}^{(h)}) $ is $\hat q_{jk}^{(j)}({\cal I}_{jk}^{(h)})$ {with $A_i=j$ switched to $A_i= k$}, 
$\hat{\sigma}^2_{jk}({\cal I}_{jk}^{(h)})$ and $\hat{\sigma}^2_{kj}({\cal I}_{jk}^{(h)})$ are the diagonal entries and $\hat{q}_{jk}({\cal I}_{jk}^{(h)}) $ is the off-diagonal entry of the sample covariance matrix 
of $(\hat\mu_{jk}(\bmX_i), \hat\mu_{kj}(\bmX_i))$'s with $i \in {\cal I}_{jk}^{(h)}$, $ \hat{\bm\Gamma}_{jk}   $ is the sample covariance matrix of $( \bar{Y}_{j}(S_i),
 \bar{Y}_{k}(S_i))$'s for $i\in {\cal I}_{jk}$, and
  $\bar{Y}_{j}(S_i)$ is the sample mean of $Y_i$'s with $i \in {\cal I}_{jk}$, $A_i=j$, and $S_i={\cal I}_{jk}^{(h)}$.

\subsection{Discussions of alternative analysis sets}

Here, we discuss the use of alternative analysis sets, defined as $\mathcal{I}_{jk}\cap \{i: g(A_i, R_i) \in \mathcal{C}\}$, where $R_i$ indicates the sub-study that individual $i$ is assigned to in sub-study platform trials, and is undefined in multi-arm platform trials, $g(\cdot)$ and $\mathcal{C}$ is respectively a function and set of fixed values that is pre-specified to define the subset. For example, in Design 3 in Figure 3, the analysis set focusing on sub-study 1 is defined as $\mathcal{I}_{jk}\cap \{i: R_i = 1\}$. The reason that we consider the analysis set of this form is that $A_i, R_i$ are randomized and therefore it can be robustly re-weighted to still consistently estimate the treatment effect in the ECE population. 

The sub-study variable $R$ and intervention assignment variable $A$ are randomized and satisfy the following condition:
\begin{condition}
For the $Z$ defined in Assumption 1, we have $   (A, R)\perp (\bmW, Y^{(1)}, \dots, Y^{(J)}) \mid \bmZ $.
\end{condition}

This analysis set $\mathcal{I}_{jk}\cap \{i: g(A_i, R_i) \in \mathcal{C}\}$ is required to satisfy a positivity condition: 
\begin{condition}
 For both $j$ and $k$, $P\big(A = j, g(A, R) \in \mathcal{C}\mid Z\big) >0$ almost surely for all values of $Z$ that have positive density within $\pi_j (Z)>0, \pi_k(Z)>0$. \label{cond: positvitiy}
\end{condition}
For example, in Design 3 of Figure 3, $Z$ denotes the enrollment window. For comparing treatments 1 and 2, the analysis set  $\mathcal{I}_{12}\cap \{i: R_i=2\}$, which subsets to the sub-study 2, violates Condition \ref{cond: positvitiy} because $P(A=2, R=2\mid Z)=0$ almost surely. However, the analysis set  $\mathcal{I}_{12}\cap \{i: R_i=1\}$, which subsets to the sub-study 1, satisfies Condition \ref{cond: positvitiy} because $P(A=2, R=1\mid Z)>0$ and $P(A=1, R=1\mid Z)>0$ almost surely. Moreover, for $j=1,2$, $P(A=j, R=1\mid Z)=P(A=j\mid R=1, Z) P(R=1\mid Z)= P(A=j\mid R=1) P(R=1\mid Z)= \frac12 P(R=1\mid Z)$, which equals $\frac12$ when $Z=\text{EW1}$ and equals $\frac16$ when $Z=\text{EW2}$. 

For the IPW approach, the identification formula based on the alternative analysis set is 
\begin{align*} 
    \theta_{jk} = E\bigg\{  \frac{I(A= j) I( g(A, R) \in \mathcal{C}) Y  }{   P(A= j, g(A, R)\in \mathcal{C} \mid Z )  }  \ \Big| \ \pi_{j}(\bmZ)>0, \pi_{k}(\bmZ)>0\bigg\} , \label{eq: IPW identification}
\end{align*}
where $  P(A= j, g(A, R)\in \mathcal{C} \mid Z ) $ is known since the joint distribution of $(A, R)\mid Z$ is known. Based on this identification, we propose the following IPW estimator, which parallels the estimator from Section 4.3:
$$
\hat\theta_{jk}^{\rm ipw} = \frac{1}{\tilde{n}_{jk}}\sum_{i\in \tilde{\cal I}_{jk}}\frac{I(A_i=j)Y_i}{\tilde{\pi}_j(Z_i)},
$$
where $\Tilde{\cal I}_{jk}$ is the analysis set ${\cal I}_{jk}\cap \{i: g(A_i,R_i)\in {\cal C}\}$, $\tilde{n}_{jk}$ denotes the size of $\tilde{\cal I}_{jk}$, and $\tilde{\pi}_j(Z_i)=P(A_i=j,g(A_i,R_i)\in{\cal C}\mid Z_i)$. Normalizing the weights $I(A_i=j)/\tilde{\pi}_j(Z_i)$ for $i\in\tilde{\cal I}_{jk}$ to sum to 1 gives the SIPW estimator.

For covariate adjustment, the AIPW estimator, which leverages covariate information from all individuals within $\tilde{\cal I}_{jk}$, can be defined as
$$
\aipw =  \ \frac{1}{\tilde{n}_{jk}}  \sum_{i \in {{\tilde{\cal I}}}_{jk}}  \frac{ I(A_i=j) \{Y_i - \tilde \mu_{jk}(\bmX_i) \} }{\tilde{\pi}_j(\bmZ_i)}   + \frac{1}{\tilde{n}_{jk}}  \sum_{i \in {{\tilde{\cal I}}}_{jk}}   \tilde \mu_{jk}(\bmX_i),
$$
where we estimate $E\left[Y\mid \bmX,\pi_j(Z)>0,\pi_k(Z)>0,A=j, g(A,R)\in{\cal C}\right]$ by $\tilde{\mu}_{jk}(\bmX)$. Similarly, one can obtain the SAIPW estimator by normalizing the weights in the AIPW estimator.

The PS-type estimators can be constructed following the approach in Section 4.4. Specifically, we divide all individuals in $\tilde{\cal I}_{jk}$ into $\tilde{\cal I}_{jk}^{(1)},\dots,\tilde{\cal I}_{jk}^{(\tilde{H}_{jk})}$, where $\tilde{H}_{jk}$ is the number of strata, such that the values of $\tilde{\pi}_j(Z_i)$ and $\tilde{\pi}_k(Z_i)$ are constant within each stratum $\tilde{\cal I}_{jk}^{(h)}$. The PS estimator of $\theta_{jk}$ can be obtained by replacing $n_{jk}$, $H_{jk}$, and ${\cal I}_{jk}$ in the PS estimator (defined by ${{{\cal I}}}_{jk}$) in Section 4.4 with $\tilde{n}_{jk}$, $\tilde{H}_{jk}$, and $\tilde{\cal I}_{jk}$, respectively:
$$
\hat\theta_{jk}^{\rm \, ps} = \frac{1}{\tilde n_{jk}} 
    \sum_{h=1}^{\tilde H_{jk}} \frac{ n({\tilde {\cal I}}_{jk}^{(h)})}{n_j({\tilde{\cal I}}_{jk}^{(h)})} \sum_{i \in {{\tilde{\cal I}}}_{jk}^{(h)}}I(A_i=j)Y_i = 
    \frac{1}{\tilde n_{jk}} 
    \sum_{h=1}^{\tilde H_{jk}}  n({\tilde{\cal I}}_{jk}^{(h)})\bar{Y}_j(\tilde {\cal I}_{jk}^{(h)}).
$$
Similarly,  further substituting $\hat\mu_{jk}$ with $\tilde{\mu}_{jk}$ in the APS estimator (defined by ${{{\cal I}}}_{jk}$) provides the corresponding APS estimator on the analysis set.

All the estimators of $\bm {\vartheta}_{jk}$ using the analysis set $\tilde{\cal I}_{jk}$ are asymptotically normal under Assumption 1 and Assumption 2  in which $\hat\mu_{jk}$ is replaced by $\tilde{\mu}_{jk}$ if covariate adjustments are applied.  Specifically, Theorem 1 still holds for the estimators based the analysis set $\tilde{\cal I}_{jk}$ when $\pi_j(\cdot),\pi_k(\cdot)$ and $\hat\mu_{jk}(\cdot)$, $\hat\mu_{kj}(\cdot)$ are replaced with $\tilde{\pi}_j(\cdot),\tilde{\pi}_k(\cdot)$ and $\tilde{\mu}_{jk}(\cdot),\tilde{\mu}_{kj}(\cdot),$ respectively. Efficiency comparisons follow similarly to those in Section 5.2.

To derive consistent estimators of the asymptotic covariance matrices for the analysis set estimators, all empirical quantities in the covariance matrix estimators  using all individuals in ${{{\cal I}}}_{jk}$ from Section \ref{sec: variance estimation} are replaced with the corresponding values from the analysis set.

\subsection{Donsker class in Assumption 2} 
{For given $j$ and $k$, there exist a class $\mathcal{F}_{jk}$ of
measurable functions of $\bmX$ and a measurable envelope $G_{jk}$
satisfying
\(
|f(\bm x)| \leq G_{jk}(\bm x)
\)
for every $f\in\mathcal{F}_{jk}$ and $\bm x$, such that
(i) $\mu_{jk}\in\mathcal{F}_{jk}$, where $\mu_{jk}$ is the working model in Assumption 2,
(ii) $P(\hat\mu_{jk}\in\mathcal{F}_{jk})\rightarrow 1$ as
$n\rightarrow\infty$, and
(iii) $E[G_{jk}^2(\bmX)]<\infty$ and
\begin{equation*}
\int_0^1 \sup_Q
\sqrt{
\log N\left(
\mathcal{F}_{jk},
\|\cdot\|_{L_2(Q)},
s\|G_{jk}\|_{L_2(Q)}
\right)
}\,\mathrm{d}s
<\infty .
\end{equation*}
Here the supremum is taken over all finitely supported probability
measures $Q$ on the range of $\bmX$ satisfying
$0<\|G_{jk}\|^2_{L_2(Q)}=\int G_{jk}^2dQ<\infty$.
For $s>0$,
$N(\mathcal{F}_{jk},\|\cdot\|_{L_2(Q)},s)$ denotes the
$s$-covering number of
$(\mathcal{F}_{jk},\|\cdot\|_{L_2(Q)})$.
Let $P_X$ denote the distribution of $\bmX$. For every $\delta>0$, define
\(
\mathcal{F}_{jk,\delta}
=
\left\{
f-g:f,g\in\mathcal{F}_{jk},
\|f-g\|_{L_2(P_X)}<\delta
\right\}
\)
and
\(
\mathcal{F}_{jk,\infty}^{\,2}
=
\left\{
(f-g)^2:f,g\in\mathcal{F}_{jk}
\right\}.
\)
We assume that $\mathcal{F}_{jk,\delta}$ and
$\mathcal{F}_{jk,\infty}^{\,2}$ are $P_X$-measurable for every
$\delta>0$
\citep[Theorem~2.5.2]{vandervaartWeakConvergenceEmpirical1996}.
}		

\section{Simulation study}\label{supp: simu}
\vspace{-3mm}

We conduct a simulation study based on the first three enrollment windows outlined in Figure \ref{fig: schema}(a) to examine the finite-sample performance of all estimators in Section 4.

In the simulation, the observed baseline covariate vector is $\bmW=(X_c, X_b, Z_{\rm sub}, Z_{\rm EW})$, where 
 $X_c$ is a continuous covariate uniformly distributed over the interval $(-3,3)$, $X_b$ is a binary covariate with $P(X_b=1)=0.5$, $Z_{\rm sub}$ is a binary indicator for disease subtype with $P(Z_{\rm sub}=1)=0.8$, and $Z_{\rm EW} \in \{1,2,3\}$ denotes the enrollment window and satisfies
$$
P(Z_{\rm EW}= t \mid X_c, X_b, Z_{\rm sub}, U) = \frac{\exp(Q_t)}{\exp(Q_1) + \exp(Q_2) + \exp(Q_3)}, \quad t=1,2,3,
$$
 $Q_1 = 0.5 + X_c + 2X_b - Z_{\rm sub} + U$, $Q_2 = 1 + 2X_c + X_b - Z_{\rm sub} + U$, $Q_3 = -0.5 + X_c + X_b + Z_{\rm sub} + U$, and the unobserved baseline variable $U$ follows a standard normal distribution and is independent of $X_c$, $X_b$, and $Z_{\rm sub}$.
{After intervention assignment, the observed outcome is generated as
\[
Y=\begin{cases}1+X_c+X_b+Z_{\rm sub}+U+\epsilon, & A=1,\\ 
1+X_c^2+X_b+Z_{\rm sub}+U+\epsilon, & A=2,\\
3+X_cX_b+Z_{\rm sub}+U+\epsilon, & A=3,\\
2+X_cZ_{\rm sub}-X_b+2U+\epsilon, & A=4,\end{cases}
\]
where $\epsilon$ follows a standard normal distribution and is independent of $\bmW$, $U$, and $A$. The observed outcome is associated with $X_c$, $X_b$, $Z_{\rm sub}$, and $U$, as well as indirectly with the enrollment window $Z_{\rm EW}$.
}

Individuals are first randomized into one of three sub-studies 1-3 depending on their enrollment windows and disease subtypes, and then to treatments within the assigned sub-study. Sub-study 1 has treatments 1
and 2, sub-study 2 has treatments 1 and 3, and sub-study 3 has treatments 1 and 4. Within each sub-study, treatments are randomized in a 1:1 ratio.  
The assignment probabilities are in Table \ref{table: prob}. 
The total sample size $n$ is 500 or 1,000. 
 When $n=500$, the expected numbers of individuals per treatment are 123, 51,  and 76 in sub-studies 1-3, respectively. 
 
\begin{table}[h]
\spacingset{1.2} 
\centering
{\footnotesize 
\caption{Assignment probabilities in the simulation study\label{table: prob}}
\begin{tabular}{cccccccccccc}
\hline
\multicolumn{2}{c}{$\bmZ$} & & \multicolumn{3}{c}{Sub-study} & & \multicolumn{4}{c}{Treatment} \\ \cline{1-2} \cline{4-6} \cline{8-11} 
$Z_{\rm EW}$ & $Z_{\rm sub}$ &  & 1 & 2 & 3 & & $\pi_1(\bmZ)$ & $\pi_2(\bmZ)$ & $\pi_3(\bmZ)$ & $\pi_4(\bmZ)$ 
\\ \hline
1 & 1 && 0.4 & 0.6 & 0 & & 0.50 & 0.20 & 0.30 & 0 \\
1 & 0 && 1 & 0 & 0 && 0.50 & 0.50 & 0 & 0 \\
2 & 1 && 0.3 & 0.3 & 0.4 & & 0.50 & 0.15 & 0.15 & 0.20 \\
2 & 0 & & 1 & 0 & 0 & & 0.50 & 0.50 & 0 & 0 \\
3 & 1 && 0.4 & 0 & 0.6 & & 0.50 & 0.20 & 0 & 0.30 \\ 
3 & 0 & & 1 & 0 & 0 & & 0.50 & 0.50 & 0 & 0 \\
\hline
\end{tabular}}
\end{table}

We consider the following estimators of the linear contrast  $\theta_{j1} - \theta_{1j}$, for $j=2,3,4.$ \vspace{-2mm}  
\begin{enumerate}
\item[(i)]
{\bf Four IPW-type estimators.} 
We consider the IPW, SIPW, and  SAIPW  estimators described in Section 4.3. For SAIPW, $\hat\mu_{jk}$ is obtained from fitting a linear regression of $Y$ on
 $(X_c, X_b, Z_{\rm sub})$, using data from treatment arm $j$ in  ${\cal I}_{jk}$.   {Since $\delta_{jk} = \delta_{kj} =0$ in this simulation, the AIPW estimator is numerically almost the same as the SAIPW estimator and thus is omitted}. We add the SAIPW  estimator with $\bmX = S$, denoted by SAIPW($S$), to check its equivalence with the PS estimator as discussed in Corollary 5.  Note that the working models are correctly specified only for treatment arm 1, as the true models for the other arms are not
  linear in $(X_c, X_b, Z_{\rm sub})$. 

 \vspace{-2mm} 
\item[(ii)]
{\bf Four PS-type estimators.} 
We consider two ways of stratification for PS and APS. 
The first way is as described in Section 4.4. Specifically, we stratify individuals according to the values that $\pi_j(Z)$ and $\pi_1(Z)$ can take. Based on the assignment probabilities in Table \ref{table: prob}, we define the following strata: ${\cal I}_{12}^{(1)},{\cal I}_{12}^{(2)},{\cal I}_{12}^{(3)}$ for estimating $\theta_{21}-\theta_{12}$, ${\cal I}_{13}^{(1)},{\cal I}_{13}^{(2)}$ for estimating $\theta_{31}-\theta_{13}$, and ${\cal I}_{14}^{(1)},{\cal I}_{14}^{(2)}$ for estimating $\theta_{41}-\theta_{14}$. The details of each stratum are provided in Table \ref{tab: strata simu}.
\begin{table}[H]
\centering
{\footnotesize
\caption{Details of strata in the simulation study}
\label{tab: strata simu}
\begin{tabular}{@{}cccc@{}}
\hline
Estimand & Stratum & Stratum Definition \\ \hline
\multirow{3}{*}{$\theta_{21}-\theta_{12}$} & ${\cal I}_{12}^{(1)}$ & $\{ i: \pi_1(\bmZ_i) = 0.5, \pi_2(\bmZ_i) = 0.5 \}$ \\
 & ${\cal I}_{12}^{(2)}$ & $\{ i: \pi_1(\bmZ_i) = 0.5, \pi_2(\bmZ_i) = 0.2  \}$ \\
 & ${\cal I}_{12}^{(3)}$ & $\{ i: \pi_1(\bmZ_i) = 0.5, \pi_2(\bmZ_i) = 0.15 \}$ \\
 \hline
\multirow{2}{*}{$\theta_{31}-\theta_{13}$} & ${\cal I}_{13}^{(1)}$ & $\{ i: \pi_1(\bmZ_i) = 0.5, \pi_3(\bmZ_i) = 0.3 \}$ \\
 & ${\cal I}_{13}^{(2)}$ & $\{ i: \pi_1(\bmZ_i) = 0.5, \pi_3(\bmZ_i) = 0.15 \}$ \\
 \hline
\multirow{2}{*}{$\theta_{41}-\theta_{14}$} & ${\cal I}_{14}^{(1)}$ & $\{ i: \pi_1(\bmZ_i) = 0.5, \pi_4(\bmZ_i) = 0.2\}$ \\
 & ${\cal I}_{14}^{(2)}$ & $\{ i: \pi_1(\bmZ_i) = 0.5, \pi_4(\bmZ_i) = 0.3 \}$ \\ \hline
\end{tabular}}
\end{table}

To see the effect of stratification, 
 the second way is to stratify on all six joint levels of $\bmZ = (Z_{\rm EW},Z_{\rm sub})$, denoted as PS($Z$) and APS($Z$).
 The same working models $\hat \mu_{jk}$ for SAIPW are used for APS.\vspace{-2mm} 
\item[(iii)] 
{\bf The naive estimator.} For comparison we include the naive estimator of $\theta_{j1} - \theta_{1j}$, $j=2,3,4$, as
described in Section 4.2. \vspace{-2mm} 
\item[(iv)]
{\bf Estimators within each sub-study.} {For comparison, we also evaluate traditional ANOVA (sample means) and ANCOVA estimators (using linear working models to adjust for $ X_c, X_b$, and $Z_{\rm sub}$) based solely on data from each individual sub-study.  Although these sub-study-specific estimators estimate the treatment effect within each sub-study, which differs from our target estimand defined in \eqref{effect1}, we can still compare their variances. 
}
\end{enumerate}

The simulation results based on $5,000$ runs are reported in Table \ref{tb: 1000} for the  eleven estimators  described in (i)-(iv). The results contain the average absolute bias,  average relative bias, standard deviation (SD), the average of the standard error (SE), and the coverage probability (CP) of the $95\%$ confidence interval. The true values of estimands are approximated using an independent simulated dataset with size $10^7$. The following is a summary of the  results in Table \ref{tb: 1000}.

\begin{table}[t]
\spacingset{1.2} 
{\centering
\caption{Simulation results based on $5,000$ simulation runs}
\label{tb: 1000}
\resizebox{\textwidth}{!}
{
\begin{tabular}{clccccccccccccccccc}
\toprule
          &               & \multicolumn{5}{c}{$\theta_{21}-\theta_{12}= 3$ }&  & \multicolumn{5}{c}{$\theta_{31}-\theta_{13}= 1.145$}&  & \multicolumn{5}{c}{$\theta_{41}-\theta_{14}=-0.886$} \\
                           \cmidrule{3-7} \cmidrule{9-13} \cmidrule{15-19}
$n$ & \multicolumn{1}{c}{Method} & {\small Abs Bias} & {\small Rel Bias (\%)}    & SD     & SE     & CP &    &  {\small Abs Bias} & {\small Rel Bias (\%)}    & SD     & SE     & CP  &   &  {\small Abs Bias} & {\small Rel Bias (\%)}   & SD     & SE     & CP     \\ 
\hline
 500 
 & Naive  & -0.231 &-7.715  & 0.320 & 0.316 & 0.874 &  & -0.185 & -16.185 & 0.342 & 0.340 & 0.916 &  & -0.205 & 23.089  & 0.384 & 0.380 & 0.911 \\
 & IPW    & -0.006 &-0.203 & 0.639 & 0.636 & 0.946 &  & 0.004 &0.390 & 0.776 & 0.777 & 0.948 &  & -0.007 & 0.782 & 0.500 & 0.497 & 0.948 \\
 & SIPW   & -0.003 &-0.084 & 0.341 & 0.336 & 0.941 &  & 0.005 &0.463 & 0.347 & 0.341 & 0.943 &  & 0.001 &-0.101  & 0.389 & 0.381 & 0.942 \\
 & SAIPW   & -0.018 & -0.603 & 0.329 & 0.317 & 0.935 &  & 0.001 &0.115 & 0.284 & 0.271 & 0.932 &  & -0.001 &0.157 & 0.297 & 0.287 & 0.933 \\
 &SAIPW$(S)$ & 0.000 &-0.009 & 0.336 & 0.327 & 0.938 &  & 0.009 &0.784 & 0.327 & 0.320 & 0.936 &  & 0.002 &-0.224 & 0.356 & 0.347 & 0.940 \\
& PS   & 0.000 &-0.009 & 0.336 & 0.335 & 0.945 &  & 0.009 & 0.784  & 0.327 & 0.330 & 0.949 &  & 0.002 &-0.224  & 0.356 & 0.356 & 0.946 \\
  & APS  & -0.013 &-0.433 & 0.329 & 0.323 & 0.939 &  & -0.001 & -0.083  & 0.286 & 0.280 & 0.941 &  & -0.002 &0.227 & 0.298 & 0.293 & 0.944 \\
 & PS$(Z)$  & \multicolumn{5}{c}{Estimate or its SE cannot} &&\multicolumn{5}{c}{Same as PS and omitted} &&\multicolumn{5}{c}{Same as PS and omitted} \\
& APS$(Z)$  & \multicolumn{5}{c}{be computed in 408 runs}&&\multicolumn{5}{c}{Same as APS and omitted} &&\multicolumn{5}{c}{Same as APS and omitted}  \\
 & ANOVA*  & 0.001 & 0.033  & 0.354 & 0.350 & 0.946 &  & 0.006 & 0.474 & 0.421 & 0.417 & 0.945 &  & 0.004 &-0.446  & 0.424 & 0.425 & 0.946 \\ \smallskip
 & ANCOVA* & -0.008 & -0.269  & 0.329 & 0.321 & 0.942 &  & -0.005 &-0.387  & 0.319 & 0.311 & 0.942 &  & 0.002 &-0.198  & 0.321 & 0.314 & 0.942 \\  \midrule
1000 
& Naive  & -0.230 &-7.667  & 0.226 & 0.224 & 0.819 &  & -0.189 & -16.479 & 0.240 & 0.239 & 0.872 &  & -0.206 & 23.201  & 0.269 & 0.268 & 0.876 \\
& IPW    & -0.001 &-0.041 & 0.453 & 0.451 & 0.947 &  & 0.012 &1.062 & 0.550 & 0.550 & 0.951 &  & 0.003 &-0.353 & 0.355 & 0.352 & 0.943 \\
& SIPW   & 0.000 & 0.009 & 0.243 & 0.239 & 0.945 &  & 0.004 & 0.345 & 0.246 & 0.243 & 0.944 &  & 0.001 & -0.148  & 0.272 & 0.270 & 0.948 \\
& SAIPW  & -0.009 & -0.299 & 0.232 & 0.227 & 0.940 &  & 0.004 & 0.360 & 0.198 & 0.195 & 0.946 &  & 0.000 & -0.019 & 0.212 & 0.205 & 0.942 \\
& SAIPW$(S)$ & 0.001 & 0.039 & 0.238 & 0.233 & 0.943 &  & 0.004 & 0.392  & 0.233 & 0.228 & 0.939 &  & 0.003 & -0.304  & 0.252 & 0.247 & 0.942  \\
& PS   & 0.001 & 0.039 & 0.238 & 0.236 & 0.948 &  & 0.004 & 0.392  & 0.233 & 0.232 & 0.944 &  & 0.003 & -0.304  & 0.252 & 0.250 & 0.947 \\
& APS  & -0.006 & -0.190 & 0.232 & 0.228 & 0.943 &  & 0.003  & 0.274 & 0.198 & 0.198 & 0.950 &  & 0.000 & -0.032 & 0.213 & 0.207 & 0.944 \\
& PS$(Z)$  & 0.001 & 0.044 & 0.238 & 0.236 & 0.948 & &\multicolumn{5}{c}{Same as PS and omitted} &&\multicolumn{5}{c}{Same as PS and omitted} \\
& APS$(Z)$  & -0.005 & -0.173 & 0.233 & 0.229 & 0.942 &&\multicolumn{5}{c}{Same as APS and omitted} &&\multicolumn{5}{c}{Same as APS and omitted} \\
& ANOVA*  &  0.003 & 0.097  & 0.251 & 0.248 & 0.947 &  & 0.007 & 0.530 & 0.295 & 0.294 & 0.944 &  & 0.002 & -0.283  & 0.301 & 0.299 & 0.948 \\ \smallskip
& ANCOVA* &  -0.003 & -0.109 & 0.233 & 0.228 & 0.947 &  & 0.005 & 0.353  & 0.222 & 0.220 & 0.947 &  & 0.001 & -0.109  & 0.227 & 0.222 & 0.945 \\ 
\bottomrule \vspace{0.1mm}
\end{tabular}%
} 
} 
{\footnotesize
 SAIPW($S$): SAIPW  with $\bmX = S$. \\
PS($Z$) and APS($Z$): post-stratification using all joint levels of $Z= (Z_{\rm EW},Z_{\rm sub})$ as strata (PS = PS($Z$) and APS = APS($Z$)  for estimating 
$\theta_{31}-\theta_{13}$ and $\theta_{41}-\theta_{14}$). \\
ANOVA* and ANCOVA* are based on data within each sub-study, estimating the within sub-study treatment effects, which are $3.054$, $1.279$, and $-0.881$ for sub-studies 1-3, respectively. The bias for these two estimators is evaluated against these within–sub-study effects.
}
\end{table}

\vspace{-4mm}
\begin{enumerate}
    \item 
    The simulation results support our asymptotic theory in Section 5. Specifically, the naive estimator is biased, while all IPW-type and PS-type estimators have negligible biases; despite the misspecification of working models, covariate adjustment improves efficiency as reflected by reductions in SD; the PS and SAIPW($S$) estimators demonstrate nearly identical performance. Additionally,  SEs are close to  SDs and CPs are close to the nominal level 95\%.
     \vspace{-3mm}
    \item No results are reported in Table 1 for PS($Z$) and APS($Z$) at $n=500$ because either the estimate of $\theta_{jk}$ or its SE cannot be computed in 408 out of 5,000 simulation runs.  This issue arises due to insufficient data in the stratum defined by $(Z_{\rm EW}, Z_{\rm sub})=(3,0)$, where the average number of individuals is 10. A similar issue occurs in 3 out of 5,000 runs at $n=1,000$, so the results in Table 1 are based on 4,997 runs. This problem does not affect the PS and APS estimators: our proposed stratification method combines three strata  $(Z_{\rm EW}, Z_{\rm sub})=(1,0), (2,0), (3,0)$ into ${\cal I}_{12}^{(1)}$ because they all have constant $\pi_1(Z) = 0.5$ and $\pi_2(Z)=0.5$. \vspace{-3mm}
    \item 
Between IPW and SIPW, it is clear that SIPW is much better in terms of SD under our simulation setting. Notably, IPW can have an even larger SD than  ANOVA within substudies, which usually does not occur with SIPW. 
    \vspace{-3mm}
    \item 
 The traditional ANOVA and ANCOVA estimators analyze the platform trials as if they were separate, stand-alone studies. Without covariate adjustment,
    the ANOVA estimator has a much larger SD than the SIPW estimator when estimating $\theta_{31}-\theta_{13}$ and $\theta_{41}-\theta_{14}$. A similar conclusion applies to the ANCOVA estimator when compared with the SAIPW or APS estimator. The efficiency difference is not as pronounced for 
 estimating $\theta_{21}-\theta_{12}$, although the ANOVA and ANCOVA estimators still have larger SDs than the corresponding stabilized {IPW-type} and PS-type estimators. The reason for this is because the traditional ANOVA and ANCOVA do not utilize data across sub-studies, for example,   when  $n=500$, about 123, 51, and 76 individuals are used under treatment arm 1 
in sub-studies 1-3, 
respectively,  whereas  about 250, 121, and 150  individuals in treatment arm 1 
 under platform trial are shared  for our proposed estimators of $\theta_{j1}-\theta_{1j}$, $j=2,3,4$, respectively. 
\end{enumerate}

\section{Technical proofs}

\subsection{Proof of the limit of the naive estimator}

Let $I_{jk}(\bmZ) = I ( \pi_j(\bmZ)>0 , \pi_k(\bmZ)>0)$.
The naive estimator described in Section 4.2 is the sample mean 
$$ \bar{Y}_{jk} = \sum_{i \in {\cal I}_{jk}} I(A_i=j)Y_i 
\Big / \sum_{i \in {\cal I}_{jk}} I(A_i=j). $$
As $n \to \infty$, by the law of large numbers, 
\begin{equation}
    \label{e10} 
\bar{Y}_{jk} \to   \frac{E\{ I( \pi_j(\bmZ)>0, \pi_k(\bmZ)>0, A = j) Y \}}{ E\{ I( \pi_j(\bmZ)>0, \pi_k(\bmZ)>0, A = j)\}} \quad \mbox{in probability.}
\end{equation} 
By consistency and randomization assumptions, the numerator of the limit in (\ref{e10}) is equal to
\begin{align*}
    E \{ I_{jk}(\bmZ) I(A=j)Y \} & = E \{ I_{jk}(\bmZ) I(A=j)Y^{(j)} \} \\
    & = E [  E \{ I_{jk}(\bmZ) I(A=j)Y^{(j)} \mid \bmZ \} ] \\
    & = E [ I_{jk}(\bmZ)  E \{  I(A=j)Y^{(j)}  \mid \bmZ \} ] \\
    & =  E [ I_{jk}(\bmZ) \pi_j(\bmZ )  E \{ Y^{(j)}  \mid \bmZ \} ] \\
    & = E [  \pi_j(\bmZ )  E \{ Y^{(j)}  \mid \bmZ \} \mid I_{jk}(\bmZ)=1 ] P \{ I_{jk}(\bmZ) =1 \} . 
\end{align*}
Similarly, the denominator of the limit in (\ref{e10}) is equal to 
$$ E \{  \pi_j(\bmZ )  \mid I_{jk}(\bmZ)=1 \} P \{ I_{jk}(\bmZ) =1 \} .  $$
Then, taking the ratio of the previous two quantities gives the limit $  \frac{E\{  \pi_j(\bmZ) E( Y^{(j)} \mid \bmZ) \mid \pi_j(\bmZ)>0, \pi_k(\bmZ)>0 \}}{ E\{  \pi_j(\bmZ)\mid  \pi_j(\bmZ)>0, \pi_k(\bmZ)>0\}} $.

\subsection{Proof of \eqref{eq: IPW identification}}

Under the consistency assumption, the right side of  \eqref{eq: IPW identification} is
\begin{align*}
    E\bigg\{  \frac{I(A= j) Y }{\pi_{j}(\bmZ)} \, \Big| \, I_{jk} (\bmZ) =1 \bigg\} 
    &= E\bigg\{   \frac{I_{jk} (\bmZ)I(A= j) Y^{(j)} }{\pi_{j}(\bmZ)} \, \Big| \, I_{jk} (\bmZ) =1 \bigg\} 
    \\
    &=  E\bigg\{   \frac{I_{jk}(\bmZ) I(A= j) Y^{(j)} }{\pi_{j}(\bmZ)}  \bigg\} \Big/ P \{ I_{jk} (\bmZ) =1 \} \\
    & =  E \left[  E\bigg\{   \frac{I_{jk}(\bmZ)I(A= j) Y^{(j)} }{\pi_{j}(\bmZ)} \, \Big| \, \bmZ \bigg\} \right] \Big/ P \{ I_{jk} (\bmZ) =1 \} \\
       & =  E \left[  \frac{I_{jk}(\bmZ) P(A=j \mid \bmZ ) E\{   Y^{(j)}\mid \bmZ \}}{\pi_{j}(\bmZ)}  \right] \Big/ P \{ I_{jk} (\bmZ) =1 \} \\
         & =  E \left[ I_{jk}(\bmZ)  E\{   Y^{(j)} \mid \bmZ \} \right] \Big/ P \{ I_{jk} (\bmZ) =1 \} \\
               & =    E\left\{I_{jk}(\bmZ)Y^{(j)}\right\}  \Big/ P \{ I_{jk} (\bmZ) =1 \} \\
                & =    E\left\{ I_{jk} (\bmZ)  Y^{(j)} \, \Big| \, I_{jk} (\bmZ) =1 \right\} \\
                           & =    E\{  Y^{(j)} \mid I_{jk} (\bmZ) =1 \} = \theta_{jk} ,
\end{align*}
where the fourth equation follows from Assumption 1 and the rest of equations follow from the properties of conditional expectations. This proves (\ref{eq: IPW identification}). 

\subsection{Lemma \ref{lemma: S}}

The following lemma is useful for asymptotic results of PS and APS estimators. 

 {\begin{lemma}
      Under Assumption 1,  for fixed $j$ and $k$ and  $S =$ the covariate for post-stratification, 
 $I(A=j),I(A=k) \perp (\bmW, Y^{(l)}) \mid S$ for any $l=1,\dots,J$ and 
     $\pi_j(Z)  = P(A=j \mid S) $,  $j=1,...,J$. \label{lemma: S}
 \end{lemma}

\noindent
{\bf Proof.} 
From Assumption 1 and the property of the propensity score \citep{rosenbaum1983central, imbens2000propensityscore}, we have that  $I(A=a) \perp (\bmW, Y^{(l)}) \mid \pi_j(\bmZ),\pi_k(\bmZ) $ for $a=j$ and $k$ and $l=1,\dots,J$, because \begin{align*}
    &P(I(A=a)=1\mid \bmW, Y^{(l)},\pi_j(\bmZ),\pi_k(\bmZ))\\
    &=E\left[P(I(A=a)=1\mid \bmW, Y^{(l)},\pi_j(\bmZ),\pi_k(\bmZ),\bmZ)\mid \bmW, Y^{(l)},\pi_j(\bmZ),\pi_k(\bmZ)\right] \\
    &=E\left[P(I(A=a)=1\mid \bmZ)\mid \bmW, Y^{(l)},\pi_j(\bmZ),\pi_k(\bmZ)\right]\\
    &=E[\pi_a(Z)\mid \bmW, Y^{(l)},\pi_j(\bmZ),\pi_k(\bmZ)]\\
    &=\pi_a(Z).
\end{align*}Then, observing that for any fixed $l=1,\dots,J$, 
\begin{align*}
    &P(I(A=j)=1,I(A=k)=1\mid \bmW,Y^{(l)},\pi_j(Z),\pi_k(Z) )=0,\\
    &P(I(A=j)=1,I(A=k)=0\mid \bmW,Y^{(l)},\pi_j(Z),\pi_k(Z) )\\
    &=P(I(A=j)=1\mid\pi_j(Z),\pi_k(Z) )\\
    &=P(I(A=j)=1,I(A=k)=0\mid\pi_j(Z),\pi_k(Z) ),
\end{align*}
and similarly $P(I(A=j)=0,I(A=k)=1\mid \bmW,Y^{(l)},\pi_j(Z),\pi_k(Z) )=P(I(A=j)=0,I(A=k)=1\mid\pi_j(Z),\pi_k(Z) )$, we conclude that $I(A=j),I(A=k) \perp (\bmW, Y^{(l)}) \mid \pi_j(\bmZ),\pi_k(\bmZ)$ for $l=1,\dots,J$. Thus, $I(A=j),I(A=k) \perp (\bmW, Y^{(l)}) \mid \pi_j(\bmZ),\pi_k(\bmZ),S $. Since $\pi_j(Z) $ is constant within each stratum defined by $S$, we obtain that $I(A=j),I(A=k) \perp (\bmW, Y^{(l)}) \mid S $.
The result  $\pi_j(S) = \pi_j (Z)$ follows from the fact that 
$\pi_j(S) = P(A = j\mid S) =  E\{ I(A = j)\mid S, \pi_j(Z)\} = E[ E\{I(A = j) \mid  S, \pi_j(Z), Z\}\mid S, \pi_j(Z)] = E[ E\{I(A = j) \mid  Z\} \mid S, \pi_j(Z)]  =  E\{ \pi_j(Z) \mid S, \pi_j(Z)\} = \pi_j (Z)$.}

\subsection{Proof of Theorem \ref{theo:1}} \label{supp: proof of thm 1}

As defined in Section 5, let $E_{jk}=E_{kj}$,  $\var_{jk}=\var_{kj}$, and $\cov_{jk}=\cov_{kj}$ denote respectively the expectation, variance, and covariance conditioned on $\pi_j(\bmZ)>0$ and $\pi_k(\bmZ)>0$ (or sometimes conditioned on $\pi_j(\bmZ_i)>0$ and $\pi_k(\bmZ_i)>0$, which should be clear from the context). 
 Also, let 
$\widetilde E_{jk}$, $\widetilde \var_{jk}$ and $\widetilde \cov_{jk}$ be expectation, variance, and covariance conditional on $\{I_{jk} (Z_i), i=1,\dots, n\}$, where $I_{jk}(\bmZ_i) = I ( \pi_j(\bmZ_i)>0 , \pi_k(\bmZ_i)>0)$. Note that conditional on  $\{I_{jk} (Z_i), i=1,\dots, n\}$,  ${\cal I}_{jk}$ and $ n_{jk}$ are both non-random.
\\
(a)
We first show that, conditioned on $\{I_{jk} (Z_i), i=1,\dots, n\}$,  $ \sqrt{n_{jk}}({\hat{ \bm \vartheta}^{\rm ipw}_{jk}} - {\bm \vartheta}_{jk} ) \xrightarrow{d}$  the bivariate normal with mean vector 0 and covariance matrix $ \bS_{jk}^{\rm \, ipw}$, where  
 $\xrightarrow{d}$ denotes convergence in distribution as $n \to \infty$. 
Then, the unconditional convergence in distribution result in Theorem \ref{theo:1}(a) follows from applying the bounded convergence theorem.   Because ${\cal I}_{jk}$ and $ n_{jk}$ are both non-random conditional on  $\{I_{jk} (Z_i), i=1,\dots, n\}$, 
$$ \widetilde E_{jk} (\hat{\theta}_{jk}^{\rm \, ipw})
= \frac{1}{n_{jk}} \sum_{i \in {\cal I}_{jk}}  \widetilde E_{jk} \left\{ \frac{I(A_i=j)Y_i}{\pi_j(\bmZ_i)}\right\}=  E_{jk} \left\{ \frac{I(A=j)Y}{\pi_j(\bmZ)}\right\}=E_{jk} \left\{ \frac{I(A=j)Y^{(j)}}{\pi_j(\bmZ)}\right\}  = \theta_{jk} , $$
where the second equality is from the  independent and identically distributed condition, the third equality comes from the consistency assumption, and the last equality is from \eqref{eq: IPW identification}. 
Since $\hat{\bm \vartheta}_{jk}^{\rm \, ipw}$ conditional on $\{I_{jk} (Z_i), i=1,\dots, n\}$ is an average of independent and identically distributed random vectors,   the convergence in distribution conditional on $\{I_{jk} (Z_i), i=1,\dots, n\}$ is directly from the central limit theorem, if we can show that $\widetilde\var_{jk}\hat{\bm \vartheta}_{jk}^{\rm \, ipw}$ conditioned on $\{I_{jk} (Z_i), i=1,\dots, n\}$ is $\bS_{jk}^{\rm \, ipw}$. 
Note that under consistency and randomization assumptions,
\begin{align*}
 &n_{jk}   \widetilde \var_{jk}(\hat{\theta}_{jk}^{\rm ipw}-\theta_{jk})\\
 &= \frac{1}{n_{jk}} \sum_{i\in {\cal I}_{jk}}\widetilde\var_{jk}\left\{ \frac{I(A_i=j)}{\pi_{j}(\bmZ_i)}Y_i\right\} \\
    &=  \var_{jk} \left\{ \frac{I(A = j)}{\pi_j(\bmZ)}Y^{(j)}\right\}\\
     & = E_{jk}\left[\var_{jk}\left\{ \frac{I(A=j)}{\pi_{j}(\bmZ)}Y^{(j)} \, \Big | \, \bmZ,Y^{(j)}\right\} \right]  +\var_{jk} \left[E_{jk}\left\{ \frac{I(A=j)}{\pi_{j}(\bmZ)}Y^{(j)} \, \Big | \,  \bmZ,Y^{(j)}\right\} \right] \\
      &=E_{jk}\left\{\frac{(Y^{(j)})^2}{\pi_j(\bmZ)}-(Y^{(j)})^2\right\}+\var_{jk}(Y^{(j)})\\
         &=  E_{jk}\left\{\frac{(Y^{(j)})^2}{\pi_j(\bmZ)}\right\} -\theta_{jk}^2, \\
         &= E_{jk}\left\{\frac{I(A=j)Y^2}{\pi_j^2(\bmZ)}\right\} -\theta_{jk}^2,
\end{align*}
which is the first diagonal entry of $ \bS_{jk}^{\rm \, ipw}$. 
Similarly, $\widetilde \var_{jk}(\hat{\theta}_{kj}^{\rm ipw}-\theta_{kj})$ is the second diagonal entry of $\bS_{jk}^{\rm \, ipw}$. The off-diagonal entry of $ \bS_{jk}^{\rm \, ipw}$ can be derived from
 \begin{align*}
  n_{jk}    \widetilde\cov_{jk}(\hat{\theta}_{jk}^{\rm \, ipw},\hat{\theta}_{kj}^{\rm \, ipw})
     &= \frac{1}{n_{jk}} \sum_{i\in{\cal I}_{jk}}\cov_{jk}\left\{\frac{I(A=j)}{\pi_{j}(\bmZ)}Y,\frac{I(A=k)}{\pi_{k}(\bmZ)}Y\right\}\\
        &=  \cov_{jk}\left\{\frac{I(A=j)}{\pi_{j}(\bmZ)}Y,\frac{I(A=k)}{\pi_{k}(\bmZ)}Y\right\}\\
        &=  \cov_{jk}\left\{\frac{I(A=j)}{\pi_{j}(\bmZ)}Y^{(j)},\frac{I(A=k)}{\pi_{k}(\bmZ)}Y^{(k)}\right\}\\
     &= 0 - E_{jk}\left\{ \frac{I(A=j)}{\pi_{j}(\bmZ)}Y^{(j)}\right\} E_{jk}\left\{ \frac{I(A=k)}{\pi_{k}(\bmZ)}Y^{(k)}\right\} \\
     &= - \theta_{jk}\theta_{kj} .
 \end{align*}

\noindent
(b) Let 
$$ \xi_{jk} = \frac{1}{n_{jk}}
\sum_{i \in {\cal I}_{jk}} \frac{I(A_i = j) (Y_i- \theta_{jk} )}{\pi_j(\bmZ_i)} \qquad \mbox{and} \qquad \eta_{jk} =\frac{1}{n_{jk} }  \sum_{i \in {\cal I}_{jk}} \frac{I(A_i = j) }{\pi_j(\bmZ_i) }. $$
Then, 
\begin{equation}
    \sipw - \theta_{jk} = \xi_{jk} / \eta_{jk}
. \label{s1}
\end{equation}
Under the consistency assumption, we have
$$
\xi_{jk} = \frac{1}{n_{jk}}
\sum_{i \in {\cal I}_{jk}} \frac{I(A_i = j) (Y_i- \theta_{jk} )}{\pi_j(\bmZ_i)}
$$
Since $\widetilde E_{jk} (\xi_{jk}) =0 $ and $\eta_{jk} $ converges to 1 in probability, $\sqrt{n_{jk}} ( \sipw - \theta_{jk}) $
 has the same asymptotic distribution as $\sqrt{n_{jk}} \xi_{jk}$ by Slutsky's theorem. Using the same argument, $\sqrt{n_{jk}} ( \hat\theta_{kj}^{\rm \, sipw} - \theta_{kj}) $
 has the same asymptotic distribution as $\sqrt{n_{jk}} \xi_{kj}$. Then,
 we can follow the steps in the proof of part (a) with $Y^{(j)}$ replaced by $Y^{(j)} - \theta_{jk}$ and $Y^{(k)}$ replaced by $Y^{(k)} - \theta_{kj}$, which concludes the proof of part (b).

\noindent
(c) Define $\bm\vartheta_{jk}^{\rm aipw} = (\theta_{jk}^{\rm \, aipw}, \theta_{kj}^{\rm \, aipw})^T$ to be 
$\hat{\bm\vartheta}_{jk}^{\rm aipw}$ with $\hat\mu_{jk}$ and $\hat\mu_{kj}$ respectively replaced by $\mu_{jk}$ and $ \mu_{kj}$ defined in Assumption 2. Then 
\begin{align*}
   \hat{\bm\vartheta}_{jk}^{\rm aipw}     -\bm\vartheta_{jk}^{\rm aipw}&= \frac{1}{n_{jk}}\sum_{i\in {\cal I}_{jk}} \left( 
   \begin{array}{c}
   \left\{ 1-\frac{I(A_i=j)}{\pi_{j}(\bmZ_i)}\right\} \{ \hat{\mu}_{jk}(\bmX_i)-\mu_{jk}(\bmX_i)\} \vspace{2mm} \\ \left\{ 1-\frac{I(A_i=k)}{\pi_{k}(\bmZ_i)}\right\} \{ \hat{\mu}_{kj}(\bmX_i)-\mu_{kj}(\bmX_i)\} \end{array} \right) . 
\end{align*}
Under Assumption 2, 
by Lemma 19.24 in \cite{vaart_1998}, 
$
\sqrt{n_{jk}} (  \hat{\bm\vartheta}_{jk}^{\rm aipw} -    \bm\vartheta_{jk}^{\rm aipw})
$ converges to 0 in probability. Hence, it suffices to show that 
$\sqrt{n_{jk}} (\bm\vartheta_{jk}^{\rm aipw} -\bm\vartheta_{jk}) \xrightarrow{d}$ the bivariate normal with mean vector 0 and covariance matrix $ \bS_{jk}^{\rm \, aipw}$.  Following the same steps as in the proofs of (a)-(b), we only need to show that $n_{jk} \widetilde\var_{jk} (\bm\vartheta_{jk}^{\rm aipw}) =\bS_{jk}^{\rm aipw} $.   Define $\varepsilon_{jk} = Y^{(j)} - \mu_{jk}(\bmX)$.  Under the consistency assumption, the first diagonal entry of $\bS_{jk}^{\rm aipw}$  is 
\begin{align*}
 n_{jk}   \widetilde\var_{jk}({\theta}_{jk}^{\rm aipw}-\theta_{jk})
 &=\var_{jk}\left\{  {\textstyle  \frac{I(A=j)}{\pi_{j}(\bmZ)}} (Y-\mu_{jk}(\bm X))  + \mu_{jk}(\bmX) \right\} \\
    &=\var_{jk}\left\{  {\textstyle  \frac{I(A=j)}{\pi_{j}(\bmZ)}} \varepsilon_{jk}  + \mu_{jk}(\bmX) \right\}\\ 
    &=\var_{jk}\left\{ Y^{(j)} +  {\textstyle \frac{I(A=j)-\pi_{j}(\bmZ)}{\pi_{j}(\bmZ)}} \varepsilon_{jk} \right\}\\
    &= \var_{jk}(Y^{(j)})+ \var_{jk} \left\{ {\textstyle\frac{I(A=j)-\pi_{j}(\bmZ)}{\pi_{j}(\bmZ)} \varepsilon_{jk}} \right\}  \\
    &= \var_{jk}(Y^{(j)}) + E_{jk} \left[ \left\{ \frac{1}{\pi_j(\bmZ)} -1 \right\} \varepsilon_{jk}^2  \right] \\
    & = E_{jk} \left\{ \frac{\varepsilon_{jk}^2}{\pi_j(\bmZ)}  \right\}  +   \var_{jk}(Y^{(j)}) - E_{jk} (\varepsilon_{jk}^2 ) \\
    & = E_{jk} \left\{ \frac{\varepsilon_{jk}^2}{\pi_j(\bmZ)}  \right\}  +   \var_{jk}(Y^{(j)})  - \var_{jk} (\varepsilon_{jk}) - \{ E_{jk} (\varepsilon_{jk} )\}^2 \\
    & = E_{jk} \left\{ \frac{I(A=j)(Y-\mu_{jk}(\bm X))^2}{\pi_j^2(\bmZ)}  \right\}  + \lambda_{jk} - \delta_{jk}^2 ,
\end{align*}
where the fourth equality is because  
\begin{align*}
    \cov_{jk}\left\{Y^{(j)},{\textstyle \frac{I(A=j)-\pi_{j}(\bmZ)}{\pi_{j}(\bmZ)}} \varepsilon_{jk}\right\}&=\cov_{jk}\left\{E_{jk}[Y^{(j)}|\bmZ,Y^{(j)},\bmX],{\textstyle E_{jk}[\frac{I(A=j)-\pi_{j}(\bmZ)}{\pi_{j}(\bmZ)}} \varepsilon_{jk}|\bmZ,Y^{(j)},\bmX]\right\}\\
     &\quad + E_{jk}\left[\cov_{jk}\left\{Y^{(j)},{\textstyle \frac{I(A=j)-\pi_{j}(\bmZ)}{\pi_{j}(\bmZ)}} \varepsilon_{jk}\mid \bmZ,Y^{(j)},\bmX\right\}\right]\\
    &=\cov_{jk}\{Y^{(j)},0\}+0=0,
\end{align*}
and the fifth equality is because 
 \begin{align*}
    &\var_{jk} \left\{ {\textstyle\frac{I(A=j)-\pi_{j}(\bmZ)}{\pi_{j}(\bmZ)} }\varepsilon_{jk} \right\}\\
    & = 
E_{jk} \left[ \left\{ {\textstyle\frac{I(A=j)-\pi_{j}(\bmZ)}{\pi_{j}(\bmZ)}} \right\}^2  \varepsilon_{jk}^2 \right]-\left\{E_{jk}\left[{\textstyle\frac{I(A=j)-\pi_{j}(\bmZ)}{\pi_{j}(\bmZ)} }\varepsilon_{jk}\right]\right\}^2 \\
&=E_{jk} \left[ E_{jk}\left[ \left\{ {\textstyle\frac{I(A=j)-\pi_{j}(\bmZ)}{\pi_{j}(\bmZ)}} \right\}^2\varepsilon_{jk}^2  |\bmZ,Y^{(j)},\bmX \right]\right] -\left\{E_{jk}\left[E_{jk}\left[{\textstyle\frac{I(A=j)-\pi_{j}(\bmZ)}{\pi_{j}(\bmZ)} }\varepsilon_{jk}|\bmZ,Y^{(j)},\bmX\right]\right]\right\}^2\\
&=E_{jk} \left[\varepsilon_{jk}^2 E_{jk}\left[ \left\{ {\textstyle\frac{I(A=j)-\pi_{j}(\bmZ)}{\pi_{j}(\bmZ)}} \right\}^2  |\bmZ \right]\right] - \left\{E_{jk}\left[\varepsilon_{jk}E_{jk}\left[{\textstyle\frac{I(A=j)-\pi_{j}(\bmZ)}{\pi_{j}(\bmZ)} }|\bmZ\right]\right]\right\}^2\\
& = 
E_{jk} \left[ \left\{ {\textstyle\frac{1}{\pi_{j}(\bmZ)} }-1\right\} \varepsilon_{jk}^2 \right].
\end{align*}
The last diagonal entry of $\bS_{jk}^{\rm aipw}$ can be obtained by swapping $j$ and $k$.
Also, 
\begin{align*}
    & \ \cov_{jk}\left\{{\textstyle\frac{I(A=j)}{\pi_j(\bmZ)}}\varepsilon_{jk} +\mu_{jk}(\bmX),{\textstyle\frac{I(A=k)}{\pi_{k}(\bmZ)}}\varepsilon_{kj} +\mu_{kj}(\bmX)\right\}\\
    = & \ \cov_{jk}\left[  E_{jk}\left\{ {\textstyle \frac{I(A=j)}{\pi_{j}(\bmZ)}}\varepsilon_{jk} +\mu_{jk}(\bmX) \Big|\bmZ,Y^{(j)},Y^{(k)}, \bmX\right\} ,  E_{jk}\left\{ {\textstyle\frac{I(A=k)}{\pi_{k}(\bmZ)}}\varepsilon_{kj} +\mu_{kj} (\bmX)\Big|\bmZ, Y^{(j)}, Y^{(k)},\bmX\right\} \right]\\
    &\ + E_{jk}\left[\cov_{jk}\left\{ {\textstyle \frac{I(A=j)}{\pi_{j}(\bmZ)}}\varepsilon_{jk} +\mu_{jk}(\bmX),{\textstyle\frac{I(A=k)}{\pi_{k}(\bmZ)}}\varepsilon_{kj} +\mu_{kj}(\bmX)\Big |\bmZ,Y^{(j)},Y^{(k)}, \bmX)\right\} \right]\\
    = & \ \cov_{jk}\{ \varepsilon_{jk} +\mu_{jk}(\bmX),\varepsilon_{kj} +\mu_{kj}(\bmX)\} +E_{jk}\left[\frac{\varepsilon_{jk}}{\pi_{j}(\bmZ)}\frac{\varepsilon_{kj} }{\pi_{k}(\bmZ)}\cov_{jk}\left\{ I(A=j),I(A=k)\mid \bmZ\right\} \right]\\
  =  &\ \cov_{jk}(Y^{(j)},Y^{(k)})-E_{jk}(\varepsilon_{jk} \varepsilon_{kj} )\\
    =  &\ \cov_{jk}(Y^{(j)},Y^{(k)})-\cov_{jk}(\varepsilon_{jk}, \varepsilon_{kj} ) - E_{jk}(\varepsilon_{jk})E_{jk}(\varepsilon_{kj})\\
 =  &\ \cov_{jk}(Y^{(j)},Y^{(k)})-\cov_{jk}\{ Y^{(j)} - \mu_{jk}(\bmX), Y^{(k)} - \mu_{kj}(\bmX)\}  - \delta_{jk}\delta_{kj}\\
  = & \ c_{jk} - \delta_{jk}\delta_{kj} ,
\end{align*}
where the third equation follows from $\cov_{jk}\left\{ I(A=j),I(A=k)\mid \bmZ\right\} = - \pi_{j}(\bmZ)\pi_{k}(\bmZ) $. 
This shows that $ n_{jk} \widetilde\cov_{jk}({\theta}_{jk}^{\rm aipw},{\theta}_{kj}^{\rm aipw})$
is equal to the off-diagonal entry of $\bS_{jk}^{\rm aipw}$.

\noindent
(d) Similar to the proof  of (c),  
we only need to show that 
$\sqrt{n_{jk}} (\bm\vartheta_{jk}^{\rm saipw} -\bm\vartheta_{jk}) \xrightarrow{d}$  the bivariate normal with mean vector 0 and covariance matrix $ \bS_{jk}^{\rm \, saipw}$,  where $\bm\vartheta_{jk}^{\rm saipw} = (\theta_{jk}^{\rm \, saipw}, \theta_{kj}^{\rm \, saipw})^T $ is $\hat{\bm\vartheta}_{jk}^{\rm saipw} $ with $\hat\mu_{jk}$ and $ \hat\mu_{kj}$ respectively replaced by  $\mu_{jk}$ and $\mu_{kj}$ defined in Assumption 2.
Define 
\begin{align*}
  \zeta_{jk} &  = \frac{1}{n_{jk}} \sum_{i \in {\cal I}_{jk}} \frac{I(A_i=j) \{ Y_i^{(j)} - \mu_{jk}(\bmX_i) - \delta_{jk}\}}{\pi_j(\bmZ_i)} +  \frac{1}{n_{jk}} \sum_{i \in {\cal I}_{jk}} [ \mu_{jk} (\bmX_i) - E_{jk} \{  \mu_{jk} (\bmX ) \}]  
\end{align*}
 and recall the definition $\eta_{jk} =\frac{1}{n_{jk} }  \sum_{i \in {\cal I}_{jk}} \frac{I(A_i = j) }{\pi_j(\bmZ_i) }$. Then, under the consistency assumption,
\begin{equation}
    \theta_{jk}^{\rm \, saipw} - \theta_{jk} = \zeta_{jk}+\left(\frac{1}{\eta_{jk}}-1\right)\frac{1}{n_{jk}} \sum_{i \in {\cal I}_{jk}} \frac{I(A_i=j) \{ Y_i^{(j)} - \mu_{jk}(\bmX_i) - \delta_{jk}\}}{\pi_j(\bmZ_i)}.
    \label{s2}
\end{equation}
Since $n_{jk}^{-1} \sum_{i \in {\cal I}_{jk}} \frac{I(A_i=j) \{ Y_i^{(j)} - \mu_{jk}(\bmX_i) - \delta_{jk}\}}{\pi_j(\bmZ_i)}=O_p(n^{-1/2})$ following from the steps in the proof of part (a) and (c), $\widetilde E_{jk} (\zeta_{jk}) =0 $ and $\eta_{jk} $ converges to 1 in probability, $\sqrt{n_{jk}} ( \theta_{jk}^{\rm saipw} - \theta_{jk} ) $
 has the same asymptotic distribution as $\sqrt{n_{kl}}\zeta_{jk}$ by Slutsky's theorem.

Let $\bm\zeta_{jk} = ( \zeta_{jk}, \zeta_{kj})^T$. Then, the asymptotic distribution of $\sqrt{n_{jk}} \bm\zeta_{jk} $ follows from the steps in the proof of part (c), with $Y_i^{(j)}$ replaced by $Y_i^{(j)} - \theta_{jk}$ and  $\mu_{jk}(\bmX_i) $ replaced by $\mu_{jk}(\bmX_i) - E_{jk} \{\mu_{jk}(\bmX_i) \} $.

\noindent
(e)  Under the consistency assumption, write
\begin{align*}
    \hat{\bm\vartheta}_{jk}^{\rm \, ps}-\bm\vartheta_{jk}&=\frac{1}{n_{jk}}\sum_{h=1}^{H_{jk}}
    n(\mathcal{I}_{jk}^{(h)}){n_j({\cal I}_{jk}^{(h)})^{-1}\choose n_k({\cal I}_{jk}^{(h)})^{-1}}
   \sum_{i \in {\cal I}_{jk}^{(h)}}  {I(A_i=j) (Y_i^{(j)}  -\theta_{jk}) \choose I(A_i=k)  (Y_i^{(k)}  -\theta_{kj}) }  = \bm{U}_{jk}  + \bm{V}_{jk} ,
\end{align*}
where 
$$ \bm{U}_{jk} = {U_{jk} \choose U_{kj} } = \frac{1}{n_{jk}} \sum_{h=1}^{H_{jk}}  n(\mathcal{I}_{jk}^{(h)}){n_j({\cal I}_{jk}^{(h)})^{-1}\choose n_k({\cal I}_{jk}^{(h)})^{-1}}  \sum_{i \in {\cal I}_{jk}^{(h)}}
  {{I(A_i=j)\{  Y_i^{(j)}  - E_{jk}(Y_i^{(j)}\mid S_i )\} \choose I(A_i=k) \{ Y_i^{(k)}  - E_{jk}(Y_i^{(k)}\mid S_i ) \} }} $$
and 
$$ \bm{V}_{jk} = {V_{jk} \choose V_{kj} } = \frac{1}{n_{jk}} \sum_{h=1}^{H_{jk}}  n(\mathcal{I}_{jk}^{(h)}){n_j({\cal I}_{jk}^{(h)})^{-1}\choose n_k({\cal I}_{jk}^{(h)})^{-1}} \sum_{i \in {\cal I}_{jk}^{(h)}} {{I(A_i=j)\{  E_{jk}(Y_i^{(j)}\mid S_i ) - \theta_{jk} \} \choose I(A_i=k) \{  E_{jk}(Y_i^{(k)}\mid S_i )-\theta_{kj}  \} }}. $$
Since 
\begin{align*}
  V_{jk}  &= \frac{1}{n_{jk}} \sum_{h=1}^{H_{jk}}  \frac{n(\mathcal{I}_{jk}^{(h)})}{n_j({\cal I}_{jk}^{(h)})}  \sum_{i \in {\cal I}_{jk}^{(h)}}  I(A_i=j)  E_{jk}(Y_i^{(j)}\mid S_i={\cal I}_{jk}^{(h)} ) - \theta_{jk} \\
    &=\frac{1}{n_{jk}} \sum_{h=1}^{H_{jk}}  \frac{n(\mathcal{I}_{jk}^{(h)})}{n_j({\cal I}_{jk}^{(h)})}  \sum_{i \in {\cal I}_{jk}^{(h)}}  I(A_i=j)  E_{jk}(Y^{(j)}\mid S={\cal I}_{jk}^{(h)} ) - \theta_{jk} \\
        &=\frac{1}{n_{jk}} \sum_{h=1}^{H_{jk}}  n(\mathcal{I}_{jk}^{(h)})  E_{jk}(Y^{(j)}\mid S={\cal I}_{jk}^{(h)} ) - \theta_{jk} \\
    & =\frac{1}{n_{jk}} \sum_{h=1}^{H_{jk}} \sum_{i \in {\cal I}_{jk}^{(h)}} E_{jk}(Y_i^{(j)} \mid S_i )- \theta_{jk}, \\
&    =  \frac{1}{n_{jk}}  \sum_{i \in {\cal I}_{jk}}    E_{jk}(Y_i^{(j)} \mid S_i )- \theta_{jk}, 
\end{align*}
we have $\widetilde{E}_{jk} (V_{jk}) = 0 $,
$\widetilde{E}_{jk} (V_{kj}) = 0 $, 
$n_{jk} \widetilde\var_{jk} ( V_{jk}) =  \var_{jk} \{ E_{jk}(Y^{(j)} \mid S ) \}  $, $n_{jk} \widetilde\var_{jk} ( V_{kj}) =  \var_{jk} \{ E_{jk}(Y^{(k)} \mid S ) \}  $,  
and 
$ n_{jk} \widetilde\cov_{jk} (V_{jk}, V_{kj}) 
=  \cov_{jk} \{E_{jk}(Y^{(j)} \mid S ), E_{jk}(Y^{(k)} \mid S) \} $. Conditional on $\{I_{jk} (Z_i), i=1,\dots, n\}$, $\bm{V}_{jk}$ is an average of independent and identically distributed random vectors and, by central limit theorem,  $\sqrt{n_{jk}}\bm{V}_{jk} \xrightarrow{d}$   bivariate normal with mean 0 and covariance matrix
\begin{align*}
 \bm{D}_V =   n_{jk} \widetilde{\var}_{jk} (\bm{V}_{jk}) = \var_{jk} 
{ E_{jk} (Y^{(j)} \mid S) \choose 
 E_{jk} (Y^{(k)} \mid S) }= \var_{jk} 
{ E_{jk} (Y \mid S,A=j) \choose 
 E_{jk} (Y \mid S, A=k) }, 
\end{align*} 
where the last equality comes from the consistency assumption and the randomization probabilities are constant within each stratum.
By the dominated convergence theorem, $\sqrt{n_{jk}}\bm{V}_{jk}\xrightarrow{d}$  the same  distribution unconditionally. 
Next, we turn to  $U_{jk}$ and $U_{kj}$. Let {${\cal A} = \{ I(A_1=j),I(A_1=k), \dots, I(A_n=j),I(A_n=k) \}$ and ${\cal S} = \{ S_1, \dots, S_n\}$}. Note that
$$
    \widetilde E_{jk} ( \bm{U}_{jk} \mid {\cal S}, {\cal A} )=\frac{1}{n_{jk}} \sum_{h=1}^{H_{jk}}  n(\mathcal{I}_{jk}^{(h)}){n_j({\cal I}_{jk}^{(h)})^{-1}\choose n_k({\cal I}_{jk}^{(h)})^{-1}}  \sum_{i \in {\cal I}_{jk}^{(h)}} { I(A_i=j) \{ E_{jk}(Y_i^{(j)}|S_i)  - E_{jk}(Y_i^{(j)}| S_i )\} \choose I(A_i=k) \{ E_{jk}(Y_i^{(k)}|S_i)  - E_{jk}(Y_i^{(k)}| S_i )\} } =0
$$
and
\begin{align*}
 n_{jk} \widetilde \var_{jk} (U_{jk}\mid {\cal S}, {\cal A} )&=  \frac{1}{n_{jk}}   \sum_{h=1}^{H_{jk}} \{n(\mathcal{I}_{jk}^{(h)})\}^2 \, \widetilde \var_{jk} \Big\{
 \bar{Y}_j({\cal I}_{jk}^{(h)}) \,
  \Big| \,{\cal S}, {\cal A} \Big\}\\
&= \frac{1}{n_{jk}}  \sum_{h=1}^{H_{jk}}\frac{\{n(\mathcal{I}_{jk}^{(h)})\}^2}{\{n_j({\cal I}_{jk}^{(h)})\}^2}
  \sum_{i \in {\cal I}_{jk}^{(h)}} I(A_i=j) \, \var_{jk}(Y_i^{(j)} \mid {\cal S}, {\cal A} ) \\
&= \frac{1}{n_{jk}}  \sum_{h=1}^{H_{jk}}\frac{\{n(\mathcal{I}_{jk}^{(h)})\}^2}{\{n_j({\cal I}_{jk}^{(h)})\}^2}
  \sum_{i \in {\cal I}_{jk}^{(h)}} I(A_i=j) \, \var_{jk}(Y^{(j)} \mid S = \mathcal{I}_{jk}^{(h)} ) \\
  &= \frac{1}{n_{jk}}  \sum_{h=1}^{H_{jk}}\frac{\{n(\mathcal{I}_{jk}^{(h)})\}^2}{n_j({\cal I}_{jk}^{(h)})}
  \var_{jk} (Y^{(j)} \mid S =\mathcal{I}_{jk}^{(h)})
\end{align*} 
where the third equality holds because $   I(A=j),I(A=k) \perp (\bmW, Y^{(j)}) \mid S $ 
from Lemma \ref{lemma: S}.  Similarly, we have 
$$  n_{jk} \widetilde \var_{jk} (U_{kj}\mid  {\cal S, A} ) = \frac{1}{n_{jk}}  \sum_{h=1}^{H_{jk}}\frac{\{n(\mathcal{I}_{jk}^{(h)})\}^2}{n_k({\cal I}_{jk}^{(h)})}
  \var_{jk} (Y^{(k)} \mid S =\mathcal{I}_{jk}^{(h)}) $$
and
\begin{align*}
\widetilde\cov_{jk}(U_{jk},U_{kj}\mid {\cal S, A})
    &=\widetilde\cov_{jk} \Bigg\{ \sum_{h=1}^{H_{jk}}\frac{n(\mathcal{I}_{jk}^{(h)})}{n_{jk}}\Bar{Y}_{j}(\mathcal{I}_{jk}^{(h)}),\,
    \sum_{h'=1}^{H_{jk}}\frac{n(\mathcal{I}_{jk}^{(h')})}{n_{jk}}\Bar{Y}_{k}(\mathcal{I}_{jk}^{(h')}) \, \Big| \, {\cal S, A}\Bigg\} \\
    &=\sum_{h,h'=1}^{H_{jk}}\frac{n(\mathcal{I}_{jk}^{(h)})n(\mathcal{I}_{jk}^{(h')})}{n_{jk}^2}\widetilde\cov_{jk}\{\Bar{Y}_{j}(\mathcal{I}_{jk}^{(h)}),\Bar{Y}_{k}(\mathcal{I}_{jk}^{(h')})\mid {\cal S,A}\}\\
    & = 0, 
\end{align*}
where the last line follows from the fact that, 
for any $h$ and $h'$, 
\begin{align*}
    &\ \widetilde\cov_{jk}\{\Bar{Y}_{j}(\mathcal{I}_{jk}^{(h)}),\Bar{Y}_{k}(\mathcal{I}_{jk}^{(h')})\mid {\cal S,A}\}\\
    =& \ \frac{1}{n_j({\cal I}_{jk}^{(h)})n_k({\cal I}_{jk}^{(h')})}\sum_{i\in{\cal I}_{jk}^{(h)}}\sum_{i'\in{\cal I}_{jk}^{(h')}}I(A_i=j,A_{i'}=k)\cov_{jk}(Y_i^{(j)},Y_{i'}^{(k)}|S_i,S_{i'})\\
    =& \ \frac{I(h=h')}{n_j({\cal I}_{jk}^{(h)})n_k({\cal I}_{jk}^{(h)})}\sum_{i\in{\cal I}_{jk}^{(h)}}I(A_i=j,A_{i}=k)\cov_{jk}(Y_i^{(j)},Y_{i}^{(k)}|S_i)\\
    =& \ 0. 
\end{align*}
Conditional on $\{I_{jk} (Z_i), i=1,\dots, n\}, {\cal S, A}$,
 $\bm{U}_{jk}$ is an average of independent random vectors. By the central limit theorem , $\sqrt{n_{jk}}\bm{U}_{jk} \mid \{I_{jk} (Z_i), i=1,\dots, n\}, {\cal S, A}\xrightarrow{d} $  bivariate normal with mean 0 and covariance matrix
\begin{align*}
  \bm{D}_{U,{\cal S,A}} =   \left(  \begin{array}{cc}
\frac{1}{n_{jk}}  \sum_{h=1}^{H_{jk}}\frac{\{n(\mathcal{I}_{jk}^{(h)})\}^2}{n_j({\cal I}_{jk}^{(h)})}
  \var_{jk} (Y^{(j)} \mid S =\mathcal{I}_{jk}^{(h)})& 0 \\
0 & \frac{1}{n_{jk}}  \sum_{h=1}^{H_{jk}}\frac{\{n(\mathcal{I}_{jk}^{(h)})\}^2}{n_k({\cal I}_{jk}^{(h)})}
  \var_{jk} (Y^{(k)} \mid S =\mathcal{I}_{jk}^{(h)})
\end{array} \right) . 
\end{align*} 
Conditional on $\{I_{jk} (Z_i), i=1,\dots, n\}, {\cal S}$, it turns out that the diagonal elements of $\bm{D}_{U,{\cal S,A}}$ 
\begin{align*}
    \frac{1}{n_{jk}}  \sum_{h=1}^{H_{jk}}\frac{\{n(\mathcal{I}_{jk}^{(h)})\}^2}{n_k({\cal I}_{jk}^{(h)})}
  \var_{jk} (Y^{(k)} \mid S =\mathcal{I}_{jk}^{(h)})=\frac{1}{n_{jk}}  \sum_{h=1}^{H_{jk}}n(\mathcal{I}_{jk}^{(h)})
  \frac{\var_{jk} (Y^{(a)} \mid S =\mathcal{I}_{jk}^{(h)})}{\pi_a({\cal I}_{jk}^{(h)})}+o_p(1)
\end{align*}
for $a=j$ and $k$. We denote the limit of $\bm{D}_{U,{\cal S,A}}\mid \{I_{jk} (Z_i), i=1,\dots, n\}, {\cal S}$ by $\bm{D}_{U,{\cal S}}$. Observing that 
\begin{align*}
   \frac{1}{n_{jk}}  \sum_{h=1}^{H_{jk}}n(\mathcal{I}_{jk}^{(h)})
  \frac{\var_{jk} (Y^{(a)} \mid S =\mathcal{I}_{jk}^{(h)})}{\pi_a({\cal I}_{jk}^{(h)})}
  &=\frac{1}{n_{jk}} \sum_{h=1}^{H_{jk}} \sum_{i \in {\cal I}_{jk}^{(h)} } \frac{ \var_{jk} (Y_i^{(a)}\mid S_i )}  {\pi_a(S_i)}
  \\
&= \frac{1}{n_{jk}}  \sum_{i \in {\cal I}_{jk}} \frac{\var_{jk} (Y_i^{(a)}\mid S_i)}  {\pi_a(S_i)}\\
&\xrightarrow{p}E_{jk}\left(\frac{\var_{jk}(Y^{(a)}\mid S)}{\pi_a(S)}\right),
\end{align*}
we have $\bm{D}_{U,\cal S}\mid \{I_{jk}(Z_i),i=1,\dots,n\}\xrightarrow{p}\bm{D}_U$, where
\begin{align*}
  \bm{D}_U =   E_{jk} \left(  \begin{array}{cc}
\frac{\var_{jk} ( Y^{(j)} \mid S )}{\pi_j(S)}& 0 \\
0 & \frac{\var_{jk} ( Y^{(k)} \mid S )}{\pi_k(S)}
\end{array} \right).
\end{align*}
By the consistency assumption and the definition of $S$, we have that
\begin{align*}
  \bm{D}_U =   E_{jk} \left(  \begin{array}{cc}
\frac{\var_{jk} ( Y \mid A=j,\, S )}{\pi_j(S)}& 0 \\
0 & \frac{\var_{jk} ( Y \mid A=k,\, S )}{\pi_k(S)}
\end{array} \right).
\end{align*}

Lastly, we show that $( \sqrt{n_{jk}}\bm U_{jk}^T, \sqrt{n_{jk}}\bm V_{jk}^T)^T\xrightarrow{d} (\bm\psi_U^T, \bm\psi_V^T)^T$, where $\bm\psi_U$ and  $\bm\psi_V$ are independent random vectors having bivariate normal distributions with mean 0 and covariance matrices $\bm{D}_U$ and $\bm{D}_V$, respectively. {Note that our previous proof shows that $\sqrt{n_{jk}}\bm U_{jk} \mid I_{jk} (Z_i), i=1,\dots, n, \, {\cal S, A} \xrightarrow{d}   \bm\psi_{U,{\cal S,A}} $,  $\sqrt{n_{jk}}\bm U_{jk} \mid I_{jk} (Z_i), i=1,\dots, n, \, {\cal S} \xrightarrow{d}   \bm\psi_{U,{\cal S}} $, $\sqrt{n_{jk}}\bm U_{jk} \mid I_{jk} (Z_i), i=1,\dots, n\xrightarrow{d}   \bm\psi_{U} $, and  $\sqrt{n_{jk}}\bm V_{jk} \mid I_{jk} (Z_i), i=1,\dots, n \xrightarrow{d}   \bm\psi_{V} $, where $\bm\psi_{*}$ has corresponding covariance matrix $\bm D_{*}$. Then, for any bivariate vectors $\bm t$ and $\bm s$, 
\begin{align*}
    &\ P(\sqrt{n_{jk}}\bm U_{jk}\leq \bm t, \sqrt{n_{jk}}\bm V_{jk}\leq \bm s) \\
    = &  \ E\left[   E \left\{I(\sqrt{n_{jk}}\bm U_{jk}\leq \bm t, \sqrt{n_{jk}}\bm V_{jk}\leq \bm s)\mid I_{jk} (Z_i), i=1,\dots, n, {\cal S, A}  \right\}\right] \\
 =   &\ E\left\{  I( \sqrt{n_{jk}}\bm V_{jk}\leq \bm s) E \{I(\sqrt{n_{jk}}\bm U_{jk}\leq \bm t ) \mid I_{jk} (Z_i), i=1,\dots, n, {\cal S, A}  \}\right\} \\ 
 =   &\ E\left\{ I( \sqrt{n_{jk}}\bm V_{jk}\leq \bm s)\left[ \, P (\sqrt{n_{jk}}\bm U_{jk}\leq \bm t  \mid I_{jk} (Z_i), i=1,\dots, n, {\cal S, A} )\right.\right.\\
 & \quad- P(\bm\psi_{U,{\cal S, A}}\leq  \bm t\mid I_{jk}(Z_i),i=1,\dots,n,{\cal S, A})  + P(\bm\psi_{U,{\cal S,A}}\leq  \bm t\mid I_{jk}(Z_i),i=1,\dots,n,{\cal S,A}) \\
 &\quad - P(\bm\psi_{U,\mathcal{S}}\leq  \bm t\mid I_{jk}(Z_i),i=1,\dots,n,{\cal S,A})+{P(\bm\psi_{U,\cal{S}}\leq  \bm t\mid I_{jk}(Z_i),i=1,\dots,n,\cal{S,A})}\\
 &\left.\left.\quad - P(\bm\psi_{U}\leq  \bm t\mid I_{jk}(Z_i),i=1,\dots,n,\cal{S,A})\right]\right\} +P( \sqrt{n_{jk}}\bm V_{jk}\leq \bm s) P(\bm\psi_U\leq  \bm t) \\ 
 \to    &\ P( \bm\psi_V \leq \bm s) P(\bm\psi_U\leq  \bm t) 
\end{align*}}
where the second equality follows from the fact that $\bm V_{jk}$ is non-random conditional on $ \{I_{jk} (Z_i), i=1,\dots, n\}$, ${\cal S, A}$, and the last step is from the bounded convergence theorem.
Consequently, 
$ \sqrt{n_{jk} } (\hat{\bm\vartheta}_{jk}^{\rm \, ps}-\bm\vartheta_{jk}) = \sqrt{n_{jk}} (\bm{U}_{jk} + \bm{V}_{jk}) \xrightarrow{d} \bm\psi_U+ \bm\psi_V$ that is bivariate normal with mean 0 and covariance matrix $\bm{D}_U+\bm{D}_V$ as  given in  (e). 

Note that $\hat{\bm\vartheta}_{jk}^{\rm \, ps}$ is an unbiased estimator of $\bm\vartheta_{jk}$ because
\[
E[\hat{\bm\vartheta}_{jk}^{\rm \, ps}-\bm\vartheta_{jk}]=E\left[\widetilde E_{jk}[\bm U_{jk}\mid {\cal S},{\cal A}]+\widetilde E_{jk}[\bm V_{jk}]\right]=(0,0)^T+(0,0)^T=(0,0)^T.
\]

\noindent
(f) Similar to the proof of (c),  
we only need to show that 
$\sqrt{n_{jk}} (\bm\vartheta_{jk}^{\rm aps} -\bm\vartheta_{jk}) \xrightarrow{d} $ the bivariate normal with mean vector 0 and covariance matrix $ \bS_{jk}^{\rm \, aps}$,  where $\bm\vartheta_{jk}^{\rm aps} = (\theta_{jk}^{\rm \, aps}, \theta_{kj}^{\rm \, aps})^T$ is $\hat{\bm\vartheta}_{jk}^{\rm aps} $ with $(\hat\mu_{jk},\hat\mu_{kj}) $ replaced by $(\mu_{jk}, \mu_{kj})$. We use the same technique in the proof of (e) to derive this asymptotic distribution. Under the consistency assumption, 
write
$ {\bm\vartheta}_{jk}^{\rm \, aps}-\bm\vartheta_{jk} = \bm{U}_{jk}^a+\bm{V}_{jk} $, 
where 
\begin{align*}
   \bm{U}_{jk}^a = { U_{jk}^a \choose U_{kj}^a }=\frac{1}{n_{jk}} \sum_{h=1}^{H_{jk}}  n(\mathcal{I}_{jk}^{(h)}){n_j({\cal I}_{jk}^{(h)})^{-1}\choose n_k({\cal I}_{jk}^{(h)})^{-1}} \sum_{i \in {\cal I}_{jk}^{(h)}} \!\! {I(A_i=j) \{ Y_i^{(j)} -\mu_{jk}(\bmX_i) +{\bar\mu_{jk}^{(h)}}  - E_{jk}(Y_i^{(j)}\mid S_i )\} \choose I(A_i=k) \{ Y_i^{(k)} -\mu_{kj}(\bmX_i) {+\bar\mu_{kj}^{(h)}}  - E_{jk}(Y_i^{(k)}\mid S_i )\}}\!, 
\end{align*}
$\bar \mu_{jk}^{(h)} = n({\cal I}_{jk}^{(h)})^{-1}\sum_{i \in {\cal I}_{jk}^{(h)}} \mu_{jk}(\bmX_i)$,   and 
$\bm{V}_{jk}$ is the same as that in the proof of (e). Note that
\begin{align*}
  \widetilde E_{jk}(U_{jk}^a\mid {\cal S,A})
   & = \frac{1}{n_{jk}} \sum_{h=1}^{H_{jk}}  \frac{n(\mathcal{I}_{jk}^{(h)})}{n_j({\cal I}_{jk}^{(h)})}  \sum_{i \in {\cal I}_{jk}^{(h)}} \! \!  I(A_i=j) E_{jk}\{ Y_i^{(j)} \!-\!\mu_{jk}(\bmX_i) +\bar\mu_{jk}^{(h)} \! -\! E_{jk}(Y_i^{(j)}\mid S_i )\mid {\cal S}\}\\
  & = \frac{1}{n_{jk}} \sum_{h=1}^{H_{jk}}  \frac{n(\mathcal{I}_{jk}^{(h)})}{n_j({\cal I}_{jk}^{(h)})} \sum_{i \in {\cal I}_{jk}^{(h)}}\!  I(A_i=j)[ \, - E_{jk}\{ \mu_{jk}(\bmX_i) \mid S_i\} +E_{jk}(\Bar{\mu}_{jk}^{(h)}\mid {\cal S}) ] \\
    & =  0, 
\end{align*}
because 
$\widetilde E_{jk}(\Bar{\mu}_{jk}^{(h)}\mid {\cal S})=n({\cal I}_{jk}^{(h)})^{-1}\sum_{i \in {\cal I}_{jk}^{(h)}}E_{jk}\{\mu_{jk}(\bmX_i)\mid S_i\}=E_{jk}\{\mu_{jk}(\bmX)\mid S={\cal I}_{jk}^{(h)}\}$ $ = E_{jk}\{\mu_{jk}(\bmX_i)\mid S_i\}$ for any $i \in {\cal I}_{jk}^{(h)}$. 
Similarly, $\widetilde E_{jk}(U_{kj}^a\mid {\cal S,A }) = 0$.

{Then, we rewrite $U_{jk}^a$ as 
\begin{align*}
   U_{jk}^a&= 
   \frac{1}{n_{jk}} \sum_{h=1}^{H_{jk}}  \frac{n(\mathcal{I}_{jk}^{(h)})}{n_j({\cal I}_{jk}^{(h)})} \sum_{i \in {\cal I}_{jk}^{(h)}} \!\! {I(A_i=j) \{ Y_i^{(j)} -\mu_{jk}(\bmX_i) +{ E_{jk}[\mu_{jk}(\bmX_i)\mid S_i]}  - E_{jk}(Y_i^{(j)}\mid S_i )\}}\\
   &\quad + \frac{1}{n_{jk}}\sum_{i\in {\cal I}_{jk}}\mu_{jk}(\bmX_i)-E_{jk}[\mu_{jk}(\bmX_i)\mid S_i]\\
   &=\frac{1}{n_{jk}}    \sum_{i \in {\cal I}_{jk}} {\frac{I(A_i=j)}{\hat\pi_j(S_i)} \{ Y_i^{(j)} -\mu_{jk}(\bmX_i) +{ E_{jk}[\mu_{jk}(\bmX_i)\mid S_i]}  - E_{jk}(Y_i^{(j)}\mid S_i )\}}\\
   &\quad + \frac{1}{n_{jk}}\sum_{i\in {\cal I}_{jk}}\mu_{jk}(\bmX_i)-E_{jk}[\mu_{jk}(\bmX_i)\mid S_i],
\end{align*}
where $\hat\pi_j(S_i)=n_j({\cal I}_{jk}^{(h)})/n(\mathcal{I}_{jk}^{(h)})$ is a constant conditional on $\{I_{jk}(Z_i),i=1\dots,n\},{\cal S,A}$ and the first equality comes from that
\begin{align*}
    \frac{1}{n_{jk}} \sum_{h=1}^{H_{jk}}  \frac{n(\mathcal{I}_{jk}^{(h)})}{n_j({\cal I}_{jk}^{(h)})} \sum_{i \in {\cal I}_{jk}^{(h)}} \!\! {I(A_i=j){ E_{jk}[\mu_{jk}(\bmX_i)\mid S_i]}} &=\frac{1}{n_{jk}} \sum_{h=1}^{H_{jk}}  \frac{n(\mathcal{I}_{jk}^{(h)})}{n_j({\cal I}_{jk}^{(h)})} \sum_{i \in {\cal I}_{jk}^{(h)}} \!\! {I(A_i=j){ E_{jk}[\mu_{jk}(\bmX_i)\mid S_i={\cal I}_{jk}^{(h)}]}} \\
    &=\frac{1}{n_{jk}} \sum_{h=1}^{H_{jk}}  \sum_{i \in {\cal I}_{jk}^{(h)}} \!\! {{ E_{jk}[\mu_{jk}(\bmX)\mid S={\cal I}_{jk}^{(h)}]}}\\
    &=\frac{1}{n_{jk}}  \sum_{i \in {\cal I}_{jk}} \!\! {{ E_{jk}[\mu_{jk}(\bmX_i)\mid S_i]}}.
\end{align*}
Following the proof of (e), since $\bm U_{jk}^a$ is the average of independent  random vectors conditional on $I_{jk}(Z_i),i=1,\dots,n,{\cal S,A}$, by Lindeberg-Feller's central limit theorem, we conclude that $\sqrt{n_{jk}}\bm U_{jk}^a\mid I_{jk}(Z_i),i=1,\dots,n,{\cal S,A}\xrightarrow{d}$ bivariate normal with mean $0$ and covariance matrix derived as follows. We hereby verify Lindeberg's condition. Denote $\frac{I(A_i=j)}{\hat\pi_j(S_i)} \{ Y_i^{(j)} -\mu_{jk}(\bmX_i) +{ E_{jk}[\mu_{jk}(\bmX_i)\mid S_i]}  - E_{jk}(Y_i^{(j)}\mid S_i )\}+ \mu_{jk}(\bmX_i)-E_{jk}[\mu_{jk}(\bmX_i)\mid S_i]$ by $\xi_{i,jk}$. Then we have that $\widetilde{E}_{jk}[\xi_{i,jk}\mid {\cal S,A}]=0$ and 
\begin{align*}
    \quad\sigma_i^2&=\widetilde{\var}_{jk}(\xi_{i,jk}\mid{\cal S,A})\\&=\frac{I(A_i=j)}{\hat\pi_j(S_i)^2}\var_{jk}(Y_i^{(j)}\mid S_i)+ 2\frac{I(A_i=j)}{\hat\pi_j(S_i)}(1-\frac{I(A_i=j)}{\hat\pi_j(S_i)})\cov_{jk}(Y_i^{(j)},\mu_{jk}(\bmX_i)\mid S_i) \\
    &\quad + (1-\frac{I(A_i=j)}{\hat\pi_j(S_i)})^2\var_{jk}(\mu_{jk}(\bmX_i)\mid S_i)\\
    &=\frac{1}{\hat\pi_j(S_i)^2}\left[I(A_i=j)\var_{jk}(Y^{(j)}\mid S)+ 2I(A_i=j)\{\hat\pi_j(S_i)-I(A_i=j)\}\cov_{jk}(Y^{(j)},\mu_{jk}(\bmX)\mid S)\right.\\
    &\quad \left.+ \{\hat\pi_j(S_i)-I(A_i=j)\}^2\var_{jk}(\mu_{jk}(\bmX)\mid S) \right]\\
    &\leq \frac{1}{\hat\pi_j(S_i)^2}\left[\var_{jk}(Y^{(j)}\mid S)+\var_{jk}\{\mu_{jk}(\bmX)\mid S\}\right].
\end{align*}
Let $s_{n_{jk}}^2=\sum_{i\in{\cal I}_{jk}}\sigma_i^2$. We observe that conditional on ${\cal S,A}$, $$\max_{i\in{\cal I}_{jk}}\sigma_i^2/s_{n_{jk}}^2\leq \frac{\var_{jk}(Y^{(j)}\mid S)+\var_{jk}\{\mu_{jk}(\bmX)\mid S\}}{s_{n_{jk}}^2}\max_{i\in{\cal I}_{jk}} \frac{1}{\hat\pi_j(S_i)^2}=o(1)$$
because $Y^{(j)}$ and $\mu(\bmX)$ have finite second moment and $\hat\pi_j(S_i)$ is bounded from 0 as $H_{jk}<\infty$. Therefore,
\begin{align*}
    \frac{1}{s_{n_{jk}}^2}\sum_{i\in{\cal I}_{jk}}E_{jk}\left[\xi_{i,jk}^2I(|\xi_{i,jk}|>\epsilon s_n)\mid {\cal S,A}\right]&=\sum_{i\in{\cal I}_{jk}}\frac{\sigma_i^2}{s_{n_{jk}}^2}E_{jk}\left[\frac{\xi_{i,jk}^2}{\sigma_i^2}I(\frac{\xi_{i,jk}^2}{\sigma_i^2}>\frac{\epsilon^2s_n^2}{\sigma_i^2})\mid {\cal S,A}\right]\\
    &\leq \max_{i\in{\cal I}_{jk}}E_{jk}\left[\frac{\xi_{i,jk}^2}{\sigma_i^2}I(\frac{\xi_{i,jk}^2}{\sigma_i^2}>\frac{\epsilon^2s_n^2}{\sigma_i^2})\mid {\cal S,A}\right]\\
    &=o(1),
\end{align*}
following from that $\frac{\xi_{i,jk}^2}{\sigma_i^2}$ has zero expectation and unit variance and $\max_{i\in {\cal I}_{jk}} \sigma_i^2/s_{n_{jk}}^2=o(1)$.}
{

Consider the variances and covariance of $U_{jk}^a$ and $U_{kj}^a$. 
Note that 
\begin{align*}
    &\ \widetilde\var_{jk}\Bigg[\frac{1}{n_j({\cal I}_{jk}^{(h)})}\sum_{i\in{\cal I}_{jk}^{(h)}}I(A_i=j)\{Y_i^{(j)}-\mu_{jk}(\bmX_i)\} \, \Big| \, {\cal S, A} \Bigg]\\
    =&\ \frac{1}{\{n_j({\cal I}_{jk}^{(h)})\}^2}\sum_{i\in{\cal I}_{jk}^{(h)}}I(A_i=j)\var_{jk}\left\{Y_i^{(j)}-\mu_{jk}(\bmX_i)\mid S_i\right\}\\
    =& \ \frac{1}{n_j({\cal I}_{jk}^{(h)})}\var_{jk}\left\{Y^{(j)}-\mu_{jk}(\bmX)\mid S= \mathcal{I}_{jk}^{(h)}\right\},
\end{align*}
\begin{align*}
    \widetilde\var_{jk}(\Bar{\mu}_{jk}^{(h)} \mid {\cal S, A}) 
    &=\frac{1}{\{n({\cal I}_{jk}^{(h)})\}^2}\sum_{i\in {\cal I}_{jk}^{(h)}}\var_{jk}\{\mu_{jk}(\bmX_i)\mid S_i\}\\
    &=\frac{1}{n({\cal I}_{jk}^{(h)})}\var_{jk}\{\mu_{jk}(\bmX)\mid S= \mathcal{I}_{jk}^{(h)}\},
\end{align*}
and 
\begin{align*}
    &\ \widetilde\cov_{jk}\Bigg\{\frac{1}{n_j({\cal I}_{jk}^{(h)})}\sum_{i\in{\cal I}_{jk}^{(h)}}I(A_i=j)\{Y^{(j)}_i-\mu_{jk}(\bmX_i)\},\, \Bar{\mu}_{jk}^{(h)} \, \Big| \,  {\cal S,A}\Bigg\}\\
    =&\ \widetilde\cov_{jk}\Bigg\{\frac{1}{n_j({\cal I}_{jk}^{(h)})}\sum_{i\in{\cal I}_{jk}^{(h)}}I(A_i=j)(Y_i^{(j)}-\mu_{jk}(\bmX_i)),\, \frac{1}{n({\cal I}_{jk}^{(h)})}\sum_{i'\in {\cal I}_{jk}^{(h)}}\mu_{jk}(\bmX_{i'})\, \Big| \, {\cal S, A}\Bigg\}\\
    =&\ \frac{1}{n_j({\cal I}_{jk}^{(h)})n({\cal I}_{jk}^{(h)})}\sum_{i\in{\cal I}_{jk}^{(h)}}\sum_{i'\in {\cal I}_{jk}^{(h)}}I(A_i=j)\cov_{jk}\left\{Y_i^{(j)}-\mu_{jk}(\bmX_i),\mu_{jk}(\bmX_{i'})\mid S_i,S_{i'}\right\}\\
    =&\ \frac{1}{n({\cal I}_{jk}^{(h)})}\cov_{jk}\left\{Y^{(j)}-\mu_{jk}(\bmX),\, \mu_{jk}(\bmX)\mid S= \mathcal{I}_{jk}^{(h)}\right\}.
\end{align*}
Combining these results gives
\begin{align*}
n_{jk}\widetilde\var_{jk}(U_{jk}^a\mid {\cal S, A})
& = \frac{1}{n_{jk}}\sum_{h=1}^{H_{jk}}\{n({\cal I}_{jk}^{(h)})\}^2 \Bigg[ \widetilde\var_{jk}\Bigg\{\frac{1}{n_j({\cal I}_{jk}^{(h)})} \! \sum_{i\in{\cal I}_{jk}^{(h)}} \!\! I(A_i=j)\{Y_i^{(j)}\!-\!\mu_{jk}(\bmX_i)\} \, \Big| \, {\cal S, A}\Bigg\}  \\
& \quad + 2
\widetilde\cov_{jk}\Bigg\{\frac{1}{n_j({\cal I}_{jk}^{(h)})}\sum_{i\in{\cal I}_{jk}^{(h)}}I(A_i=j)\{Y^{(j)}_i-\mu_{jk}(\bmX_i)\},\, \Bar{\mu}_{jk}^{(h)} \, \Big| \,  {\cal S, A}\Bigg\} \\
& \quad +  \widetilde\var_{jk}(\Bar{\mu}_{jk}^{(h)} \mid {\cal S, A}) \Bigg]  \\
& = \frac{1}{n_{jk}}\sum_{h=1}^{H_{jk}} n({\cal I}_{jk}^{(h)})\Bigg[\frac{n({\cal I}_{jk}^{(h)})}{n_{j}({\cal I}_{jk}^{(h)}) }\var_{jk}\{Y^{(j)}-\mu_{jk}(\bmX)\mid S= \mathcal{I}_{jk}^{(h)} \} \\
& \quad +2 \cov_{jk}\{Y^{(j)}-\mu_{jk}(\bmX),\mu_{jk}(\bmX)\mid S= \mathcal{I}_{jk}^{(h)}\} \\
    & \quad + \var_{jk}\{\mu_{jk}(\bmX)\mid S= \mathcal{I}_{jk}^{(h)}\}
     \Bigg]
\end{align*}
$n_{jk}\widetilde\var_{jk}(U_{kj}^a\mid {\cal S, A})$ can be obtained by swapping $j$ and $k$. 
Let  $\Bar{\mu}_{jk}({\cal I}_{jk}^{(h)};a)$ be the sample mean of $\mu_{jk}(\bmX_i)$'s  for units in ${\cal I}_{jk}^{(h)}$ with $A_i = a$. 
From the same argument in the proof of part (e), 
$$ \widetilde\cov_{jk}\{\Bar{Y}_j({\cal I}_{jk}^{(h)})-\Bar{\mu}_{jk}({\cal I}_{jk}^{(h)};j),\Bar{Y}_k({\cal I}_{jk}^{(h)})-\Bar{\mu}_{kj}({\cal I}_{jk}^{(h)};k)\mid {\cal S, A}\} = 0.$$
In addition,
\begin{align*}
    & \ \widetilde\cov_{jk}\{\Bar{Y}_j({\cal I}_{jk}^{(h)})-\Bar{\mu}_{jk}({\cal I}_{jk}^{(h)};j),\Bar{\mu}_{kj}^{(h)}\mid {\cal S, A}\}\\
    =& \ \frac{1}{n_j({\cal I}_{jk}^{(h)})n({\cal I}_{jk}^{(h)})}\sum_{i\in{\cal I}_{jk}^{(h)}}\sum_{i'\in{\cal I}_{jk}^{(h)}}I(A_i=j)\cov_{jk}\{Y_i^{(j)}-\mu_{jk}(\bmX_i),\mu_{kj}(\bmX_{i'})\mid S_i,S_{i'}\}\\
    =&\ \frac{1}{n_j({\cal I}_{jk}^{(h)})n({\cal I}_{jk}^{(h)})}\sum_{i\in{\cal I}_{jk}^{(h)}}I(A_i=j)\cov_{jk}\{Y_i^{(j)}-\mu_{jk}(\bmX_i),\mu_{kj}(\bmX_{i})\mid S_i\}\\
    =&\ \frac{1}{n({\cal I}_{jk}^{(h)})}\cov_{jk}\{Y^{(j)}-\mu_{jk}(\bmX),\mu_{kj}(\bmX)\mid S={\cal I}_{jk}^{(h)}\},
\end{align*}
and
\begin{align*}
    \widetilde\cov_{jk}\{\Bar{\mu}_{jk}^{(h)},\Bar{\mu}_{kj}^{(h)} \mid {\cal S, A}\}&=\frac{1}{\{n({\cal I}_{jk}^{(h)})\}^2}\sum_{i\in{\cal I}_{jk}^{(h)}}\cov_{jk}\{\mu_{jk}(\bmX_i),\mu_{kj}(\bmX_i)\mid S_i\}\\
    &= \ \frac{1}{n({\cal I}_{jk}^{(h)})}\cov_{jk}\{\mu_{jk}(\bmX),\mu_{kj}(\bmX)\mid S={\cal I}_{jk}^{(h)}\}.
\end{align*}
Putting these together shows that
\begin{align*}
  & \  n_{jk}\widetilde\cov_{jk}(U_{jk}^a,U_{kj}^a \mid {\cal S, A})\\
=  &\ \frac{1}{n_{jk}}\sum_{h=1}^{H_{jk}}\{n({\cal I}_{jk}^{(h)})\}^2\Bigg[ \widetilde\cov_{jk}\{\Bar{Y}_j({\cal I}_{jk}^{(h)})-\Bar{\mu}_{jk}({\cal I}_{jk}^{(h)};j),\Bar{Y}_k({\cal I}_{jk}^{(h)})-\Bar{\mu}_{kj}({\cal I}_{jk}^{(h)};k)\mid {\cal S, A}\} \\
   &\ +\widetilde\cov_{jk}\{\Bar{Y}_j({\cal I}_{jk}^{(h)})-\Bar{\mu}_{jk}({\cal I}_{jk}^{(h)};j),\Bar{\mu}_{kj}^{(h)}\mid {\cal S, A}\} +\widetilde\cov_{jk}\{\Bar{Y}_k({\cal I}_{jk}^{(h)})-\Bar{\mu}_{kj}({\cal I}_{jk}^{(h)};k),\Bar{\mu}_{jk}^{(h)}\mid {\cal S, A}\}\\
    &\ + \left.\widetilde\cov_{jk}\{\Bar{\mu}_{jk}^{(h)},\Bar{\mu}_{kj}^{(h)}\mid {\cal S, A}\}\right\}\\
=& \ \frac{1}{n_{jk}} \sum_{h=1}^{H_{jk}}n({\cal I}_{jk}^{(h)}) \left[\cov_{jk}\left\{Y^{(j)}-\mu_{jk}(\bmX),\mu_{kj}(\bmX)\mid S={\cal I}_{jk}^{(h)} \right\}\right.\\
 &\quad + \, \cov_{jk}\left\{Y^{(k)}-\mu_{kj}(\bmX),\mu_{jk}(\bmX)\mid S={\cal I}_{jk}^{(h)}\right\} \\
    &\quad\left.+ \, \cov_{jk}\left\{\mu_{jk}(\bmX),\mu_{kj}(\bmX)\mid S={\cal I}_{jk}^{(h)} \right\} \right]\\
    =& \ \frac{1}{n_{jk}} \sum_{i \in {\cal I}_{jk} } c_{jk}(S_i) .
\end{align*}
The rest of  proof follows the proof of (e). 
 }

\subsection{Equivalence between IPW-type estimators and PS-type estimators}
We show that when $\bmZ$ is discrete, replacing the known
randomization probabilities in the IPW-type estimators by the sample proportions within each level of $S$ yields
the PS-type estimators exactly. 

We take AIPW estimator as an example. Replacing $\pi_j(\bmZ_i)$ with $\hat\pi_j(S_i)$ gives
\begin{align*}
&\ \frac{1}{n_{jk}}  \sum_{i \in {{{\cal I}}}_{jk}}  \frac{ I(A_i=j) \{Y_i - \hat \mu_{jk}(\bmX_i) \} }{\hat \pi_j(S_i)}   + \frac{1}{n_{jk}}  \sum_{i \in {{{\cal I}}}_{jk}}   \hat \mu_{jk}(\bmX_i)\\
&=\frac{1}{n_{jk}}  \sum_{h=1}^{H_{jk}}\sum_{i \in {{{\cal I}}}_{jk}^{(h)}}  \frac{ I(A_i=j) \{Y_i - \hat \mu_{jk}(\bmX_i) \} }{\hat \pi_j(S_i)}   + \frac{1}{n_{jk}}  \sum_{i \in {{{\cal I}}}_{jk}}   \hat \mu_{jk}(\bmX_i)\\
&=\frac{1}{n_{jk}}  \sum_{h=1}^{H_{jk}}\frac{n({\cal I}_{jk}^{(h)})}{n_j({\cal I}_{jk}^{(h)})}\sum_{i \in {{{\cal I}}}_{jk}^{(h)}} I(A_i=j) \{Y_i - \hat \mu_{jk}(\bmX_i) \}   + \frac{1}{n_{jk}}  \sum_{i \in {{{\cal I}}}_{jk}}   \hat \mu_{jk}(\bmX_i)\\
&=\hat\theta_{jk}^{\rm aps}.
\end{align*}
As noted in Section 4.4, the weights are
already normalized. Therefore, stabilization does not change the
estimators: SAIPW using the sample proportions also equals APS, while
SIPW using the sample proportions equals PS. The IPW result follows
from the calculation above by setting $\hat\mu_{jk}=0$. The same
arguments apply after interchanging $j$ and $k$.

\subsection{Derivation of Semiparametric Efficiency Lower Bound}
Suppose we are interested in comparing treatment effects between $j$ and $k$. Note that $\pi_j(\bmZ)=\pi_j(\bm W)$. Namely, let $\pi_j(\bm W)=p(j\mid \bm W)$ and $R=r(\bm W) = I(\pi_j(\bm W)>0,\pi_k(\bm W)>0)$ be a deterministic function of $\bm W$. Given two arms $j$ and $k$, the following proof rely on a study of the statistical functional
$$
\theta_{jk}=\Theta_{jk}(P)=E_P[Y^{(j)}\mid R=1]=E_P[\frac{I(A=j,R=1)Y}{\pi_j(W)}]/P(R=1).
$$
For a given distribution $P_0\in {\cal M}$, denote the density of $P_0$ with respect to some dominating measure $\nu$ by $p_0$. For bounded $g\in L_2(P_0)$, we define the parametric submodel $p_t=(1+tg)p_0$, which is valid for small enough $t$ and has score $g$ for $t$ at $t=0$. Given that the model $\cal M$ is nonparametric and some $\phi^*_{jk}\in L_2(P_0)$ with $P_0\phi^*_{jk}=0$, we will show that $\Theta_{jk}(P_0)$ is pathwise differentiable with respect to $\cal M$ at $P_0$ with efficient influence function (EIF) $\phi^*_{jk}$ following from that for any $P_0\in\mathcal M$, we have
\begin{align*}
    \frac{\partial}{\partial t}\Theta_{jk}(P_t)\big|_{t=0}=\langle \phi^*_{jk},g\rangle_{P_0},
\end{align*}
where $\langle\cdot,\cdot\rangle_{P_0}$ is the inner product in $L_2(P_0)$.

The statistical functional can be written as
$$
\theta_{jk,t}=\Theta_{jk}(P_t)=\int \frac{I(a=j,r(w)=1)y}{p_t(j\mid w)p_t(r(w)=1)}p_t(w,y,a)dwdyda,
$$
and we can see that
\begin{align*}
    \Theta_{jk}(P_t)-\Theta_{jk}(P_0)&=\int I(a=j,r=1)y\left\{\underbrace{\frac{p_t(w,y,a)}{p_t(j\mid w)p_t(R=1)}-\frac{p_t(w,y,a)}{p_0(j\mid w)p_t(R=1)}}_{u_1}\right.\\
    &+\underbrace{\frac{p_t(w,y,a)}{p_0(j\mid w)p_t(R=1)}-\frac{p_t(w,y,a)}{p_0(j\mid w)p_0(R=1)}}_{u_2}\\
    &+\left.\underbrace{\frac{p_t(w,y,a)}{p_0(j\mid w)p_0(R=1)}-\frac{p_0(w,y,a)}{p_0(j\mid w)p_0(R=1)}}_{u_3}\right\}dwdyda.
\end{align*}

\textbf{Analysis of $u_3$.} It is straightforward that
\begin{align*}
    u_3 = \int \frac{I(a=j,r=1)y}{p_0(j\mid w)p_0(R=1)}t p_0(w,y,a)g(w,y,a)dwdyda=t\langle \frac{I(a=j,r(w)=1)y}{p_0(j\mid w)p_0(R=1)}, g\rangle_{P_0}.
\end{align*}

\textbf{Analysis of $u_2$.} Some calculations show that
\begin{align*}
    p_t(R=1)&=p_t(r(W)=1)=\int I(r(w)=1)p_t(w)dw\\
    &=\int I(r(w)=1)p_t(w,y,a)dwdyda\\
    &=p_0(R=1)+t\int I(r(w)=1)p_0(w,y,a)g(w,y,a)dwdyda.
\end{align*}
It follows that
\begin{align*}
    \frac{1}{p_t(R=1)}=\frac{1}{p_0(R=1)}-t\int \frac{I(r=1)}{p_0(R=1)^2}p_0(w,y,a)g(w,y,a)dwdyda+o(t).
\end{align*}
Therefore, it turns out that
\begin{align*}
    u_2&=\int I(a=j,r=1)y\frac{p_0(w,y,a)+O(t)}{p_0(j\mid w)}(\frac{1}{p_t(R=1)}-\frac{1}{p_0(R=1)})dwdyda\\
    &=\theta_{jk,0}p_0(R=1)(\frac{1}{p_t(R=1)}-\frac{1}{p_0(R=1)}) + o(t)\\
    &=t\langle -\theta_{jk,0}\frac{I(r(w)=1)}{p_0(R=1)},g\rangle_{P_0} + o(t).
\end{align*}

\textbf{Analysis of $u_1$.} Some calculations show that
\begin{align*}
    &\quad \frac{1}{p_t(j\mid w)}=\frac{p_t(w)}{p_t(a=j,w)}=\frac{\int p_t(w,y',a')dy'da'}{\int p_t(a=j,w,y')dy'}\\
    &=\frac{p_0(w)+t\int p_0(w,y',a')g(w,y',a')dy'da'}{p_0(a=j,w)+t\int p_0(a=j,w,y')g(a=j,w,y')dy'}\\
    &=\frac{p_0(w)}{p_0(a=j, w)}+t\frac{\int p_0(w,y',a')g(w,y',a')dy'da'}{p_0(a=j,w)}-t\frac{p_0(w)\int p_0(a=j,w,y')g(a=j,w,y')dy'}{p_0(a=j,w)^2}+o(t)\\
    &=\frac{1}{p_0(j\mid w)}+t\frac{\int p_0(w,y',a')g(w,y',a')dy'da'}{p_0(a=j,w)}-t\frac{\int p_0(a=j,w,y')g(a=j,w,y')dy'}{p_0(j\mid w)p_0(a=j,w)}+o(t).
\end{align*}
We can rewrite $u_1$ as 
\begin{align*}
    u_1&=\int I(a=j,r=1)y(\frac{1}{p_t(j\mid w)}-\frac{1}{p_0(j\mid w)})(\frac{p_0(w,y,a)}{p_0(R=1)}+O(t))dwdyda\\
    &=t\int\frac{I(a=j,r=1)y}{p_0(a=j,w)p_0(R=1)}p_0(w,y,a)dyda\int p_0(w,y',a')g(w,y',a')dy'da'dw\\
    &\quad-t\int \frac{I(a=j,r=1)yp_0(w,y,a)}{p_0(j\mid w)p_0(a=j,w)p_0(R=1)}dyda\int p_0(a=j,w,y')g(a=j,w,y')dy'dw+o(t)\\
    &=t\int\frac{I(r=1)yp_0(y\mid w,a=j)}{p_0(R=1)}dy\int p_0(w,y',a')g(w,y',a')dy'da'dw\\
    &\quad-t\int \frac{I(r=1)yp_0(y\mid w,a=j)}{p_0(j\mid w)p_0(R=1)}dy\int p_0(a=j,w,y')g(a=j,w,y')dy'dw+o(t)\\
    &=t\int\frac{I(r=1)}{p_0(R=1)}\mu_{jk}(w) p_0(w,y',a')g(w,y',a')dy'da'dw\\
    &\quad-t\int \frac{I(r=1)}{p_0(j\mid w)p_0(R=1)}\mu_{jk}(w) p_0(a=j,w,y')g(a=j,w,y')dy'dw+o(t)\\
    &=t\langle \frac{I(r=1)}{p_0(R=1)}(\mu_{jk}(w)-\frac{I(a=j)}{p_0(j\mid w)}\mu_{jk}(w)),g\rangle_{P_0}+o(t),
\end{align*}
where $\mu_{jk}(w)=E[Y\mid \bm W=w,A=j, r(W)=1]$ and we used that $\mu_{jk}(w)=E[Y\mid \bm W=w,A=j]$ under $r(\bm W)=1$. 

Putting them together gives the following
\begin{align*}
    \lim_{t\rightarrow0}\frac{\Theta_{jk}(P_t)-\Theta_{jk}{(P_0)}}{t}=\langle \frac{I(r=1)}{p_0(R=1)}\left\{\frac{I(a=j)}{p_0(j\mid w)}(y-\mu_{jk}(w))+\mu_{jk}(w)-\theta_{jk,0}\right\},g\rangle_{P_0}.
\end{align*}
Therefore, the EIF $\phi_{jk}^*(w,y,a)$ in a nonparametric model $\cal M$ is
\begin{align*}
   \frac{I(r=1)}{p_0(R=1)}\left\{\frac{I(a=j)}{p_0(j\mid w)}(y-\mu_{jk}(w))+\mu_{jk}(w)-\theta_{jk,0}\right\}+c.
\end{align*}
It is obvious that $c=0$ since under consistency assumption,
$$P_0\phi_{jk}^*=E[E[\phi_{jk}^*(\bm W,Y,A)\mid \bm W]]=E[\frac{I(R=1)Y^{(j)}}{p_0(R=1)}]-\theta_{jk,0}+c=c=0.
$$
Therefore, the EIF in a nonparametric model $M$, where we have no constraints,  is
\begin{align*}
    \phi_{jk}^*(w,y,a)=\frac{I(r=1)}{p_0(R=1)}\left\{\frac{I(a=j)}{p_0(j\mid w)}(y-\mu_{jk}(w))+\mu_{jk}(w)-\theta_{jk,0}\right\},
\end{align*}
Writing $p(w,y,a)=p(y\mid w,a)p(a\mid w)p(w)$, we note that $\theta_{jk}$ does not depend on $p(a\mid w)$. Denote $M_0\subset M$ the submodel where the intervention assignment mechanism $p(a\mid w)$ is known. The tangent space $T_{M_0}(P)$ can be written as $L_{0,Y\mid W,A}^2(P)+L^2_{0,W}(P)$, where 
\begin{align*}
    L_0^2(P) &= \{s\in L_2(P): Ps=0\},\\
    L_{0,Y\mid W,A}^2(P)&=\{s\in L_0^2(P):E[s(\bm W,Y,A)\mid \bm W=w,A=a]=0 \text{ for all }w,a\},\\
    L_{0,W}^2(P)&=\{s\in L_0^2(P):s(w,y_1,a_1)=s(w,y_2,a_2) \text{ for all }y_1,y_2,a_1,a_2\}.
\end{align*}
As $\frac{I(r=1)}{p_0(R=1)}\left\{\frac{I(a=j)}{p_0(j\mid w)}(y-\mu_{jk}(w))\right\}\in L_{0,Y\mid W,A}^2(P)$ and $\frac{I(r=1)}{p_0(R=1)}(\mu_{jk}(w)-\theta_{jk,0})\in L_{0,W}^2(P)$, it yields $\phi_{jk}^*\in T_{M_0}(P)$. 

The semiparametric efficiency bound for $(\theta_{jk},\theta_{kj})$ is given by $\Sigma_{jk}^{\rm EIF}=\var((\phi^*_{jk}(\bm W,Y,A),\phi_{kj}^*(\bm W,Y,A))$. The first diagonal entry of $\Sigma_{jk}^{\rm EIF}$ is
\begin{align*}
    \var(\phi_{jk}^*(\bm W,Y,A))&=\var(E[\phi_{jk}^*(\bm W,Y,A)\mid \bm W,Y^{(j)}])+E[\var(\phi_{jk}^*(\bm W,Y,A)\mid \bm W,Y^{(j)})]\\
    &=\var(\frac{I(R=1)}{p(R=1)}(Y^{(j)}-\theta_{jk}))+E[\frac{I(R=1)}{p(R=1)^2}\frac{1-\pi_j(Z)}{\pi_j(Z)}(Y^{(j)}-\mu_{jk}(\bm W))^2]\\
    &=\frac{1}{p(R=1)}\left\{\var_{jk}(Y^{(j)})+E_{jk}[(\frac{1}{\pi_j(Z)}-1)(Y^{(j)}-\mu_{jk}(\bm W))^2]\right\}.
\end{align*}
The last diagonal entry can be obtained by swapping $j$ and $k$. The off-diagonal entry is that
\begin{align*}
    &\quad\cov(\phi^*_{jk}(\bm W,Y,A),\phi_{kj}^*(\bm W,Y,A))\\
    &=\cov\left\{E[\phi^*_{jk}\mid \bm W,Y^{(j)},Y^{(k)}],E[\phi^*_{kj}\mid \bm W,Y^{(j)},Y^{(k)}]\right\}+E[\cov(\phi^*_{jk},\phi_{kj}^*\mid \bm W,Y^{(j)},Y^{(k)})]\\
    &=\cov\left\{\frac{I(R=1)}{p(R=1)}(Y^{(j)}-\theta_{jk}),\frac{I(R=1)}{p(R=1)}(Y^{(k)}-\theta_{kj})\right\}\\
    &\quad + E[\frac{I(R=1)}{p(R=1)^2}(Y^{(j)}-\mu_{jk}(\bm W))(Y^{(k)}-\mu_{kj}(\bm W))\frac{\cov(I(A=j),I(A=k)\mid \bm W)}{\pi_j(Z)\pi_k(Z)}]\\
    &=\frac{1}{p(R=1)}\left\{\cov_{jk}(Y^{(j)},Y^{(k)})-E_{jk}[\varepsilon_{jk}\varepsilon_{kj}]\right\}.
\end{align*}
Since $\phi^*_{jk}\in L_2(P)$, we conclude that the functional $\Theta_{jk}(P)$ are pathwise differentiable at each $P\in {\cal M}_0$ relative to ${\cal M}_0$ with EIF $\phi^*_{jk}$. It also holds in model $\cal M$.

The EIF implies the following one-step estimator
\begin{align*}
    \hat\theta_{jk}=\frac{1}{n}\sum_{i=1}^n \frac{I(R_i=1)}{\hat p_R}\left\{\frac{I(A_i=j)}{\pi_j(Z_i)}(Y_i-\hat\mu_{jk}(\bm W_i))+\hat\mu_{jk}(\bm W_i)\right\},
\end{align*}
where $\hat p_R=\frac{1}{n}\sum_{i=1}^nI(R=1)$. This estimator coincides with the AIPW estimator when $\bm X=\bm W$. Moreover, recalling that $\eta_{jk}=\frac{1}{n_{jk}}\sum_{i\in {\cal I}_{jk}}\frac{I(A_i=j)}{\pi_j(Z_i)}$, the $\saipw$ is obtained by replacing $\pi_j(Z_i)$ with $\eta_{jk}\pi_j(Z_i)$ for each $i$ in ${\cal I}_{jk}$. By Corollary 2, $\saipw$ and $\aipw$ are asymptotically equivalent given { $\mu_{jk}(\bmX)=E_{jk}[I(A=j)Y/\pi_j(\bmZ)\mid \bm X]$ and $\mu_{kj}(\bmX)=E_{jk}[I(A=k)Y/\pi_k(\bmZ)\mid \bm X]$}. Therefore, $p(R=1)\Sigma_{jk}^{EIF}=\Sigma_{jk}^{\rm aipw}=\Sigma_{jk}^{\rm saipw}$. As $n_{jk}/n$ converges to $p(R=1)$ in probability, we conclude that $\aipw$ and $\saipw$ achieve the semiparametric efficiency lower bound provided that the efficient augmentation functions are $\bm X$-measurable, i.e., for $a=j$ and $k$,
\[
E(Y\mid \bm W,A=a,R=1)=E(Y\mid \bm X, A=a,R=1) \mbox{ almost surely.}
\]
More generally, if the working outcome regression models $\hat\mu_{jk}$ and $\hat\mu_{kj}$ are misspecified, the AIPW and SAIPW estimators remain consistent due to the randomization assumption, although they do not generally attain the semiparametric efficiency bound.

\subsection{Consistency of Variance Estimators}\label{supp: consistency of var estimator}
We demonstrate the consistency of the variance estimators $\hat\Sigma_{jk}^{\rm saipw}$ and $\hat\Sigma_{jk}^{\rm aps}$ by showing it converges in probability to $\Sigma_{jk}^{\rm saipw}$ and $\Sigma_{jk}^{\rm aps}$. The consistency of the remaining variance estimators follows analogously.

As shown in the proof of Theorem 1(c) in Section \ref{supp: proof of thm 1}, replacing the estimated model $\hat\mu_{jk}$ with its limit $\hat\mu_{jk}$ does not change the limits of $\hat\Sigma_{jk}^{\rm saipw}$ and $\hat\delta_{jk}$ in probability. Denote $\hat\delta_{jk}$ with $\hat\mu_{jk}$ replaced with $\mu_{jk}$ by $\bar\delta_{jk}$. Since $\bar\delta_{jk}$ conditional on $\{I_{jk}(Z_i),i=1,\dots,n\}$ is an average of independent and identically distributed random vectors, the convergence in probability conditional on $\{I_{jk}(Z_i),i=1,\dots,n\}$ is directly from the law of large number. Therefore, $\bar\delta_{jk}$ converges in probability to $\delta$, following from the bounded convergence theorem and
\begin{align*}
    \tilde{E}_{jk}[\bar\delta_{jk}]=E_{jk}[\frac{I(A=j)\{Y-\mu_{jk}(\bmX)\}}{\pi_j(Z)}]=\delta_{jk.}
\end{align*}

Let $\bar\Sigma_{jk}^{\rm saipw}$ be $\hat\Sigma_{jk}^{\rm saipw}$ with $\hat\mu_{jk}$ and $\hat\delta_{jk}$ replaced by $\mu_{jk}$ and $\delta_{jk}$. Therefore, $\hat\Sigma_{jk}^{\rm saipw}-\bar\Sigma_{jk}^{\rm saipw}$ converges in probability to 0 . Following the same steps, we can see that $\frac{1}{n_{jk}}\sum_{i\in{\cal I}_{jk}}\frac{I(A_i=j)\{Y_i-\mu_{jk}(\bmX_i)-\delta_{jk}\}^2}{\pi_j^2(Z_i)}$, $\hat q_{jk}^{(j)}$, $\hat \sigma_{jk}^2$, and $\hat q_{jk}$ converge in probability to $E_{jk}[\frac{I(A=j)\{Y-\mu_{jk}(\bmX)\}^2}{\pi_j^2(Z)}]$, $\cov_{jk}\{I(A=j)Y/\pi_j(\bmZ),\mu_{jk}(\bmX)\}$, $\var_{jk}(\mu_{jk}(\bmX))$, and $\cov_{jk}\{\mu_{jk}(\bmX),\mu_{kj}(\bmX)\}$, respectively.
Therefore, $\hat\Sigma_{jk}^{\rm saipw}$ converges in probability to $\Sigma_{jk}^{\rm saipw}$ by swapping $j$ and $k$ in the above derivation and applying Slutsky's lemma.

For $\hat\Sigma_{jk}^{\rm aps}$, we show the consistency of $\bar\Sigma_{jk}^{\rm aps}$, which is defined the same as $\bar\Sigma_{jk}^{\rm saipw}$ and additionally we replace $\hat\pi_j({\cal I}_{jk}^{(h)})$ with $\pi_j({\cal I}_{jk}^{(h)})$ because $\hat\pi_j({\cal I}_{jk}^{(h)}) = \pi_j({\cal I}_{jk}^{(h)})+o_p(1)$. Conditional on $\{I_{jk}(Z_i),i=1,\dots,n\},{\cal S},{\cal A}$, it can be seen $\bar q_{jk}^{(j)}({\cal I}_{jk}^{(h)})=q_{jk}^{(j)}({\cal I}_{jk}^{(h)})+o_p(1)$, and therefore, we have that
\begin{align*}
    \bar\Sigma_{jk}^{\rm aps}& = \sum_{h=1}^{H_{jk}}
  \frac{n({\cal I}_{jk}^{(h)})}{n_{jk}} \, \Big[ {\rm diag} \! \left\{ {\textstyle 
\frac{\tau_j^2 ( {\cal I}_{jk}^{(h)}) }{\pi_j({\cal I}_{jk}^{(h)})}  , \, 
\frac{\tau_k^2 ( {\cal I}_{jk}^{(h)}) }{\pi_k({\cal I}_{jk}^{(h)})}  } \right\} + {\bm\Lambda}_{jk}({\cal I}_{jk}^{(h)}) \Big] + {\bm\Gamma}_{jk} + o_p(1)\\
&=\frac{1}{n_{jk}}\sum_{h=1}^{H_{jk}}\sum_{i\in{\cal I}_{jk}^{(h)}}
   \, \Big[ {\rm diag} \! \left\{ {\textstyle 
\frac{\tau_j^2 ( S_i) }{\pi_j(S_i)}  , \, 
\frac{\tau_k^2 ( S_i) }{\pi_k(S_i)}  } \right\} + {\bm\Lambda}_{jk}(S_i) \Big] + {\bm\Gamma}_{jk} + o_p(1)\\
&=E_{jk}\Big[{\rm diag} \! \left\{ {\textstyle 
\frac{\tau_j^2 (S) }{\pi_j(S)}  , \, 
\frac{\tau_k^2 (S) }{\pi_k(S)}  } \right\} + {\bm\Lambda}_{jk}(S) \Big]+ {\bm\Gamma}_{jk} + o_p(1)\\
&=\Sigma_{jk}^{\rm aps} + o_p(1),
\end{align*}
where  $\tau_j^2({\cal I}_{jk}^{(h)})=\var_{jk}\{Y-\mu_{jk}(\bmX)\mid A=j, S={\cal I}_{jk}^{(h)}\}$, $\bm\Gamma_{jk}=\var_{jk}\{E_{jk}(Y\mid A=j, S), E_{jk}(Y\mid A=k,S)\}$, and we followed the steps in \ref{supp: proof of thm 1}.

\subsection{Proof of Corollary 1 and Related Results} 

Under the consistency and randomization assumptions, the difference between the first diagonal elements of $\bS_{jk}^{\rm ipw}$ and $\bS_{jk}^{\rm sipw}$ is
\begin{align*}
E_{jk} \left\{I(A=j) \frac{Y^2 - (Y-\theta_{jk})^2}{\pi_j^2(\bmZ)}\right\} - \theta_{jk}^2
     & = E_{jk} \left\{ \frac{(Y^{(j)})^2 - (Y^{(j)}-\theta_{jk})^2}{\pi_j(\bmZ)}\right\} - \theta_{jk}^2 \\
     & =  E_{jk} \left\{ \frac{2\theta_{jk}Y^{(j)}-\theta_{jk}^2}{\pi_j(\bmZ)}\right\} - \theta_{jk}^2
    \\
    & = 2 \theta_{jk} E_{jk} \left\{ \frac{Y^{(j)}-\theta_{jk}}{\pi_j(\bmZ)}\right\} + \theta_{jk}^2 E_{jk} \left\{ \frac{1}{\pi_j(\bmZ)}-1 \right\} \\
    & = 2 \theta_{jk} \cov_{jk} \left\{ Y^{(j)}, \frac{1}{\pi_j(\bmZ)}\right\} + \theta_{jk}^2 E_{jk} \left\{ \frac{1}{\pi_j(\bmZ)}-1 \right\} \\
    & = 2 \theta_{jk} \cov_{jk} \left\{ \frac{I(A=j)Y}{\pi_j(\bmZ)}, \frac{1}{\pi_j(\bmZ)}\right\} + \theta_{jk}^2 E_{jk} \left\{ \frac{1}{\pi_j(\bmZ)}-1 \right\}.
\end{align*} 
The difference between the second diagonal elements of $\bS_{jk}^{\rm ipw}$ and $\bS_{jk}^{\rm sipw}$ can be similarly obtained. It is clear that the off-diagonal element of $\bS_{jk}^{\rm ipw}- \bS_{jk}^{\rm sipw}$ is $- \theta_{jk}\theta_{kj}$. This proves Corollary 1.

From Corollary 1, $\bS_{jk}^{\rm \, ipw} - \bS_{jk}^{\rm \, sipw}$ is a sum of two matrices. The first matrix is always positive semidefinite because its diagonal elements are non-negative and its determinant is equal to
\begin{align*}
& \ \theta_{jk}^2\theta_{kj}^2\left[E_{jk} \left\{ \frac{1}{\pi_j(\bmZ)} - 1\right\}E_{jk} \left\{ \frac{1}{\pi_k(\bmZ)} - 1\right\}-1\right]\\
\geq &\ \theta_{jk}^2\theta_{kj}^2\left[E_{jk} \left\{ \frac{\pi_k(\bmZ)}{\pi_j(\bmZ)} \right\}E_{jk} \left\{ \frac{\pi_j(\bmZ)}{\pi_k(\bmZ)}\right\}-1\right]\\
\geq & \ 0,
\end{align*}
where the first inequality follows from $\pi_j(\bmZ)+\pi_k(\bmZ)\leq1$ almost surely and the second inequality results from Cauchy-Schwarz inequality. When $ \cov_{jk} \left\{ \frac{I(A=j)Y}{\pi_j(\bmZ)}, \frac{\theta_{jk}}{\pi_j(\bmZ)}\right\} $ and $\cov_{jk} \left\{ \frac{I(A=k)Y}{\pi_k(\bmZ)}, \frac{\theta_{kj}}{\pi_k(\bmZ)}\right\}  $ are both non-negative, 
the second matrix  is positive semidefinite and, hence, $\bS_{jk}^{\rm \, ipw} - \bS_{jk}^{\rm \, sipw}$ is positive semidefinite.

Finally, we construct an example under which IPW is asymptotically more efficient than SIPW for estimating $ \theta_{jk} - \theta_{kj}$. Suppose that $Z$ is binary with $P(Z=1)=0.5$ and $P(Z=0)=0.5$,  $E_{jk}(Y^{(j)}\mid \bmZ=1) = -6$,  $E_{jk}(Y^{(j)}\mid \bmZ=0) = 4$, and $\theta_{kj}=0$. Then, $\theta_{jk} = E_{jk}(Y^{(j)}) = -3+2=-1$. Also suppose that $\pi_j(Z=1) = 0.5$ and $\pi_j(Z=0)=1/3$. Then,
\[
E_{jk}\left\{\frac{1}{\pi_j(\bmZ)} - 1 \right\} = 0.5\times (2-1) + 0.5\times(3 - 1) =1.5, 
\]
$E_{jk}[Y^{(j)}/\pi_j(\bmZ)]=0.5\times (-12)+0.5\times 12=0$, and
\[
\cov_{jk}\left\{ Y^{(j)}, \frac{\theta_{jk}}{\pi_j(\bmZ)} \right\} =  - \theta_{jk}^2 E_{jk}\left\{  \frac{1}{\pi_j(\bmZ)} \right\} = -2.5. 
\]
Putting it together, we have 
\begin{align*}
    &(1,-1) (  \bS_{jk}^{\rm \, ipw} - \bS_{jk}^{\rm \, sipw} ) {1 \choose -1}  \\
    &=\theta_{jk}^2 E_{jk} \left\{ \frac{1}{\pi_j(\bmZ)} - 1\right\} + 2 \cov_{jk} \left\{ Y^{(j)}  , \frac{\theta_{jk}}{\pi_j(\bmZ)}\right\} +0 \\
    &= 1.5+ 2\times ( -2.5 ) \\
    &= -3.5< 0. 
\end{align*}

\subsection{Proof of Corollary 2} 

The proof of Corollary 2 is the same as the proof of Corollary 1, with $\theta_{jk}$ and $Y^{(j)}$ replaced by $\delta_{jk}$ and $Y^{(j)} - \mu_{jk}(\bmX)$, respectively.

\subsection{Proof of Corollary 3} 

Under condition (\ref{gura2}), 
$c_{jk} = \cov_{jk} \{ \mu_{jk}(\bmX),  \mu_{kj}(\bmX)\}$, which shows the result for the off-diagonal element of $  \bS_{jk}^{\rm \, sipw} - \bS_{jk}^{\rm \, saipw}$.
 Also, under (\ref{gura}), 
$ \lambda_{jk}  = \var_{jk} \{ \mu_{jk}(\bmX)\} 
= E_{jk} \{ \mu_{jk}(\bmX)-\mu_{jk}\}^2$. 
Note that $\theta_{jk} = \mu_{jk} + \delta_{jk}$ and $\pi_j(Z) = \pi_j(S)$. 
Hence, the first diagonal entry of  $\bS_{jk}^{\rm sipw} - \bS_{jk}^{\rm saipw}$ is 
\begin{align*}
 & \  E_{jk} \left[I(A=j) \frac{(Y-\theta_{jk})^2 - \{Y-\mu_{jk}(\bmX)-\delta_{jk}\}^2}{\pi_j^2(S)} \right]  - \lambda_{jk} \\
 = & E_{jk} \left[\frac{(Y^{(j)}-\theta_{jk})^2 - \{Y^{(j)}-\mu_{jk}(\bmX)-\delta_{jk}\}^2}{\pi_j(S)} \right]  - \lambda_{jk}\\
 = &  \ E_{jk} \left[ \frac{2 \{ \mu_{jk}(\bmX)-\mu_{jk}\} ( Y^{(j)}-\theta_{jk}) - \{ \mu_{jk}(\bmX)-\mu_{jk}\}^2 }{\pi_j(S)} \right]  - E_{jk} \{  \mu_{jk}(\bmX)-\mu_{jk}\}^2 \\
= & \  E_{jk} \left[ \frac{\{ \mu_{jk}(\bmX)-\mu_{jk}\}^2 }{\pi_j(S)} \right]  - E_{jk} \{  \mu_{jk}(\bmX)-\mu_{jk}\}^2 ,
\end{align*}
where the last equality follows from  that, under (\ref{U})-(\ref{gura}), 
\begin{align*}
    & \ E_{jk} \left[\{ \mu_{jk}(\bmX)-\mu_{jk}\} ( Y^{(j)}-\theta_{jk})\mid S\right]\\
    =&\ E_{jk} \left[\{ \mu_{jk}(\bmX)-\mu_{jk}\} ( Y^{(j)}-\mu_{jk}(\bmX)-\delta_{jk})\mid S \right]+E_{jk}\left[\{  \mu_{jk}(\bmX)-\mu_{jk}\}^2\mid S\right]\\
    =&\ E_{jk} \left\{\mu_{jk}(\bmX)-\mu_{jk}\mid S \right\} E_{jk}\left\{ Y^{(j)}-\mu_{jk}(\bmX)-\delta_{jk}\mid S \right\}+E_{jk}\left[\{  \mu_{jk}(\bmX)-\mu_{jk}\}^2\mid S \right]\\
    =&\ E_{jk}\left[\{  \mu_{jk}(\bmX)-\mu_{jk}\}^2\mid S\right]. 
\end{align*}
This proves the result for the first diagonal element of  $\bS_{jk}^{\rm sipw} - \bS_{jk}^{\rm saipw}$. The proof for the second diagonal element of  $\bS_{jk}^{\rm sipw} - \bS_{jk}^{\rm saipw}$ is the same. 

To prove that $\bS_{jk}^{\rm sipw} - \bS_{jk}^{\rm saipw}$ is positive definite or semidefinite,  note that its diagonal elements are non-negative and its determinant is equal to
\begin{align*}
    &\ E_{jk}\left[\left\{\mu_{jk}(\bmX)-\mu_{jk}\right\}^2\left\{\frac{1}{\pi_j(S)}-1\right\}\right]
    E_{jk}\left[\left\{\mu_{kj}(\bmX)-\mu_{kj}\right\}^2\left\{\frac{1}{\pi_k(S)}-1\right\}\right] \\
    &\quad\quad - \left[\cov_{jk}\left\{\mu_{jk}(\bmX),\mu_{kj}(\bmX)\right\}\right]^2\\
    \geq & \ E_{jk}\left[\left\{\mu_{jk}(\bmX)-\mu_{jk}\right\}^2\frac{\pi_k(S)}{\pi_j(S)}\right]E_{jk}\left[\left\{\mu_{kj}(\bmX)-\mu_{kj}\right\}^2\frac{\pi_j(S)}{\pi_k(S)}\right]\\
    &\quad\quad -\left\{\cov_{jk}\left\{\mu_{jk}(\bmX),\mu_{kj}(\bmX)\right\}\right\}^2\\
    \geq & \ \left[E_{jk}\left\{\mu_{jk}(\bmX)-\mu_{jk}\right\}\left\{\mu_{jk}(\bmX)-\mu_{jk}\right\}\right]^2-\left\{\cov_{jk}\left\{\mu_{jk}(\bmX),\mu_{kj}(\bmX)\right\}\right\}^2\\
    = & \ 0,
\end{align*}
where the first inequality follows from $\pi_j(S)+\pi_k(S)\leq1$ and the second inequality results from the Cauchy-Schwarz inequality. Hence, $\bS_{jk}^{\rm sipw} - \bS_{jk}^{\rm saipw}$ is positive definite and is positive semidefinite if and only if the previous two inequalities are equalities, which are true if and only if %
either one of  $\var_{jk}\{ \mu_{kj}(\bmX)\}$ and $\var_{jk}\{ \mu_{kj}(\bmX)\}$ is 0 or $\pi_j(S)+\pi_k(S)=1$ and
the correlation between $\mu_{jk}(\bmX)$ and $\mu_{kj}(\bmX)$ is $\pm 1$. 
This completes the proof of Corollary 3. 

\subsection{Proof of Corollary 4}
 Under conditions \eqref{U} and \eqref{gura2}, $E_{jk}[c_{jk}(S)]=E_{jk}[\cov_{jk}\{\mu_{jk}(\bmX),\mu_{kj}(\bmX)\mid S\}]$, which shows the result for the off-diagonal element of $  \bS_{jk}^{\rm \, ps} - \bS_{jk}^{\rm \, aps}$. Also, under condition \eqref{gura}, $\lambda_{jk}(S)=\var_{jk}\{\mu_{jk}(\bmX)\mid S\}$. Hence, the first diagonal entry of $\bS_{jk}^{\rm ps}-\bS_{jk}^{\rm aps}$ is
 \begin{align*}
     &\ E_{jk}\left[\frac{\var_{jk}(Y\mid S,A=j)-\var_{jk}(Y-\mu_{jk}(\bmX)\mid S, A=j)}{\pi_j(S)}-\lambda_{jk}(S)\right]\\
     =&\ E_{jk}\left[\left\{\frac{1}{\pi_j(S)}-1\right\}\var_{jk}\left\{\mu_{jk}(\bmX)\mid S\right\}\right].
 \end{align*}
This proves the result for the first diagonal element of  $\bS_{jk}^{\rm ps} - \bS_{jk}^{\rm aps}$. The proof for the second diagonal element of  $\bS_{jk}^{\rm ps} - \bS_{jk}^{\rm aps}$ is the same. 

To prove that $\bS_{jk}^{\rm ps} - \bS_{jk}^{\rm aps}$ is positive definite or semidefinite, note that its diagonal elements are non-negative and  its determinant is equal to
\begin{align*}
    &\ E_{jk}\left[\left\{\frac{1}{\pi_j(S)}-1\right\}\var_{jk}\left\{\mu_{jk}(\bmX)\mid S\right\}\right]E_{jk}\left[\left\{\frac{1}{\pi_k(S)}-1\right\}\var_{jk}\left\{\mu_{kj}(\bmX)\mid S\right\}\right]\\
    &\quad\quad-\left\{E_{jk}\left[\cov_{jk} \left\{ \mu_{jk} (\bmX), \mu_{kj} (\bmX) \mid S \right\}\right]\right\}^2\\
    &\geq E_{jk}\left[\frac{\pi_k(S)}{\pi_j(S)}\var_{jk}\left\{\mu_{jk}(\bmX)\mid S\right\}\right]E_{jk}\left[\frac{\pi_j(S)}{\pi_k(S)}\var_{jk}\left\{\mu_{kj}(\bmX)\mid S\right\}\right]\\
    &\quad\quad-\left\{E_{jk}\left[\cov_{jk} \left\{ \mu_{jk} (\bmX), \mu_{kj} (\bmX) \mid S \right\}\right]\right\}^2\\
    &\geq \left\{E_{jk}\left[\sqrt{\var_{jk}\{\mu_{jk}(\bmX)\mid S\}\var_{jk}\{\mu_{kj}(\bmX)\mid S\}}\right]\right\}^2-\left\{E_{jk}\left[\cov_{jk} \left\{ \mu_{jk} (\bmX), \mu_{kj} (\bmX) \mid S \right\}\right]\right\}^2\\
    &\geq \left\{E_{jk}\left|\cov_{jk}\{\mu_{jk}(\bmX),\mu_{kj}(\bmX)\mid S\}\right|\right\}^2-\left\{E_{jk}\left[\cov_{jk} \left\{ \mu_{jk} (\bmX), \mu_{kj} (\bmX) \mid S \right\}\right]\right\}^2\\
    &\geq 0,
\end{align*}
where the first inequality follows from $\pi_j(S)+\pi_k(S)\leq 1$  and the second to the third inequalities result from the Cauchy-Schwarz inequality. Hence, $\bS_{jk}^{\rm ps}-\bS_{jk}^{\rm aps}$ is positive definite and is semidefinite if and only if 
either one of $\var_{jk}\{ \mu_{kj}(\bmX)\mid S\}$ and $\var_{jk}\{ \mu_{kj}(\bmX)\mid S\}$ is 0 or $\pi_j(S)+\pi_k(S)=1$ and
the correlation between $\mu_{jk}(\bmX)$ and $\mu_{kj}(\bmX)$  condtioned on $S$ is $\pm 1$. 
This completes the proof of Corollary 4.

\subsection{More Discussions on Conditions \eqref{U}-\eqref{gura2}}
We show that conditions (\ref{U})-(\ref{gura2}) hold when applying a linear ANHECOVA working model that includes $S$ and its interactions with $\bmX$ \citep{ye2021better}. Specifically, denote the obtained model limit as $\mu_{jk}(\bmX) = \sum_{h=1}^{H_{jk}}  I(S=\mathcal{I}_{jk}^{(h)}) \{\alpha_{jk} (h) + \gamma_{jk}(h)^T\bmX\} $, where $(\alpha_{jk}(h), \gamma_{jk}(h)^T)$ is the probability limit of the coefficient vector from the ANHECOVA working model. The population score equations ensure that $\delta_{jk}=0$ and  $ E_{jk} [  \{Y^{(j)} - \mu_{jk}(\bmX) \} I(S= {\cal I}_{jk}^{(h)})] =0$ for all $h$ and, thus, $ E_{jk} \{ Y^{(j)} - \mu_{jk}(\bmX)  \mid S\} =0$, 
\begin{align*}
   E_{jk} [ \{ Y^{(j)} - \mu_{jk}(\bmX) \} \mu_{kj}(\bmX)  ]
   &=     \sum_{h=1}^{H_{jk}} E_{jk} [ \{ Y^{(j)} - \mu_{jk}(\bmX) \} I(S= {\cal I}_{jk}^{(h)})  ]  \alpha_{kj}(h)\\
   & \quad +   \sum_{h=1}^{H_{jk}} E_{jk} [ \{ Y^{(j)} - \mu_{jk}(\bmX) \} I(S= {\cal I}_{jk}^{(h)}) \bmX^T  ]  \gamma_{kj}(h)  \\
   & = 0,
\end{align*}
 and $ E_{jk} [ \{ Y^{(j)} - \mu_{jk}(\bmX) \} \mu_{jk}(\bmX) ] = 0$ for the same reason.    Lastly, 
 \begin{align*}
     & \  E_{jk} [ \{ Y^{(j)} - \mu_{jk}(\bmX) \} \mu_{jk}(\bmX)  I(S= {\cal I}_{jk}^{(h)}) ] \\
     = & \        E_{jk} [ \{ Y^{(j)} - \mu_{jk}(\bmX) \} I(S= {\cal I}_{jk}^{(h)}) \{\alpha_{jk}(h) + \gamma_{jk} (h)^T \bmX  \}] \\
     = & \ 0
 \end{align*}
 for all $h$ and, thus, $        E_{jk} [ \{ Y^{(j)} - \mu_{jk}(\bmX) \} \mu_{jk}(\bmX)  \mid S]  =0$. These results also hold if swapping $j$ and $k$. Thus, this proves that conditions \eqref{U}-\eqref{gura2} hold.

        The joint calibration strategy by \cite{bannick2023general} can be proved in the same way.

\subsection{Proof of Corollary 5} 

Consider SAIPW with $\bmX = S$ and 
 $\mu_{jk} (S) = E_{jk}(Y\mid S,\, A=j )$. Under the consistency and randomization assumptions, $\mu_{jk} (S) = E_{jk}(Y^{(j)}\mid S)$.
Since 
\begin{align*}
    \cov_{jk} \{ Y^{(j)} , E_{jk}(Y^{(k)} \mid S)\} & = E_{jk} \{ Y^{(j)}  E_{jk}(Y^{(k)} \mid S )\} - E_{jk} ( Y^{(j)}) E_{jk} \{ E_{jk}(Y^{(k)} \mid S)\} \\
   & = E_{jk} \{ E_{jk}(Y^{(j)}\mid S)  E_{jk}(Y^{(k)} \mid S)\} - \theta_{jk} \theta_{kj} \\
   & = \cov_{jk} \{ Y^{(k)} , E_{jk}(Y^{(j)} \mid S )\} \\
   & = \cov_{jk} \{ E_{jk}(Y^{(j)} \mid S ) ,  E_{jk}(Y^{(k)} \mid S ) \} ,
\end{align*}
we obtain that $c_{jk} = \cov_{jk} \{ E_{jk}(Y^{(j)} \mid S) ,  E_{jk}(Y^{(k)} \mid S) \}$, i.e., the off-diagonal elements of $\bS_{jk}^{\rm ps}$ and $\bS_{jk}^{\rm saipw} $ are the same. In this special case, $\delta_{jk}=0$ and $\lambda_{jk} = \var_{jk} \{ E_{jk}(Y^{(j)}\mid S ) \}$. Then the  first diagonal entry of $\bS_{jk}^{\rm ps}- \bS_{jk}^{\rm saipw}$ is
\begin{align*}
   E_{jk} \left\{ \frac{ \var_{jk} (Y^{(j)}\mid S ) }{\pi_j(S )} \right\} 
-E_{jk} \left[ \frac{\{Y^{(j)}-\mu_{jk}(S)\}^2 }{\pi_j(S )}\right] = 0. 
\end{align*} 
Similarly, 
the last diagonal entry of $\bS_{jk}^{\rm ps}- \bS_{jk}^{\rm saipw}$ is 0. Hence, $\bS_{jk}^{\rm ps}= \bS_{jk}^{\rm saipw}$.

\subsection{Proof of Corollary 6} 

Under condition (\ref{U}), 
\begin{align*}%
E_{jk} \left[ \frac{\{Y^{(j)}-\mu_{jk}(\bmX)-\delta_{jk}\}^2}{\pi_j(S)} \right]
& = E_{jk}  \left\{ E_{jk}\left[ \frac{\{Y^{(j)}-\mu_{jk}(\bmX)-\delta_{jk}\}^2}{\pi_j(S)} \, \Big| \, S \right] \right\} \\
& =  E_{jk} \left[ \frac{\var_{jk}\{Y^{(j)}-\mu_{jk}(\bmX) \mid S \}}{\pi_j(S)}   \right] 
\end{align*}
and 
\begin{align*}
    \lambda_{jk} & = \var_{jk} (Y^{(j)}) - \var_{jk} \{ Y^{(j)}-\mu_{jk}(\bmX)\} \\    
    & = E_{jk} \{ \var_{jk} (Y^{(j)} \mid S )\}
    + \var_{jk} \{ E_{jk} (Y^{(j)} \mid S )\} \\
    & \quad - E_{jk} [ \var_{jk} \{ Y^{(j)}-\mu_{jk}(\bmX)\mid S \} ] - \var_{jk} [ E_{jk}\{ Y^{(j)}-\mu_{jk}(\bmX)\mid S \} ]\\   
    & = E_{jk} \{ \var_{jk} (Y^{(j)} \mid S )\}
    + \var_{jk} \{ E_{jk} (Y^{(j)} \mid S )\} 
    - E_{jk} [ \var_{jk} \{ Y^{(j)}-\mu_{jk}(\bmX)\mid S \} ]\\
  &  =  E_{jk} \{ \lambda_{jk}(S) \} + \var_{jk} \{ E_{jk} (Y^{(j)} \mid S )\}.
\end{align*} 
This shows that the first diagonal elements of 
$\bS_{jk}^{\rm \, aps}$ and $\bS_{jk}^{\rm \, saipw}$ are the same. We can similarly show that the second  diagonal elements of 
$\bS_{jk}^{\rm \, aps}$ and $\bS_{jk}^{\rm \, saipw}$ are the same. For the off-diagonal entry, 
\begin{align*}
    c_{jk} & = \cov_{jk}\{Y^{(j)}, \mu_{kj}(\bmX)\} + 
    \cov_{jk}\{Y^{(k)}, \mu_{jk}(\bmX)\} - \cov_{jk} \{ \mu_{jk}(\bmX), \mu_{kj}(\bmX)\} \\
    & = \cov_{jk}\{Y^{(j)}-\mu_{jk}(\bmX), \mu_{kj}(\bmX)\} + 
    \cov_{jk}\{Y^{(k)}-\mu_{kj}(\bmX), \mu_{jk}(\bmX)\} + \cov_{jk} \{ \mu_{jk}(\bmX), \mu_{kj}(\bmX)\} \\
    & = E_{jk} [ \cov_{jk}\{Y^{(j)}- \mu_{jk}(\bmX), \mu_{kj}(\bmX) \mid S \}] + 
    E_{jk}[ \cov_{jk}\{Y^{(k)}-\mu_{kj}(\bmX) , \mu_{jk}(\bmX) \mid S \}]\\
    & \quad + E_{jk}[ \cov_{jk} \{ \mu_{jk}(\bmX), \mu_{kj}(\bmX) \mid S \} ]  + \cov_{jk}[ E_{jk}\{ \mu_{jk}(\bmX) \mid S \} , E_{jk} \{ \mu_{kj}(\bmX) \mid S \}]  \\
       & = E_{jk} [ \cov_{jk}\{Y^{(j)}, \mu_{kj}(\bmX) \mid S \}] + 
    E_{jk}[ \cov_{jk}\{Y^{(k)}, \mu_{jk}(\bmX) \mid S \}]\\
    & \quad - E_{jk}[ \cov_{jk} \{ \mu_{jk}(\bmX), \mu_{kj}(\bmX) \mid S \} ]  + \cov_{jk}\{ E_{jk}( Y^{(j)} \mid S ) , E_{jk}(Y^{(k)}\mid S )\} \\
    & = E_{jk} \{ c_{jk}(S) \} +  \cov_{jk}\{ E_{jk}( Y^{(j)} \mid S ) , E_{jk}(Y^{(k)}\mid S )\} ,
\end{align*}
where the third and fourth equations follow from (\ref{U}). 
This proves the result.

\subsection{Comparison between SAIPW with $\bmX \supset S$ and  SAIPW with $\bmX = S$} 

Under (\ref{U})-(\ref{gura2}), asymptotically, 
SAIPW is equivalent to APS (Corollary 6), APS is more efficient than PS (Corollary 4),  PS is equivalent to SAIPW with $\bmX = S$ (Corollary 5) and, therefore,  SAIPW is more efficient than SAIPW with $\bmX = S$.

\spacingset{1.2}
\bibliographystyle{apalike}
\bibliography{reference}